\documentclass[twocolumn]{aastex631}
\usepackage{amsmath}
\usepackage{threeparttable}

\newcommand{\targ}{COOL\,J1153$+$0755}
\newcommand{\targA}{COOL\,J1153A}
\newcommand{\targB}{COOL\,J1153B}

\newcommand{\HII}{H{\sc ~ii}}
\newcommand{\Halpha}{H$\alpha$}
\newcommand{\Hbeta}{H$\beta$}
\newcommand{\NII}{N{\sc ~ii}}

\newcommand{\OIII}{O{\sc ~iii}}

\definecolor{mycustomcolor}{HTML}{198f8b}

\newcommand{\sarcs}{.\!\!^{\prime\prime}}

\vspace{-15pt}
\submitjournal{ApJ}
\vspace{-5pt}

\shorttitle{COOL-LAMPS IX}
\shortauthors{Solhaug et al.}

\begin{document}

\title{COOL-LAMPS IX: A Rare Duo of Quasars Each Lensed by a Single Massive Galaxy Cluster}

\correspondingauthor{Erik Solhaug}
\email{eriksolhaug@uchicago.edu}

\author[0000-0003-3302-0369]{Erik Solhaug}
\affiliation{Department of Astronomy and Astrophysics, University of Chicago, 5640 South Ellis Avenue, Chicago, IL 60637, USA}
\affiliation{Kavli Institute for Cosmological Physics, University of Chicago, 5640 South Ellis Avenue, Chicago, IL 60637, USA}
\author[0000-0003-1370-5010]{Michael D. Gladders}
\affiliation{Department of Astronomy and Astrophysics, University of Chicago, 5640 South Ellis Avenue, Chicago, IL 60637, USA}
\affiliation{Kavli Institute for Cosmological Physics, University of Chicago, 5640 South Ellis Avenue, Chicago, IL 60637, USA}
\author[0009-0006-7664-877X]{Andi M. Kisare}
\affiliation{Department of Astronomy and Astrophysics, University of Chicago, 5640 South Ellis Avenue, Chicago, IL 60637, USA}
\author[0000-0002-5573-9131]{Simon D. Mork}
\affiliation{School of Earth and Space Exploration, Arizona State University, 781 Terrace Mall, Tempe, AZ 85287, USA}
\author[0000-0003-1074-4807]{Matthew B. Bayliss}\affiliation{Department of Physics, University of Cincinnati, Cincinnati, OH 45221, USA}
\author[0000-0001-9978-2601]{Aidan P. Cloonan}
\affiliation{Department of Astronomy, University of Massachusetts, 710 North Pleasant Street, Amherst, MA 01003-9305, USA}
\author[0000-0003-2200-5606]{H\r{a}kon Dahle}
\affiliation{Institute of Theoretical Astrophysics, University of Oslo, P.O. Box 1029, Blindern, NO-0315 Oslo, Norway}
\author[0000-0003-0896-8502]{Isaiah R. Escapa}
\affiliation{Department of Astronomy and Astrophysics, University of Chicago, 5640 South Ellis Avenue, Chicago, IL 60637, USA}
\author[0000-0001-5097-6755]{Michael K. Florian}\affiliation{Steward Observatory, University of Arizona, 933 North Cherry Ave., Tucson, AZ 85721, USA}
\author[0000-0002-3475-7648]{Gourav Khullar}
\affiliation{Department of Astronomy, University of Washington, Physics-Astronomy Building, Box 351580, Seattle, WA 98195-1700, USA}
\affiliation{eScience Institute, University of Washington, Physics-Astronomy Building, Box 351580, Seattle, WA 98195-1700, USA}
\author[0000-0003-3266-2001]{Guillaume Mahler}\affiliation{STAR Institute, Quartier Agora - All\'ee du six Ao\^ut, 19c B-4000 Li\`ege, Belgium}
\author[0000-0002-5825-7795]{Natalie Malagon}
\affiliation{Department of Astronomy and Astrophysics, University of Chicago, 5640 South Ellis Avenue, Chicago, IL 60637, USA}
\author[0000-0003-4470-1696]{Kate Napier}\affiliation{Kavli Institute for Particle Astrophysics and Cosmology, Department of Physics, Stanford University, Stanford, CA 94309, USA}
\affiliation{SLAC National Accelerator Laboratory, Menlo Park, CA 94025, USA}
\author[0000-0003-1832-4137]{Allison Noble}
\affiliation{School of Earth and Space Exploration, Arizona State University, 781 Terrace Mall, Tempe, AZ 85287, USA}
\author[0000-0002-7627-6551]{Jane R. Rigby}
\affiliation{Astrophysics Science Division, Code 660, NASA Goddard Space Flight Center, 8800 Greenbelt Road, Greenbelt, MD 20771, USA}
\author[0000-0001-7905-2134]{Riley Rosener}
\affiliation{Department of Astronomy and Astrophysics, University of Chicago, 5640 South Ellis Avenue, Chicago, IL 60637, USA}
\author[0000-0002-9204-3256]{T. Emil Rivera-Thorsen}\affiliation{The Oskar Klein Centre, Department of Astronomy, Stockholm University, AlbaNova, 10691 Stockholm, Sweden}
\author[0000-0002-7559-0864]{Keren Sharon}
\affiliation{Department of Astronomy, University of Michigan, 1085 S. University Ave, Ann Arbor, MI 48109, USA}
\author[0000-0002-2718-9996]{Antony A. Stark}
\affiliation{Center for Astrophysics | Harvard \& Smithsonian, 60 Garden St, Cambridge, MA 02138, USA}
\author[0009-0008-6557-2065]{Kabelo Tsiane}
\affiliation{Department of Astronomy and Astrophysics, University of Chicago, 5640 South Ellis Avenue, Chicago, IL 60637, USA}
\author[0000-0003-0295-875X]{Grace C. Wagner}
\affiliation{Department of Astronomy and Astrophysics, University of Chicago, 5640 South Ellis Avenue, Chicago, IL 60637, USA}
\author[0000-0003-1815-0114]{Brian Welch}
\affiliation{International Space Science Institute, Hallerstrasse 6, 3012 Bern, Switzerland}
\author[0009-0006-4143-1159]{Yifan ``Megan'' Zhao}
\affiliation{Department of Astronomy and Astrophysics, University of Chicago, 5640 South Ellis Avenue, Chicago, IL 60637, USA}





\begin{abstract}

Wide-separation lensed quasars (WSLQs) are rare systems that arise from the chance alignment of two objects: a galaxy cluster and a background quasar. After two decades, only seven WSLQs have been found. Here, we report the discovery of \targ\ by the COOL-LAMPS collaboration in DECaLS imaging and its confirmation with follow-up observations with the Magellan Telescopes and the Nordic Optical Telescope. This system features \textit{two} multiply-imaged quasars each lensed into four images by the same $z=0.4301$ cluster: a classic broad-line Type I quasar at $z=1.524$ (\targA) and a dust-obscured Type II quasar at $z=1.939$ (\targB), with maximum image separations of $25\sarcs6$ and $26\sarcs0$,  respectively. We construct a lens model to estimate a projected cluster mass of $\,\,\,\,\,\,\,\,\,\,\,\,\,\,\,M(<500\,{\rm kpc})\sim3.3\times10^{14}{\rm M}_{\odot}$ and relative time delays between the three brightest images of each quasar of $\Delta t_{\rm \,A3,A1}\sim800$, $\Delta t_{\rm \,A2,A1}\sim1200$, $\Delta t_{\rm \,B1,B3}\sim800$, and $\Delta t_{\rm \,B2,B3}\sim1000$ days. \targA\ resides in a dense environment with three nearby galaxies, two of which are also strongly lensed. We identify \targ\ without making a morphological cut in the DECaLS catalog; none of its multiple images are classified as point sources in those data, implying that morphology-based selection would miss such systems. \targ\ expands the WSLQ sample from 7 to 8 systems (9 individual quasars), adding two powerful laboratories for probing black hole-galaxy co-evolution at Cosmic Noon and for time-delay cosmography constraints on the Hubble constant, $H_0$.

\end{abstract}

\vspace{-50pt}


\vspace{5mm}

\section{Introduction} \label{sec:intro}

Wide-separation lensed quasars (WSLQs) are an exceptionally rare subset of lensed quasars, in which a foreground lens -- a massive galaxy cluster -- produces lensed image separations significantly larger than $\sim 10$ arcseconds. After 20 years of searching, seven WSLQs have been found. These are SDSS\,J1004$+$4112 \citep[][]{Inada:2003}, SDSS\,J1029$+$2623 \citep[][]{Inada:2006}, SDSS\,J2222$+$2745 \citep[][]{Dahle:2013}, SDSS\,J0909$+$4449 \citep[][]{Shu:2018}, SDSS\,J1326$+$4806 \citep[][]{Shu:2019}, COOL\,J0542$-$2125 \citep[][]{Martinez:2023}, and COOL\,J0335$-$1957 \citep[][]{Napier:2023a}.

To illustrate just how rare WSLQs are, consider that among the $\sim 10^6$ quasars known \citep[][]{Flesch:2023}, about $\sim 300$ are gravitationally lensed into multiple  images \citep[see e.g.,][]{Anguita:2018, Lemon:2019, Lemon:2023, Dux:2024}, and only seven of these are WSLQs. While WSLQ systems generally have a characteristic $\gtrsim10''$ separation between images, the quality that unambiguously separates them from other lensed quasars is the presence of a foreground galaxy cluster acting as the gravitational lens. The large image separations are hence due to the deep gravitational potentials of foreground cluster-scale halos whose typical masses exceed $\sim 10^{14} \, \mathrm{M}_{\odot}$.

These WSLQ systems are exceptionally valuable. Brightness variations of multiple images of the same quasar provide a means to independently measure the Hubble constant, $H_0$, through strong lensing time delays using the Refsdal method \citep[][]{Refsdal:1964, Wong:2020}. Time delays of WSLQs have the potential to reach a 1\% precision on $H_0$ with a sufficient number of image pairs ($\gtrsim 100$) and accurate lens models of the foreground clusters \citep[see e.g.,][]{Napier:2023b}. Additionally, WSLQs benefit from having long time delays, which translate into much smaller fractional uncertainties on the time delays once measured to a typical absolute uncertainty of a few days. This comes with the caveat that cluster lenses are more complex than galaxy lenses, which typically require more complex lens models than their galaxy-scale counterparts. At present, the seven known WSLQs provide 14 ground-accessible image pairs -- a conservative count, as some systems have more images that could be monitored with improved imaging. With the discovery of additional WSLQs, these systems offer a promising, independent avenue for measuring $H_0$ via time-delay cosmography \citep[][]{Suyu:2017, Wong:2020}, and may provide a meaningful path toward resolving the persistent Hubble tension between early- and late-Universe measurements \citep[][]{Planck:2020, Riess:2022}.

Deep potential wells of the foreground cluster create strong gravitational lensing that magnifies the gas and stars in the quasar's host environment. Since the quasar appears as a point source while the host galaxy's angular size is increased by lensing, the host galaxy's light can more readily be separated from that of the bright quasar. For galaxy-scale lensed quasars, this is a greater challenge as the smaller image separations can lead to substantial contamination from the foreground lens, and the magnification is generally smaller than found in larger-separation lenses. WSLQs circumvent these limitations, enabling spatially resolved measurements of the stellar mass–black hole mass relation at high redshift, which is critical for understanding active galactic nuclei (AGN) feedback, galaxy evolution, and the baryon cycle \citep[e.g.,][]{Ross:2009, Cloonan:2024}. The high magnification of the host galaxy enables studies into the complex escape of Ly$\alpha$ and ionizing photon escape from quasars through the surrounding cloudy neutral hydrogen in the interstellar medium beyond the nearby universe \citep[][who measured the spatial extent of the Ly$\alpha$ halo of the WSLQ SDSS J2222+2745 at $z = 2.801$]{Bayliss:2017}. In that system, the $\sim 700$ days time delay between the quasar images allowed \cite{Williams:2021,Williams:2021b} to perform reverberation mapping of the two brightest lensed images of the WSLQ, providing the best-constrained black hole mass beyond the nearby universe.

The wide spatial separations of WSLQ images, typically $10-50 \, {\rm kpc}$ in the lens plane \citep{Stark:2013, Sharon:2017, Napier:2023a, Dutta:2024}, provide multiple sightlines through the extended gaseous envelopes of foreground galaxies that probe a broad range of impact parameters. With sufficient physical separation between the sightlines, the spatial coherence and kinematics of outflows into the circumgalactic medium of the quasar host itself can be measured \citep[][]{Misawa:2013, Misawa:2014, Misawa:2016}. Observations like these offer a window into the galactic baryon cycle and feedback at a range of scales.

WSLQs are significantly underrepresented in current samples. 
Simulations show that for a lensed quasar at a typical redshift of $z = 2$, cluster-scale lenses should make up 25\% of the total lensing cross-section, and the true fraction of quasars lensed by galaxy clusters should be an order of magnitude larger than observed \citep[][]{Robertson:2020}. An ideal algorithm for finding WSLQs should therefore be able to uncover the \textit{missing} WSLQ population even in extant data. Approaches that rely exclusively on identifying point-source-like morphology, however, risk overlooking systems whose images appear extended, e.g., either because of signal-to-noise limitations in the survey's morphological classification pipeline, because the survey's fitting engine, while efficient for survey-wide processing, cannot adequately handle crowding from foreground cluster galaxies, because the quasar host galaxy has been magnified into view by the foreground cluster, or because of some yet-to-be-identified cause.

Yet, previous WSLQ searches have focused on point sources near massive foreground structures \citep[see, e.g.,][]{Oguri:2006}, and recent work indeed demonstrates that some WSLQs are not cataloged as point sources in public survey imaging \citep{Napier:2023a}. As we will highlight in this paper, at least some WSLQs can be found without reference to source morphology using a purely photometric analysis. Notably, two of the seven already known WSLQs have \textit{at most} one image that is morphologically classified as a point source in the corresponding best available survey data, DECaLS \citep[][]{Dey:2019}. Nonetheless, the systematics that complicate current searches remain poorly characterized, underscoring the need for improved search strategies.

The discovery pace of WSLQs is expected to accelerate in the next decade. Projections show that Rubin \citep[][]{Ivezic:2019} will lead to the discovery of $\sim 80$ WSLQs from the ground \citep[see e.g.][]{Abe:2025, Robertson:2020}, and space-based infrared flagship NASA and ESA missions will further expand these efforts. As data from Euclid \citep[][]{Scaramella:2022} and Roman \citep[][]{Akeson:2019} begin to flow, we expect significant numbers of WSLQs to be found, particularly probing to fainter apparent magnitudes, and higher redshifts, than current ground-based searches. Another prospect is the development of machine learning algorithms to identify lensed quasars, although for now, the available training set is limited to the seven found WSLQs, which is much smaller than the order $\gtrsim 100$ objects in training sets used for training current neural networks to find quasars lensed by single galaxies or groups of galaxies \citep[][]{Sweeney:2026}. Regardless, addressing the known limitations on current searches for WSLQs will be necessary for interpreting the growing population of WSLQs and deriving meaningful statistics as new missions are expected to expand the known sample.

Here, we present the discovery of the lensing galaxy cluster \targ, which, as we will show, bears compelling evidence for lensing not one, but two WSLQs, \targA\ and \targB. 
In Section~\ref{sec:discovery} we describe the discovery of \targ\ and summarize the imaging and spectroscopic data used in this work. In Section~\ref{sec:measurements}, we describe our methodology for surface-brightness modeling and spectral-line measurements. Section~\ref{sec:analysis} details our analysis based on these measurements, including SED analysis and lens modeling. In Section~\ref{sec:space} we introduce space-based imaging data obtained later in the development of this manuscript and how these data complement and expand our results from the ground-based data. In Section~\ref{sec:discussion}, we discuss the broader implications of the discovery of \targA\ for studies of wide-separation lensed quasars and possible directions for future work and, finally, in Section~\ref{sec:conclusions} we provide a summary of our main conclusions.

Throughout this paper, we assume a flat cosmology with $H_0\!=\!70\, \mathrm{km}\,\mathrm{s}^{-1}\,\mathrm{Mpc}^{-1}$, $\Omega_{\rm M}\!=\!0.3$, and $\Omega_{\Lambda}\!=\!0.7$. If not specified, we use the AB magnitude for all reported magnitudes. In images, north is up and east is left unless otherwise specified.

\section{Discovery, Follow-Up Observations and Archival Data}
\label{sec:discovery}

We present the discovery and multi-wavelength follow-up observations of \targ, a WSLQ system identified by the ChicagO Optically selected strong Lenses -- Located At the Margins of Public Surveys (COOL-LAMPS) collaboration. Our dataset spans optical and near-infrared imaging and spectroscopy from the Magellan Telescopes and the Nordic Optical Telescope, complemented by archival data from Herschel, Spitzer, and ALMA. This section details the discovery process, the follow-up observations, and the archival datasets used for analyzing the system.

\subsection{DECaLS discovery}

\begin{figure}[ht!]
    \centering
    \includegraphics[width=0.9\columnwidth]{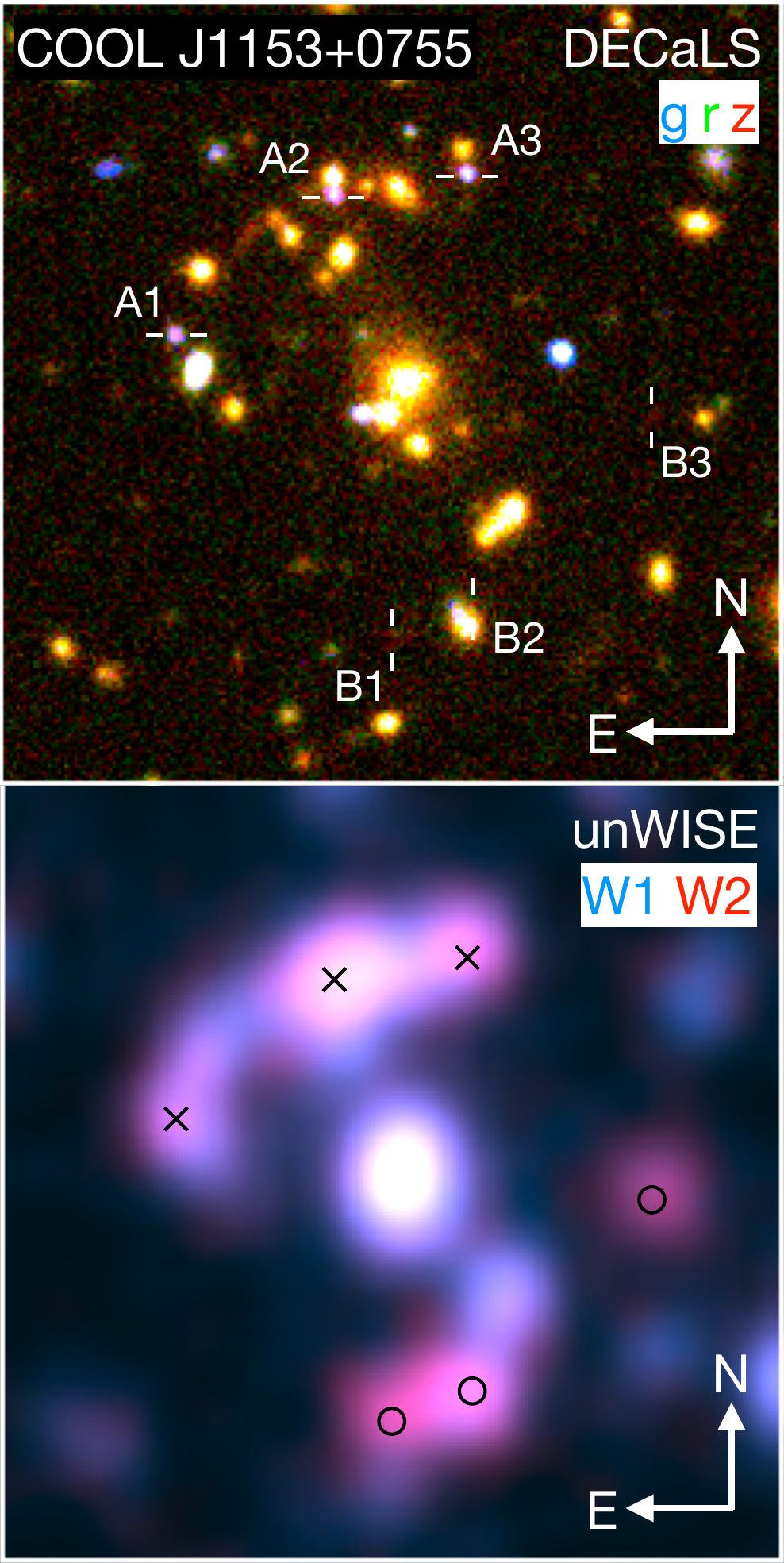}

    \caption{Discovery imaging of \targ\ from DECaLS \textit{(top)} and unWISE \textit{(bottom)}. The $1'\times1'$ cutouts show DECaLS DR9 $g$, $r$, and $z$ imaging \textit{(top)} and unWISE W1 and W2 imaging \textit{(bottom)}, revealing infrared emission consistent with two candidate multiply-imaged quasars in \targ\ (A and B), which motivated the deeper follow-up imaging shown in Fig.~\ref{fig:grz_JHKs_I1I2}. We mark the locations of \targA\ ($z=1.524$, crosses) and \targB\ ($z=1.939$, circles). This figure demonstrates the value of combining DECaLS optical data with complementary WISE infrared imaging, especially for \targB, which is only marginally detected in the \textit{grz} bands.}
     
    \label{fig:discimages}
    
\end{figure}

\targ\ was identified through the COOL-LAMPS collaboration's search \citep[][]{Khullar:2021} for WSLQs in archival DECaLS data, continuing the efforts that led to the discoveries of COOL J0542-2125 \citep[][]{Martinez:2023} and COOL J0335-1957 \citep[][]{Napier:2023a}. Our WSLQ search strategy 
(modified from \citealt{Martinez:2023}) targets sources in the DECaLS Data Release 9 \citep[DR9;][]{Dey:2019}, which covers 19,000 deg$^2$ of the sky down to apparent $z$-band magnitudes of $z_{\mathrm{mag}} \sim 22.5$. We identify promising widely separated ($\gtrsim 10''$) lensed quasar candidates by examining both point sources and extended sources located near high densities of luminous red galaxies (LRGs). The initial expectation is that multiply-imaged quasars appear as point-sources in ground-based imaging. However, the discovery of COOL J0335-1957 revealed that two out of three of its lensed quasar images are classified as extended sources in the DECaLS DR9 catalog, which motivated us to broaden our search beyond strict point-source classification to capture quasars that may be blended or marginally resolved from their host galaxies.
As noted in the DR9 documentation, morphological classifications can vary for blended or marginally resolved sources, and the classification algorithm of the survey is not optimized for all objects.

Accordingly, our WSLQ search strategy (modified from \citealt{Martinez:2023}) comprises these steps:

\begin{enumerate}
    \item We first identify LRGs in the DECaLS survey data by selecting galaxies brighter and redder than a redshift track in color-magnitude space for an early-forming stellar population of some fiducial mass. This is done in $g-r$ versus $r$ for notional redshifts less than 0.45 and in $r-z$ versus $z$ for notional redshifts greater than 0.35, and the lists are then combined. For the searches executed by the COOL-LAMPS collaboration, the fiducial mass limit is approximately M$^*$. We then compute a red-sequence galaxy overdensity (a richness) around each selected candidate LRG considering galaxies down to M$^*$+2, to a radius of 0.5 Mpc. A cut is then made in the rank-ordered richness list to select the most overdense lines of sight; that cut is made in redshift slices, to ensure that high-redshift overdensities are not inappropriately downweighted due to incompleteness in the input photometric catalogs. This process selects cluster-scale overdensities of red-sequence galaxies across a wide range of redshifts; both the chosen initial LRG mass cut (i.e., $\sim$M$^*$) and the later richness cut ensures selection of a complete sample of clusters and groups to well below the mass threshold of interest for finding WSLQs.

    \item Along these cluster sightlines, we identify likely quasars by performing a selection in $g-r-z-W1-W2$. Like \cite{Martinez:2023} this is done by perturbing the measured photometry of individual sources by the measured uncertainties, and comparing the positions of the resulting objects (over several thousand instances) to the notional locations of stars, galaxies, and quasars in the multi-dimensional color-color space.  The comparison samples are drawn from spectroscopically identified objects of all three classes from the SDSS \citep{York:2000}. Objects which align best with the sample of notional quasars are then considered further, and sightlines
    containing multiple such likely candidates are flagged and carried through to the last step.
    \item Finally, we visually inspect all identified lines of sight ($\sim$10$^3$) to select WSLQ candidates suitable for follow-up observations, with \targ\ being one of the high-priority candidates that was spectroscopically observed.
\end{enumerate}

\vspace{10pt}
\subsection{Magellan Observations}

\begin{figure}
    \centering
    \includegraphics[width=\columnwidth]{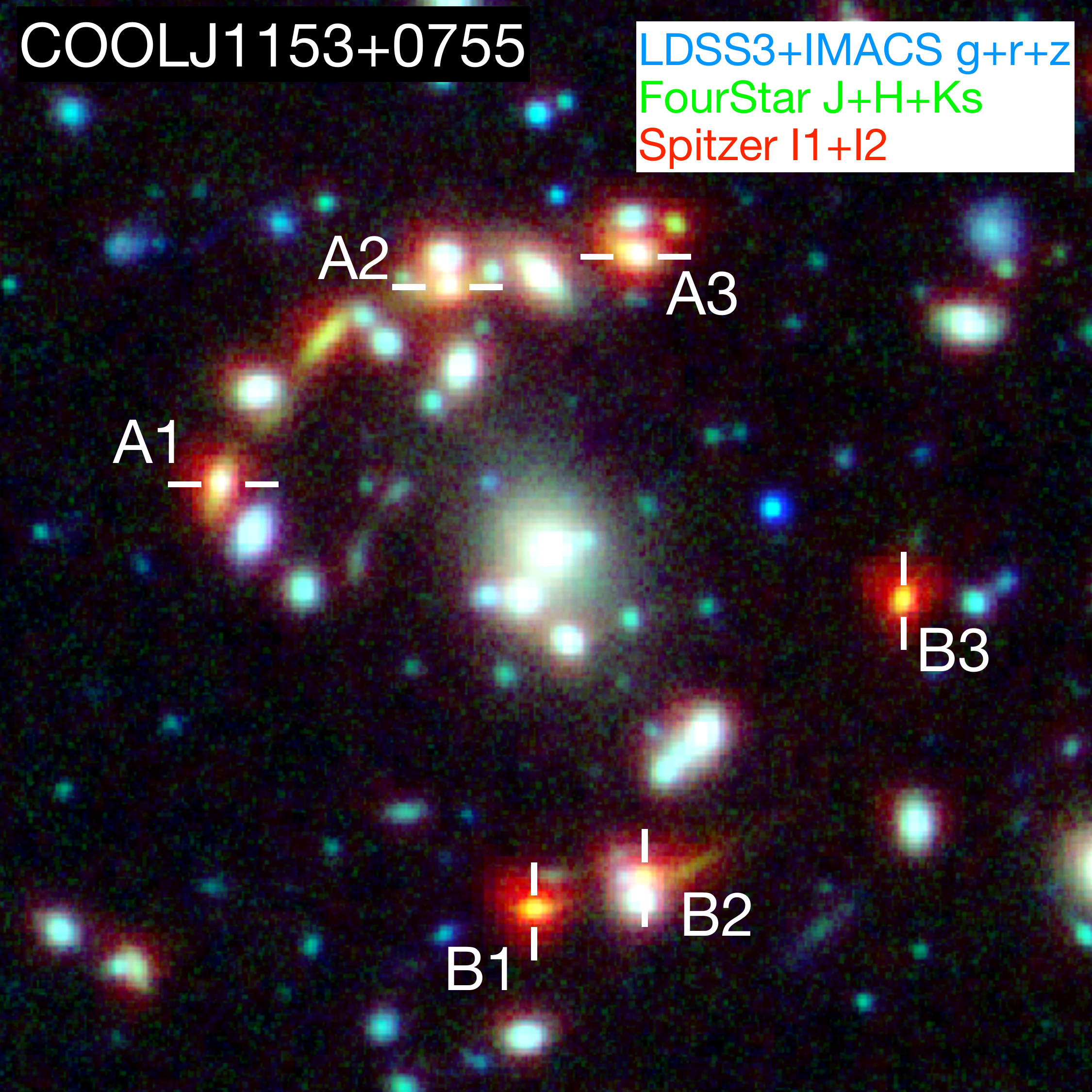}
    \caption{Deeper Magellan follow-up imaging and archival infrared Spitzer imaging reveal \textit{two} lensed quasars, each with three images: \targA\ and \targB. Shown is a $1'\times1'$ multi-wavelength composite image using Magellan LDSS3 and IMACS $grz$ imaging (blue), FourStar $JHK_s$ imaging (green), and archival Spitzer/IRAC channels I1 and I2 (red).}
    \label{fig:grz_JHKs_I1I2}
\end{figure}

To confirm the nature of \targ\ as a WSLQ system, we carried out a series of follow-up observations using the 6.5-m Magellan Telescopes at Las Campanas Observatory in Chile. These observations were designed to obtain high-quality imaging and spectroscopy, allowing us to characterize the system's morphology, firmly identify multiple quasar images, and measure redshifts. The following subsections detail the imaging and spectroscopic follow-up performed with FourStar (Section \ref{sec:fourstar}), FIRE (Section \ref{sec:fire}), LDSS3 (Section \ref{sec:ldss3}), and IMACS (Section \ref{sec:imacs}). An overview of all single-slit spectra of \targ\ can be found in Table \ref{tab:spectra_obs}. A composite RGB-image made with the stacked LDSS3 and IMACS $grz$-bands (blue) and FourStar $JHK_s$-bands (green), along with archival Spitzer/IRAC channels I1 and I2 imaging (red, see Section~\ref{sec:archivaldata}), is presented in Fig.~\ref{fig:grz_JHKs_I1I2}.

\subsubsection{Magellan/FourStar}

\label{sec:fourstar}

We obtained near-infrared imaging of \targ\ in February and March 2023 with FourStar \citep[][]{Persson:2013} on the 6.5-m Magellan Baade Telescope, using the $J$, $H$ and $K_s$ filters and a dice-5 dither pattern. Total exposure times for the three bands are 3429~s for the $J$ band, 3493~s for the $H$ band, and 3493~s for the $K_s$ band. The data were reduced using the \texttt{FSRED} pipeline, which generates astrometry solutions using Gaia DR3 \citep[][]{Gaia:2016b, Gaia:2023j} and computes photometric zero points using 2MASS archival magnitudes \citep[][]{2MASS:2006}. The median seeing in the reduced images, estimated from the FWHM of known unsaturated stars in the field, is $0\sarcs51$ in $J$, $0\sarcs49$ in $H$, and $0\sarcs60$ in $K_s$.


\subsubsection{Magellan/FIRE}

\label{sec:fire}

We acquired near-infrared spectroscopy of \targ\ with the Folded port InfraRed Echellette (FIRE) spectrograph \citep[][]{Simcoe:2008} on the Magellan Baade Telescope, in February, March and April 2023 and March 2024, with a typical exposure time of 1204.7~s for each object targeted in the field. FIRE covers a continuous bandpass across $0.82-2.51 \, \mu \mathrm{m}$. The raw FIRE data were reduced using PypeIt version 1.13.1 \citep[][]{pypeit:joss_arXiv, pypeit:joss_pub, pypeit:zenodo}.

\begin{table*}
\centering
\caption{Coordinates, Exposure Times, and Instruments for Sources Targeted with Magellan Single-Slit Spectroscopy}
\label{tab:spectra_obs}
\renewcommand{\arraystretch}{1.2}
\centering
\begin{tabular}{llcccc}
\hline
\hline
ID & Object & R.A. & Decl. & Exposure Time & Instrument \\
& & (J2000) & (J2000) & (s) & \\
\hline
\makebox[30mm][l]{A1} & \makebox[35mm][l]{\targA} & 11:53:20.4584 & +7:56:00.783 & 4800 & LDSS3 \\
\makebox[30mm][l]{A2} & \makebox[35mm][l]{\targA} & 11:53:19.6334 & +7:56:11.452 & 1600 & LDSS3 \\
\makebox[30mm][l]{A3} & \makebox[35mm][l]{\targA} & 11:53:18.9537 & +7:56:13.152 & 4800 & LDSS3 \\
\makebox[30mm][l]{5.1, 6.1} & \makebox[35mm][l]{Blue arc} & 11:53:19.9617 & +7:55:56.824 & 1204.8 & FIRE \\
\makebox[30mm][l]{5.3, 6.3} & \makebox[35mm][l]{Blue arc (counterimage)} & 11:53:18.3864 & +7:56:09.485 & 2409.4 & FIRE \\
\makebox[30mm][l]{B1} & \makebox[35mm][l]{\targB} & 11:53:19.3245 & +7:55:37.829 & 4818.8 & FIRE \\
\hline
\end{tabular}

\vspace{2mm}
\raggedright
\textbf{Notes.} A complementary set of LDSS3 multislit and longslit observations of foreground cluster members and background sources was also conducted and is included in Appendix~\ref{app:slit_redshifts}.
\end{table*}

\subsubsection{Magellan/LDSS3}

\label{sec:ldss3}

We obtained $g$, $r$, and $z$-band imaging and longslit and multislit spectroscopy using the Low Dispersion Survey Spectrograph-3 (LDSS3) on the Magellan Clay Telescope. Our two longslit spectra each have exposure times of $4\times1200$s. We acquired the imaging in January 2023 and collected spectroscopic observations during April 2023. The longslit spectra were reduced in the usual manner, using a set of custom IRAF and IDL scripts. We also took multislit exposures with two masks to confirm the redshift of the foreground cluster members associated with the lensing potential of \targ. The multislit spectra were reduced using the COSMOS2 data reduction package \citep[][]{Oemler:2017}.

\subsubsection{Magellan/IMACS}

\label{sec:imacs}

We used the Inamori-Magellan Areal Camera and Spectrograph (IMACS; \citealt{Dressler:2011}) on the Magellan Baade Telescope to obtain additional imaging and spectroscopy. For imaging, we employed the IMACS f/4 camera mode and collected images in the $g$, $r$, and $z$-bands with an exposure time of $4\times180$~s in each band. Both IMACS and LDSS3 contributed imaging in these bands, and the final $g$-, $r$-, and $z$-band stacks used in the lens model combine imaging data from both instruments. The seeing values, quantified by the median of stellar FWHM measurements across the images, are $0\sarcs80$, $0\sarcs78$, and $0\sarcs71$ in $g$, $r$, and $z$, respectively.

\vspace{15pt}
\subsection{Nordic Optical Telescope/ALFOSC Observations}

\label{sec:not/alfosc}

Monitoring data in the $g$-band were acquired using the 2.56-m Nordic Optical Telescope (NOT) at the Roque de los Muchachos Observatory in La Palma, Spain. The imaging observations were carried out using the Alhambra Faint Object Spectrograph and Camera (ALFOSC), which in imaging mode provides a $6.\!\!'4 \times 6.\!\!'4$ field of view.

\subsection{Archival Data}

\label{sec:archivaldata}

We complement the survey imaging data and dedicated observations with archival data from Herschel, Spitzer, WISE, and ALMA, as follows.

\targ\ was first identified as an object of interest in a Herschel Space Observatory program executed in July 2012 (PID: OT2\_eegami\_6, PI: E. Egami) \citep[][]{Pillbratt:2010}. The program used Herschel's Spectral and Photometric Imaging Receiver (SPIRE) to conduct the SPIRE Snapshot Survey II. They targeted gravitational lenses by selecting clusters from the South Pole Telescope (SPT) cluster catalog, which identifies clusters using the Sunyaev-Zel'dovich (SZ) effect \citep[][]{Bleem:2015}, and from the COnstrain Dark Energy with X-ray (CODEX) survey, which combines ROSAT All-Sky Survey (RASS) X-ray and Sloan Digital Sky Survey (SDSS) optical data.

We make use of this archival far-infrared imaging from Herschel, to sample the infrared–submillimeter part of the spectral energy distribution (SED), essential for constraining the far-infrared dust emission at the redshift(s) of \targ. We use the Level 2 pipeline products, which are fully reduced and science-ready. The observations of \targ\ were taken on 2012 July 12.
SPIRE provides imaging in three photometric bands centered at 250 $\mu {\rm m}$ (short-wavelength), 350 $\mu {\rm m}$ (medium-wavelength), and 500 $\mu {\rm m}$ (long-wavelength) with beam sizes of ${\rm FWHM}=18\sarcs1, 25\sarcs2, 36\sarcs6$ respectively \citep[][]{Griffin:2010}. These data provide the lowest-resolution imaging of \targ\ among the available datasets, with sampling of $6''/{\rm pixel}$, $10''/{\rm pixel}$ and $14''/{\rm pixel}$ in the SW, MW, and LW channels respectively. The three Herschel bandpasses each have total exposure times of $5{,}700$s.

We also use WISE (Wide-field Infrared Survey Explorer) archival photometry. Bands W1 (3.4\,$\mu$m) and W2 (4.6\,$\mu$m) are from unWISE (\citet[][]{Lang:2014, Meisner:2017a, Meisner:2017b};\dataset[10.26131/IRSA524]{http://dx.doi.org/10.26131/IRSA524}), a reprocessing of WISE data with improved depth and spatial resolution (see bottom panel of Fig.~\ref{fig:discimages}). Bands W3 (12\,$\mu$m) and W4 (22\,$\mu$m) are from AllWISE (\citet[][]{Wright:2019};\dataset[10.26131/IRSA1]{http://dx.doi.org/10.26131/IRSA1}).

We include archival mid-infrared imaging from the Spitzer Space Observatory, taken on 2013 August 12 in two channels: 3.6\,$\mu{\rm m}$ (channel I1) and 4.5\,$\mu{\rm m}$ (channel I2) (\citet[][]{Werner:2004};\dataset[10.26131/IRSA543]{http://dx.doi.org/10.26131/IRSA543}). We use Level 2 mosaics, fully processed and calibrated for science use, with a pixel scale of $0\sarcs6 / {\rm pixel}$ in both bands.

Finally, we make use of millimeter data from the Atacama Large Millimeter Array (ALMA). The field containing \targ\ was first observed with ALMA in Band 6 in 2016 \citep[PI: Egami, Eiichi; Project Code: 2016.1.00372.S;][]{Sun:2021} as part of a survey targeting lensed submillimeter galaxies in 20 clusters selected from the Herschel Lensing Survey \citep[HLS;][]{Egami:2010}. Follow-up observations were obtained in Band 7 in 2019 (PI: Sun; Project Code: 2019.2.00040.S). This object is cataloged in \citet{Sun:2021} as \textit{CODEX 52909}, though they reported a non-detection and did not pursue further analysis of the system. In Appendix ~\ref{app:alma}, we present the detection of \targ\ in ALMA Bands 6 and 7.

We note that the lensing cluster is detected in the eROSITA catalog of galaxy clusters \citep[][]{Predehl:2021, Merloni:2024}.
The eROSITA imaging point-spread function (PSF) is comparable to the maximum image separation between the lensed quasars ($\gtrsim 25''$), which precludes detailed X-ray analysis of the individual lensed sources.

\section{Measurements} \label{sec:measurements}

In this section, we present quantitative measurements derived from the archival and follow-up data described above, including photometry and spectral line measurements.

\subsection{Surface Brightness Decomposition and Photometric Extraction}
\label{sec:photometry}

\begin{figure*}[ht!]
\centering
    \includegraphics[width=\linewidth]{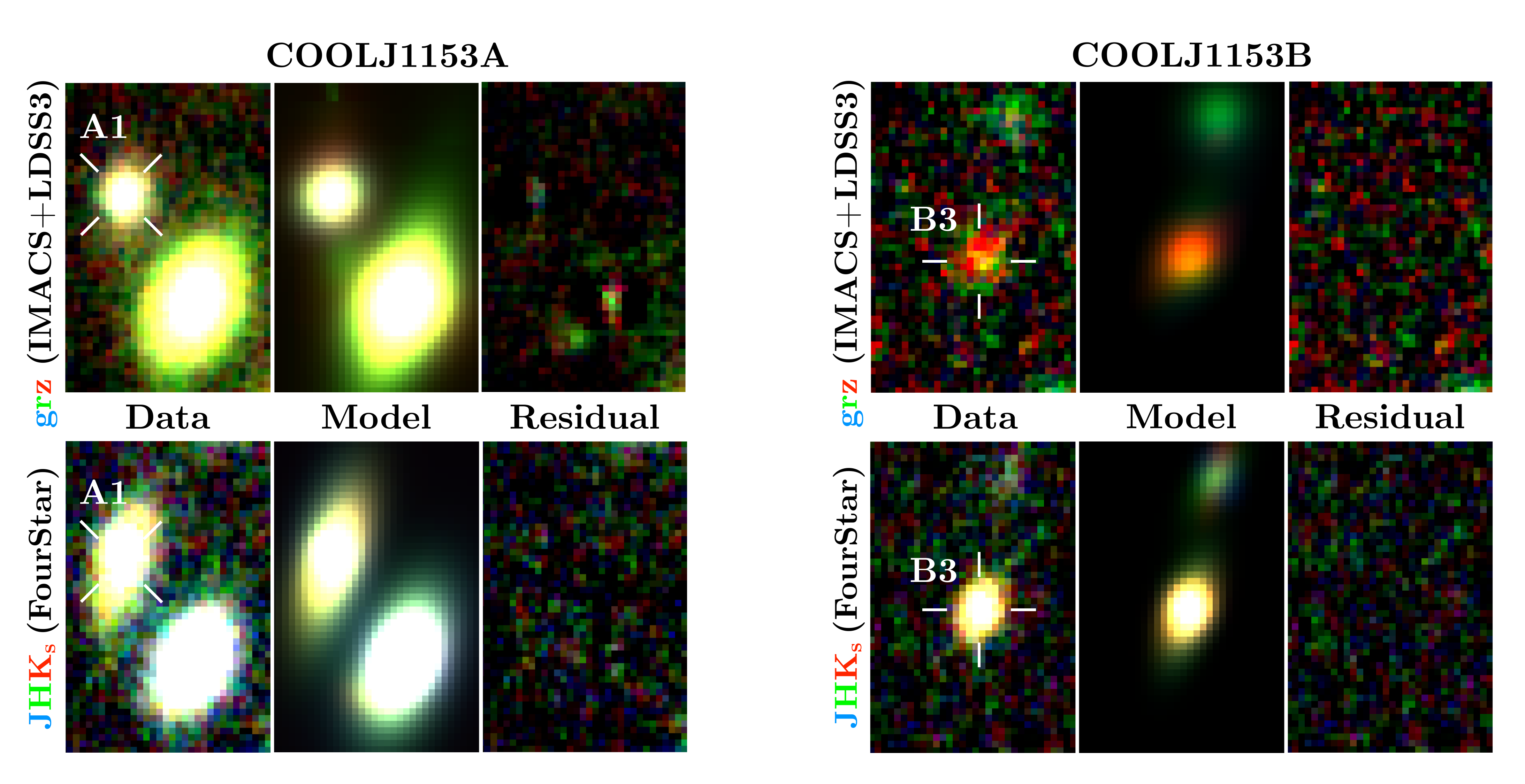}
    \caption{Surface brightness profile modeling of \targA\ (image A1) and \targB\ (image B3). For each system, we show the observed image, the best-fitting model, and the residual image in both the $grz$ and $JHK_s$ bands. The RGB composites are constructed using $g$/$J$ as blue, $r$/$H$ as green, and $z$/$K_s$ as red. White crosshairs in the leftmost panels mark the positions of \targA\ and \targB. The nearby objects, a foreground cluster member and a multiply-imaged galaxy (image 7.3 in Fig.~\ref{fig:rgbimage}) are likewise modeled and subtracted.}
\label{fig:galfit}
\end{figure*}

We perform surface brightness profile modeling and photometry using \texttt{GALFIT} \citep{Peng:2002, Peng:2010} on the most isolated images of each source, specifically A1 (from \targA) and B3 (from \targB), to minimize contamination from nearby galaxies. For A1, we explore a range of parametric surface-brightness profiles and find that a model consisting of a single S\'ersic component plus a central PSF provides the best fit in the $g$-, $r$-, $z$-, $K_s$-, ${\rm I1}$-, and ${\rm I2}$-imaging and an additional S\'ersic component in the $J$- and $H$-imaging. Throughout this work, we adopt the total flux of this combined model when constructing the SED. A more detailed decomposition separating the point-source and extended emission will be presented in a forthcoming paper. Applying the same procedure to B3 yields a similar best-fitting S\'ersic+PSF model. A visual summary of the morphological fits is provided in Fig.~\ref{fig:galfit}.

To estimate the uncertainty on the measured photometry, we characterize the sky background by identifying $N\gtrsim100$ regions free of objects and noise artifacts in each band, add this background noise to the data, and refit the models $N$ times to generate a set of Monte Carlo realizations. The photometric uncertainty is taken as the $1\sigma$ dispersion of the resulting distribution of fluxes. The fits have negligible structured residual, so the profile fits are dominated by Poisson noise from the background.

In addition to the resolved optical and near-infrared photometry, we incorporate far-infrared measurements from archival Herschel SPIRE imaging. Owing to the large point-spread function, the multiple lensed images of each quasar are not well separated, and instead appear as partially blended emission associated with each system. To measure these data, we construct a model of the lensing cluster, quasar images, and surrounding galaxies using \texttt{GALFIT} \citep{Peng:2002} where each object in the model is treated as an unresolved \textit{PSF} element, and the appropriate flux is solved for. This simplified approach works well because the Herschel PSF is much larger than any of the individual distant galaxies considered. The model was built iteratively, with components initially added on just the quasar images and the central cluster galaxy, and with the quasar images assigned fixed flux ratios matching that measured in the $K_s$ band. 
Other galaxies were then added at larger radii, when needed as indicated by residuals visible in the Herschel data minus the model; the added additional model elements were placed on galaxies with red colors, as indicated by the NIR and Spitzer data.  As we are only interested in removing the flux of these additional objects to ensure a robustly measured background level, the precise choices of component locations on scales much smaller than the PSF matters little.  

We also include millimeter flux measurements from archival ALMA imaging in Bands 6 and 7. As the sources are unresolved at ALMA resolution (but not blended like in the Herschel imaging), we measure fluxes for the most isolated images (A1 and B3) by fitting two-dimensional Gaussian profiles to the emission using \texttt{CARTA} \citep[][]{Comrie:2021}. The fits are performed within regions several times larger than the synthesized beam to ensure accurate recovery of the integrated flux. Uncertainties are estimated from the RMS of emission-free regions surrounding each source. These measurements are further documented in Appendix~\ref{app:alma} and the photometry across all bands are in Appendix~\ref{app:photometry}.

\subsection{Spectral Line Measurements}

Here, we analyze emission and absorption features in the spectra of \targA\ and \targB\ obtained with LDSS3 and FIRE. The two individual LDSS3 longslit spectra were designed to cover likely foreground cluster galaxies and all three images of \targA. The FIRE $1'' \times 6''$ slit covers image B1 of \targB.

\subsubsection{Spectral Fitting of LDSS3 Spectra of \targA}

\begin{figure*}[ht!]
    \centering
    \includegraphics[width=\linewidth]{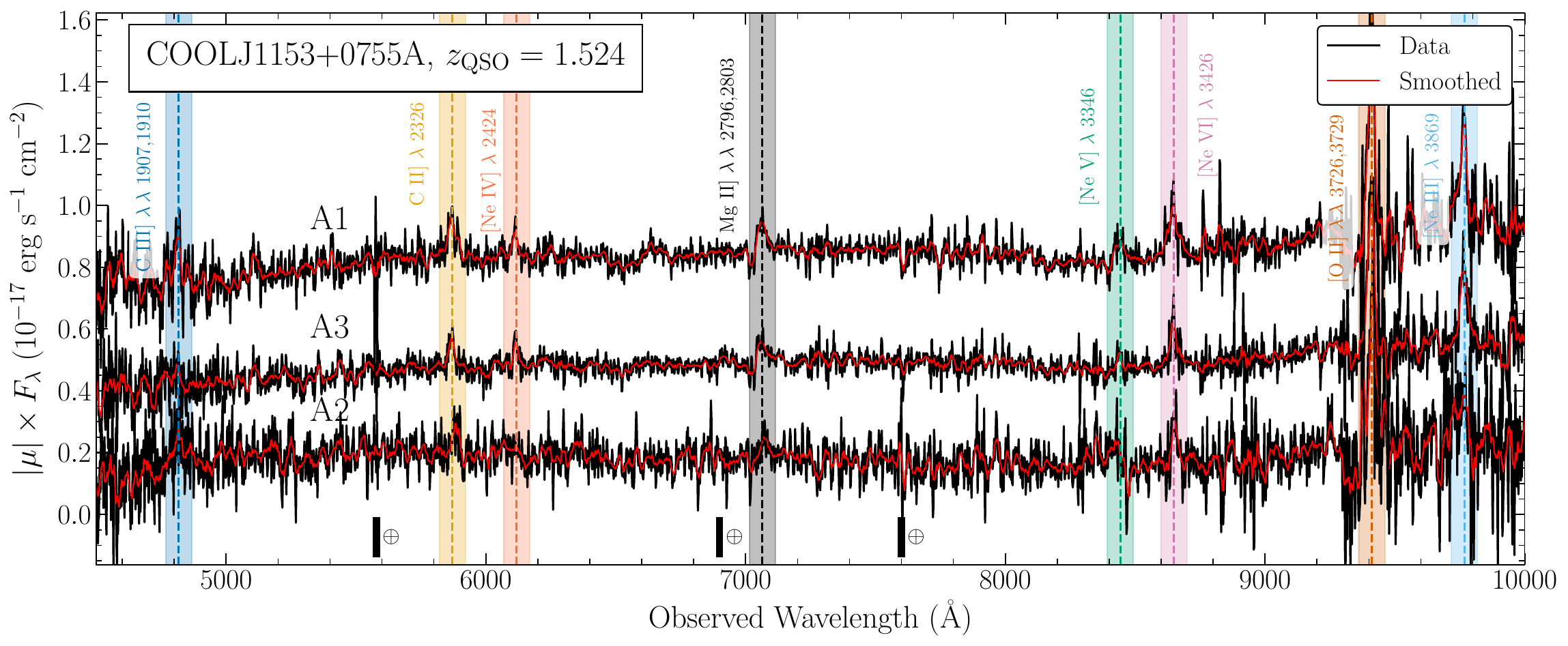}
    \caption{LDSS3 spectra of the three quasar images of \targA\ confirm identical redshifts and spectral features. The A1 and A3 spectra have been vertically offset for visual clarity. Major telluric line residuals are marked with $\oplus$.
    Mg \textsc{ii}], a line from the quasar broad-line region, experiences significant foreground absorption. We present a fit of the broad Mg \textsc{ii}] emission and foreground absorption in Appendix~\ref{app:fits}, Fig.~\ref{fig:ldss3_COOLJ1153A1_MgII}.}
    \label{fig:ldss3_spectra}
\end{figure*}

Within the \targ\ strong lensing region, we identify three multiple images of \targA\ and we confirm their redshifts from their identical spectra (see Fig.~\ref{fig:ldss3_spectra}). All three images of \targA\ exhibit multiple emission lines characteristic of AGN. We employ the Python package \texttt{pyspeckit} \citep[][]{Ginsburg:2022, Ginsburg:2011} to fit the continuum and spectral lines, and we perform all subsequent line fits with this software unless otherwise specified. 

To derive the systemic redshift of \targA, and characterize the broad emission line properties, we perform a simultaneous multi-component Gaussian fit of the full LDSS3 spectrum across $4000{–}10{,}000$ Å. We fit the strongest observed emission lines: C \textsc{ii}] $\lambda 2326$, [Ne \textsc{iv}] $\lambda 2424$ (not included in fit for image A2 due to poor signal), [Ne \textsc{vi}] $\lambda 3426$, [O \textsc{ii}] $\lambda \lambda3726, 3729$, and [Ne \textsc{iii}] $\lambda 3869$. We exclude C \textsc{iii}] $\lambda \lambda 1907, 1910$, and [Ne \textsc{v}] $\lambda 3346$ from the fit since their centroids are not well-fitted across all three quasar images. We also exclude the Mg \textsc{ii}] $\lambda \lambda2796, 2803$ doublet region (6850–7250 \AA\ rest-frame) and perform a separate fit (Appendix~\ref{app:fits}, Fig.~\ref{fig:ldss3_COOLJ1153A1_MgII}) due to the presence of multiple Mg \textsc{ii}] absorption lines along the line of sight. All fitted line components vary freely within broad limits. For the [O \textsc{ii}] $\lambda \lambda3726, 3729$ doublet, we fix the wavelength ratio, tie the velocity widths, and constrain the amplitude to the theoretical value in the optically thin limit $W_{3729}/W_{3726} = 1.5$ \citep[][]{Pradhan:2006}. From the fitted line centers, we calculate individual redshifts and derive a weighted mean redshift for \targA\ using inverse variance weighting. We report the redshifts for each quasar spectrum in Table~\ref{tab:spectra_z}. The weighted mean redshift using all three images of \targA\ is $z_{\rm QSO} = 1.52352 \pm 0.00050$.

In \targA, we detect two line-of-sight Mg \textsc{ii}] $\lambda \lambda 2786, 2803$ absorbers on top of the broad Mg \textsc{ii}] quasar emission line whose redshifts are consistent with the redshifts of the ``blue arc'' and a singly-imaged galaxy northeast of the cluster center (Section~\ref{sec:quasar_environments}). We therefore fit a multimodal Voigt and Gaussian model comprising three components: one broad emission system and two absorption systems grouped into three kinematic systems (each containing a doublet pair, see Appendix~\ref{app:fits}, Fig.~\ref{fig:ldss3_COOLJ1153A1_MgII}). A \textit{Voigt} profile captures the dynamical broadening of the line (Gaussian component) and the broad wings from radiative and collisional broadening (Lorentzian component). It is therefore an appropriate choice for fitting the broad emission component. The narrow absorption line systems are fitted with Gaussian profiles. For each doublet pair, the secondary line is tied to the primary line center via the ratio of their rest wavelengths.
The amplitude ratio of the Mg \textsc{ii}] doublet is set by atomic physics to $\sim 2.0$ in the optically thin limit \citep[e.g.][]{Lawther:2012}. We therefore fix the ratio $W_{2796}/W_{2803} = 2.0$ for our analysis while noting that this value can vary depending on the line-of-sight neutral gas column density \citep[][]{Chisholm:2020}. The broad doublet Voigt component is allowed independent Gaussian width ($\sigma$) and damping parameters ($\gamma$). For the narrow-line systems, the Gaussian width is individually tied for each doublet, but not to each other. The broad emission system is fixed to the measured redshift of \targA\ ($z=1.52352$). The absorption system centers are constrained to maintain kinematic coherence but allowed to vary in redshift. 
We compute a velocity width for the broad component of ${\rm FWHM} \simeq 1200 {\rm km} \, {\rm s}^{-1}$ (by accounting for instrumental broadening, which we measure as $R\sim 750$ from the Mg \textsc{ii}] absorption lines).
This line is in a region with multiple foreground absorbers, so the line width of the Mg \textsc{ii}] emission lines is not well constrained.

\subsubsection{Spectral Fitting of FIRE Data for \targB}

\label{sec:fire_fits}

We analyze the rest-frame optical emission lines of \targB\ using the FIRE spectrograph, focusing on the H$\alpha$ $\lambda 6563$ \AA, [N \textsc{ii}] $\lambda \lambda 6549, 6585$ \AA, H$\beta$ $\lambda 4861$ \AA, and the [O \textsc{iii}] $\lambda \lambda \, 4959, 5007$ \AA\ doublet. The infrared FIRE spectrum is corrected for telluric atmospheric absorption by dividing it by a telluric transmission model interpolated to the wavelength grid of the science spectrum; the inverse variance is propagated accordingly. The spectrum is then detrended using a median filter (via the \texttt{median}\_\texttt{smooth} function in the \texttt{specutils} library). As we do for the LDSS3 spectrum, we perform spectral fitting using the \texttt{pyspeckit} package. For the H$\alpha$ region, we fit a three-component Gaussian model to the H$\alpha$, and [N \textsc{ii}] $\lambda \lambda 6549, 6585$ lines. The H$\alpha$ and $\mathrm{[N\,II]}$ line centers are tied through their rest-frame wavelength ratio. All three lines are allowed independent amplitudes.

We simultaneously fit H$\alpha$, [N \textsc{ii}] $\lambda \lambda 6548,6583$, and [O \textsc{iii}] $\lambda 5007$ with single Gaussian profiles for each line with tied velocity widths (we refer to this as method (a)). However, literature on AGN spectroscopy \citep[e.g.][]{Greene:2005} shows that even in Type II AGN, broad emission line components can sometimes be detected, particularly Balmer lines, if there is any unobscured path to the broad-line region or if light is scattered into our line of sight. Since our spectrum is too noisy to cleanly distinguish whether a broad H$\alpha$ component is present, we account for this possibility by fitting a broad H$\alpha$ component in addition to the aforementioned narrow Gaussian components (method (b)). We let the width of this component vary freely while the other component widths are tied to a narrow H$\alpha$ component (Appendix~\ref{app:fits}).

The region of the spectrum where we find H$\beta$ and $\mathrm{[O\,III]}$ is fitted using a two-component Gaussian model. The $\mathrm{[O\,III]} \,4959$ and $5007$ lines are tied to each other by their rest wavelengths. The $\mathrm{[O\,III]} \,5007$ amplitude is constrained to be 3.0 times the $\mathrm{[O\,III]} \,4959$ amplitude, reflecting their intrinsic intensity ratio. We tie all line components to share the same line width. H$\beta$ is not detected in the spectrum; we therefore compute a $1\sigma$ upper limit on its flux using the standard deviation of the noise in the H$\beta$ region and derive a lower limit on the $\mathrm{[O\,III]} \,5007$/H$\beta$ flux ratio for BPT diagnostics, from this upper limit line amplitude.

For a Gaussian line, the line flux $F$ is given by:
\begin{equation}
    F = A \cdot \sigma \cdot \sqrt{2\pi}.
\end{equation}

where $A$ is the amplitude and $\sigma$ is the standard deviation of the Gaussian profile. Since the constant $\sqrt{2\pi}$ cancels out in flux ratios, we compute the diagnostic line ratios as:
\begin{equation}
    \log\left( \frac{\mathrm{[N\,II]}}{\mathrm{H\alpha}} \right) = \log \left( \frac{A_{\mathrm{[N\,II]}} \cdot \sigma_{\mathrm{[N\,II]}}}{A_{\mathrm{H\alpha}} \cdot \sigma_{\mathrm{H\alpha}}} \right),
\end{equation}

Uncertainties on the logarithmic flux ratios are derived from the measured uncertainties on $A$ and $\sigma$. We present a Baldwin, Phillips \& Terlevich (BPT) ionization diagram for \targB\ in Section \ref{sec:evidence_two_agn}. Note that we mark the upper limit for H$\beta$ with an arrow, since the line is not detected among the noise, which gives a lower limit for the $\log\left( \frac{\mathrm{[O\,III]}}{\mathrm{H\beta}} \right)$.

\subsubsection{Obtaining A Redshift for the ``Blue Arc''}

We also measure the redshift of another lensed galaxy with detectable line flux in our FIRE spectra. From the ground-based imaging available at the time, we identified a bright ``blue arc’’ and posited that it might have a counterimage located to the northwest of the BCG. The FIRE spectrum of the bright arc shows a single emission line, which we identify as H$\alpha$, yielding a redshift of $z_{\rm spec} = 1.524626 \pm 0.000013$ based on the $1\sigma$ uncertainty of the centroid measurement. This places the blue arc within the same group environment as the \targA\ WSLQ. Because this redshift is derived from a single emission line, it carries additional systematic uncertainty, as single-line identifications are less robust than identifications based on multiple lines.

Guided by our initial lens model, which predicted the position of a counterimage, we obtained a second FIRE spectrum by placing the slit on a source near the model-predicted location. This spectrum confirms the existence of a single line, H$\alpha$, at the same redshift as the arc, providing confirmation of its counterimage.
We present additional evidence for this identification in Section~\ref{sec:space} from space-based imaging obtained during the final stages of developing this manuscript. Redshifts of the quasars and other lensed sources are summarized in Table~\ref{tab:spectra_z}.

\begin{table}
\caption{Spectroscopic Redshifts of Primary Objects}
\label{tab:spectra_z}
\renewcommand{\arraystretch}{1.2}
\centering
\begin{tabular}{llc}
\hline
\hline
ID & Object & $z_{\mathrm{spec}}$ \\
\hline
A1 & \targA & $1.52297 \pm 0.00094$ \\
A2 & \targA & $1.52450 \pm 0.00101$ \\
A3 & \targA & $1.52333 \pm 0.00073$ \\
5.1, 6.1 & Blue arc & $1.52463 \pm 0.00001$ \\
5.3, 6.3 & Blue arc (counterimage) & 1.525 \\
B1 & \targB\ & $1.93888 \pm 0.00008$ \\
\hline
\end{tabular}

\vspace{1mm}
\raggedright
\textbf{Notes.} Reported uncertainties correspond to $1\sigma$ where available.
\end{table}

\subsubsection{Multislit Spectroscopy of Cluster Members}

We use LDSS3 multislit spectroscopy to efficiently identify cluster members and their redshifts. Each individual galaxy redshift is estimated from absorption lines (Ca H and Ca K for early-type galaxies) or emission lines (Balmer series, [O \textsc{iii}] or [O \textsc{ii}] for late-type galaxies) to produce a redshift distribution. These redshifts are listed in Appendix~\ref{app:slit_redshifts}. From this, we derive a cluster redshift of $z=0.4301 \pm 0.0016$ and a cluster velocity dispersion of $\sigma_v=1298 \pm 330 \, {\rm km} \, {\rm s}^{-1}$, based on 16 galaxy redshifts, considering only the galaxies that were also determined to be cluster members using the red-sequence technique \citep[][]{Gladders:2000}.

We identify a second cluster southwest of the primary cluster from LDSS3 multislit spectroscopy, with a redshift of $z=0.5766 \pm 0.0007$ derived from nine galaxy redshifts. The secondary cluster is located at a projected separation of $\sim3'$ from the primary cluster.

The multislit spectroscopy also reveals a blue galaxy (marked with a white diamond in Fig.~\ref{fig:rgbimage}) located at $z=1.5058$, placing it in the neighborhood of the lensed quasar. This galaxy is not multiply imaged, consistent with its position relative to the foreground lens being outside the strong lensing region.

\section{Analysis}
\label{sec:analysis}

The multi-wavelength imaging and spectroscopy of \targ\ reveal a rich lensing galaxy cluster that produces widely separated images of background quasars ($25\sarcs6$ and $26\sarcs0$), along with multiple lensed galaxies in their source-plane environments. While the discovery initially revealed the bright blue lensed quasar \targA, follow-up observations unveiled a second quasar candidate that is multiply imaged by the same lensing potential, \targB. In this section, we present evidence for a compact group environment surrounding \targA\ at $z=1.524$ and classify \targB\ using a BPT diagram. In sum, these results provide a window into AGN activity and galaxy evolution in dense environments at Cosmic Noon.

We classify \targA\ and \targB\ as quasars based on a brightness threshold, which distinguishes them from lower-luminosity Seyfert galaxies. The distinction between quasars and Seyfert galaxies has traditionally been qualitative: Seyferts are more commonly found at low redshift and exhibit lower luminosities while quasars represent the high-luminosity, high-Eddington-ratio end of the AGN population. 
A common definition in the literature distinguishes quasars by their bolometric luminosity $L_{\rm bol} > 10^{45} \, \mathrm{erg\,s^{-1}}$
\citep{Hopkins:2007, Yuan:2016}. Adopting this delineation, we find that both \targA\ and \targB\ are quasars (see Sections~\ref{sec:sed_modeling} and \ref{sec:slmodel}).
We will therefore use the term \textit{quasar} for these objects throughout the paper. We use the term \textit{AGN} to refer to the broader class of active galactic nuclei that also encompasses quasars.


\vspace{15pt}
\subsection{Evidence of Two Independent Lensed Quasars}

\label{sec:evidence_two_agn}

We now classify each quasar further by their spectroscopic features. The broad emission lines in the spectrum of \targA\ establish it as a typical broad-line Type I quasar. For \targB, the absence of broad lines motivates the use of rest-frame optical emission-line diagnostics to determine whether star-formation or AGN radiation is the primary power source of the narrow-line emission. The BPT diagram \citep[][]{Baldwin:1981} is used to distinguish between AGN and star-forming galaxies based on emission-line ratios of two pairs of strong nebular emission lines \citep[e.g.,][]{Veilleux:1987}. This classification method, further refined by other works such as \citep[][]{Kauffman:2003, Schawinski:2007}, provides insight into the ionization mechanisms at play in different galaxy populations, where elevated [\NII] and [\OIII] emission can arise from shocks and/or strong ionization from the AGN.

\begin{figure}[ht!]
    \includegraphics[width=0.95\columnwidth]{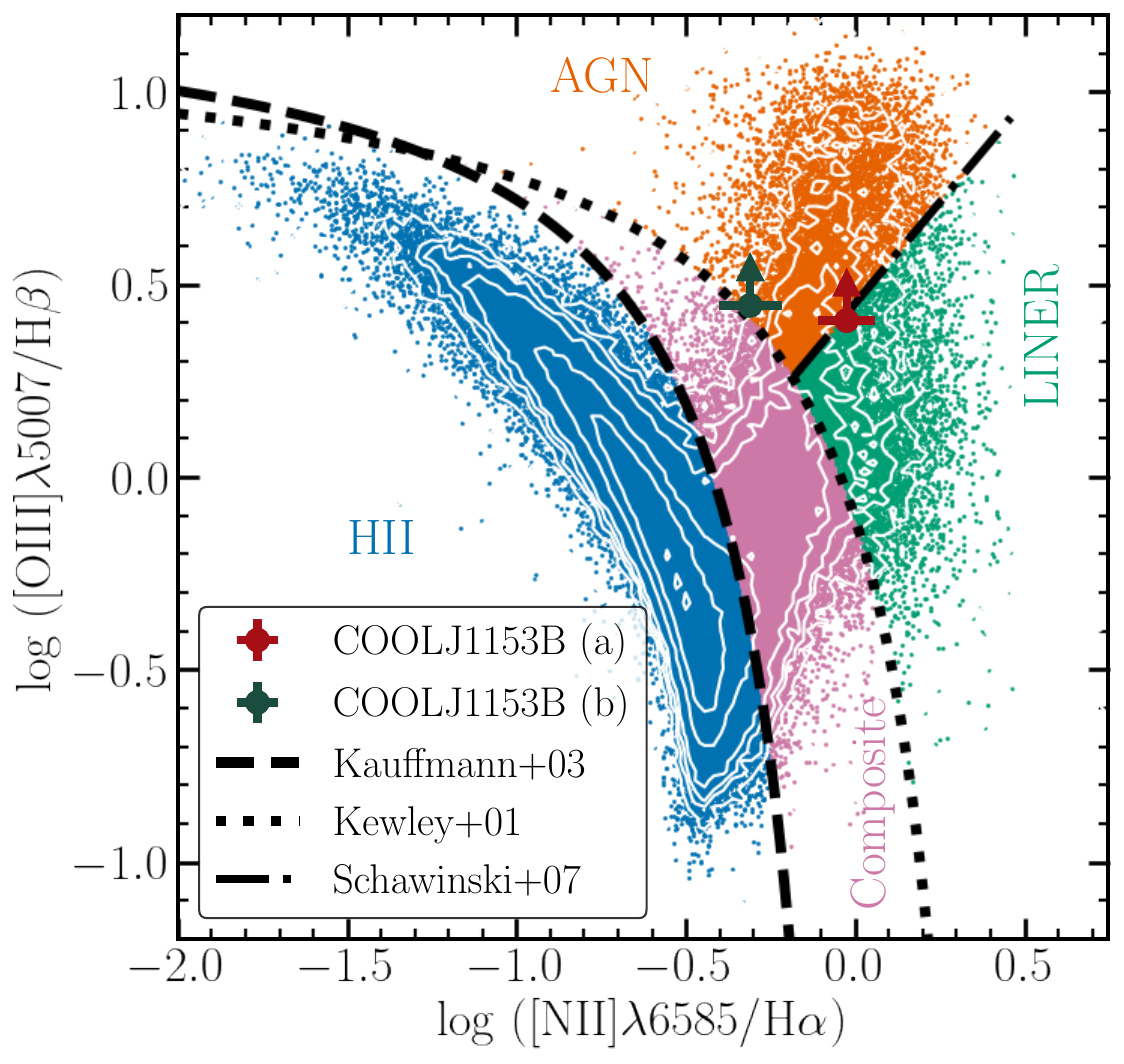}
    \caption{BPT diagram showing the positions of \targB\ (image B1) from (a) one Gaussian component for each line, and (b) one broad and narrow component for H$\alpha$ with the other lines tied to the narrow component. We measure the line fluxes of [\OIII] $\lambda 5007$ and [\NII] $\lambda 6584$ and relative to the Balmer emission lines H$\alpha$ and H$\beta$. The multi-colored distribution are galaxies from SDSS DR16 \citep[][]{Ahumada:2020} with $0.02 < z < 0.2$, color-coded by their classifications: star–forming (``H\,\textsc{ii},'' blue), Composite (magenta), LINER (green), and Seyfert (orange). The dashed line indicates the AGN-star-forming boundary from \cite{Kauffman:2003}, the dotted line indicates the theoretical ``maximum starburst line'' from \cite{Kewley:2001} hollowing out the region of transition-like objects, and the dashdotted line is the demarcation between Seyfert and LINER AGN from \cite{Schawinski:2007}. The SDSS DR16 catalog \citep[][]{Ahumada:2020} and its classifications were accessed using the notebooks from \cite{Juneau:2021}.}
    \label{fig:bpt}
\end{figure}

In Fig.~\ref{fig:bpt}, we present a BPT diagram \citep{Baldwin:1981}, plotting [\NII] $\lambda 6584$ / \Halpha\ and [\OIII] $\lambda 5007$ / \Hbeta\ on the abscissa and ordinate axes derived using the FIRE spectroscopy of \targB\ (image B1). The multi-colored distribution of points are nearby galaxies from the Sloan Digital Sky Survey \citep[SDSS,][]{Blanton:2017, York:2000}, with classifications based on the classification scheme from \citet{Kewley:2001}, \citet{Kauffman:2003}, and \citet{Schawinski:2007}.
We mark the AGN-star-forming galaxy boundary defined by \cite{Kauffman:2003} with a black dashed line.

The distribution of points in Fig.~\ref{fig:bpt} reflects the ionization properties of different galaxy types in the local universe \citep[see e.g.][for a review]{Kewley:2006, Kewley:2019}, as drawn from the SDSS. Star-forming galaxies (blue) trace a curved sequence on the lower-left side of the diagram, where line ratios are consistent with ionization by young, massive stars that ionize the surrounding gas to produce \HII\ regions that source the nebular lines and is therefore marked as ``HII'' in the diagram. Galaxies that fall between the \HII\ and AGN regions are classified as composites, reflecting a mix of ionization sources like star formation, AGN-induced shocks, and/or a hard incident radiation field. AGNs appear toward the upper-right, with elevated [\NII] $\lambda 6584$ / \Halpha\ and [\OIII] $\lambda 5007$ / \Hbeta\ ratios that reflect ionization by harder AGN spectra. Low-ionization nuclear emission-line regions (LINERs) typically occupy the lower-right region of the diagram and are associated with older stellar populations or weak AGN activity. \targB, using either measurement (a) and (b), is clearly offset from the star-forming sequence and falls in the AGN region of the diagram and could plausibly but less likely fall within the LINER classification.

We derive a probabilistic assessment of whether \targB\ belongs to a particular diagnostic domain through Monte Carlo analysis. We do this by generating $10^5$ realizations of $\big(\log([\mathrm{N\,II}]/\mathrm{H}\alpha),\,\log([\mathrm{O\,III}]/\mathrm{H}\beta)\big)$ based on the measured constraints. For the x-coordinate, we draw samples from a Gaussian distribution centered on the measured value $\log([\mathrm{N\,II}]/\mathrm{H}\alpha)$ 
with $\sigma$ equal to the $1\sigma$ measurement uncertainty. 
For the y-coordinate, since H$\beta$ is not detected and only an upper limit is available, $\log([\mathrm{O\,III}]/\mathrm{H}\beta)$ represents a lower limit. We therefore sample from a normal distribution centered at this lower limit value and then truncate the distribution at the lower limit, ensuring that all realizations satisfy $\log([\mathrm{OIII}]/\mathrm{H}\beta)$ is greater than the lower limit value.
Each realization is then classified according to the \citet{Kauffman:2003} and \citet{Kewley:2001} demarcations, with AGN subclasses distinguished using the \citet{Schawinski:2007} line. 
For the two line fitting methods described in Section~\ref{sec:fire_fits} we obtain for method (a): $P_{\mathrm{\ HII}}=0.000$, $P_{\mathrm{composite}}=0.000$, $P_{\mathrm{AGN}}=0.565$, and $P_{\mathrm{LINER}}=0.435$, and for method (b): $P_{\mathrm{\ HII}}=0.000$, $P_{\mathrm{composite}}=0.225$, $P_{\mathrm{AGN}}=0.775$, and $P_{\mathrm{LINER}}=0.000$. Both line fitting methods support that the nebular emission from \targB\ is powered by an AGN. These probabilities represent a conservative estimate, as the true $\log([\mathrm{OIII}]/\mathrm{H}\beta)$ ratio exceeds the measured lower limit. An improved constraint on the $\log([\mathrm{OIII}]/\mathrm{H}\beta)$ ratio from a higher-signal-to-noise spectrum would further strengthen the quasar classification.

\subsection{Spectral Energy Distribution Modeling}

\label{sec:sed_modeling}

Having established that \targA\ and \targB\ are two independent lensed quasars of Type I and II, respectively, we next model their SEDs. We employ the SED fitting code \texttt{CIGALE} \citep[v2025.0,][]{Boquien:2019} to unpack the properties of the stellar population, dust attenuation and re-emission, and the AGN contribution of \targA\ and \targB. We use photometry from 12 bands that we fitted in Section ~\ref{sec:photometry} spanning the rest-frame ultraviolet through the far-infrared (LDSS3+IMACS $g$, $r$, and $z$ and FourStar $J$, $H$, and $K_S$, Spitzer IRAC $3.6\,\mu{\rm m}$ and $4.5\,\mu{\rm m}$, three far-infrared bands from Herschel SPIRE ($250\,\mu$m, $350\,\mu$m, $500\,\mu$m), and ALMA Band 7). For \targB, there is also a detection in ALMA Band 6 (see Appendix \ref{app:alma}). The optical-to-near-infrared bands primarily constrain the stellar population properties and dust attenuation; the far-infrared observations further constrain the dust emission.

The \texttt{CIGALE} SED modeling uses a modular approach to optimize a multi-component model to the measured photometry. We adopt a delayed $\tau$ star formation history (\texttt{sfhdelayed}) module without bursts (setting the optional burst fraction to zero) to perform a fit to a model for the  stellar emission. The stellar population synthesis is computed using the \texttt{cb19} templates \citep[][]{Coelho:2020}, adopting a Chabrier initial mass function \citep[][]{Chabrier:2003} and solar metallicity ($Z = 0.02$). Nebular emission is accounted for with the \texttt{nebular} module, setting the gas metalicity equal to the stellar metalicity. We let dust attenuation follow the \citet{Calzetti:2000} dust law which in the module \texttt{dustatt}\_\texttt{modified}\_\texttt{starburst} can be modified by multiplication with a power-law of slope $\delta_{\rm dust}$. This module is further parameterized by the reddening of emission lines $E(B{-}V)_{\rm lines}$ and a scaling factor relating the reddening of the continuum to the reddening of the emission lines $f = \frac{E(B{-}V)_{\rm continuum}}{E(B{-}V)_{\rm lines}}$. The \texttt{dl2014} module models the absorbed UV and visible energy as two re-emitted dust components: one modeling the diffuse dust heated by the stellar population and one modeling dust linked to star-forming regions. The former component treats the dust as being irradiated by a single radiation field $U_{\rm min}$. The latter component models a radiation field ranging from $U_{\rm min}$ to $U_{\rm max} = 10^7$ following a power-law index $\alpha$ where $\gamma$ sets the mass fraction of the dust in star-forming regions, thereby regulating the relative contribution of the two dust emission components ($1-\gamma$ is the mass fraction of diffuse dust). The parameter ($q_{\rm PAH}$) sets the mass fraction of polycyclic aromatic hydrocarbons producing the strong features seen around $8\mu$m (in the restframe).

AGN emission is modeled using the clumpy torus model \texttt{SKIRTOR2016} \citep[][]{Stalevski:2012, Stalevski:2016}, which provides a physically motivated description of the obscuring dust geometry and the wavelength-dependent AGN visibility. We fix the torus half-opening angle to ${\rm oa} = 40^\circ$ for both sources, a fiducial value consistent with typical AGN torus geometries, and select a disk-type spectrum of 1, corresponding to the \citet{Schartmann:2005} accretion disk prescription. We also fix the outer-to-inner radius ratio to $R = 20$, the fraction of dust mass in clumps to $M_{\rm cl} = 0.97$, the polar-angle dust-density index $pl=1$, and the power-law modification to the equatorial disk spectrum $q=1$. We use the SMC extinction law for the polar dust. The inclination angle is set by the AGN type: $i = 30^\circ$ for \targA\ (Type I view, face-on: $i=[0^{\circ}, 90^{\circ}-\mathrm{oa})$), and $i = 60^\circ$ for \targB\ (Type II view, edge-on: $i=[90^{\circ}-\mathrm{oa}, 90^\circ]$). 

An important constraint in our SED analysis is the relative brightness between the quasar and host galaxy. Since \targA\ is significantly bluer and less dust-obscured than \targB, we use morphological measurements from our \texttt{GALFIT} models to constrain the AGN fraction, ${\rm frac}_{\rm AGN}$, grid explored with \texttt{CIGALE}. The AGN fraction is defined as ${\rm frac}_{\rm AGN} = \frac{L_{\rm dust, AGN}}{L_{\rm dust, AGN} + L_{\rm dust, galaxy}}$, where $L_{\rm dust, AGN}$ and $L_{\rm dust, galaxy}$ are AGN and galactic dust luminosity integrated over the wavelength range $(\lambda_{\rm min}, \,\lambda_{\rm max})$ \citep[][]{Yang:2022}. We compute ${\rm frac}_{\rm AGN} = \frac{F_{\rm PSF}}{F_{\rm PSF} + F_{{\rm S}\acute{e}rsic}}$ for \targA\ using the measured PSF flux as $L_{\rm dust, \, AGN}$ and the S\'ersic component flux as $L_{\rm dust, \, galaxy}$ from our surface brightness decomposition in the $J$, $H$, and $K_s$ FourStar bands ($\lambda_{\rm min} = 1.13~\mu{\rm m}$, $\lambda_{\rm max} = 2.3~\mu{\rm m}$), yielding ${\rm frac}_{\rm AGN} = (0.1925, 0.2145, 0.2105)$ for ($J$, $H$, $K_s$). These measurements suggest a quasar contribution around $\sim20\%$ in the infrared bands. Accordingly, in \texttt{CIGALE} we restrict the AGN fraction grid for \targA\ to ${\rm frac}_{\rm AGN} \in \{0.10, 0.15, 0.20, 0.25, 0.30\}$ to explore values bracketing the measured photometric estimates over the same wavelength range. For \targB, which is much dustier and has a lower expected AGN fraction, we default to letting ${\rm frac}_{\rm AGN}$ be integrated across the full SED. The remaining AGN parameters -- dust optical depth $\tau$, radial dust-density gradient, $\delta_{\rm AGN}$, polar dust extinction $E(B{-}V)_{\rm AGN}$, and dust temperature $T_d$ -- are allowed to vary.

\begin{figure*}[ht!]
\centering
    \includegraphics[width=\textwidth]{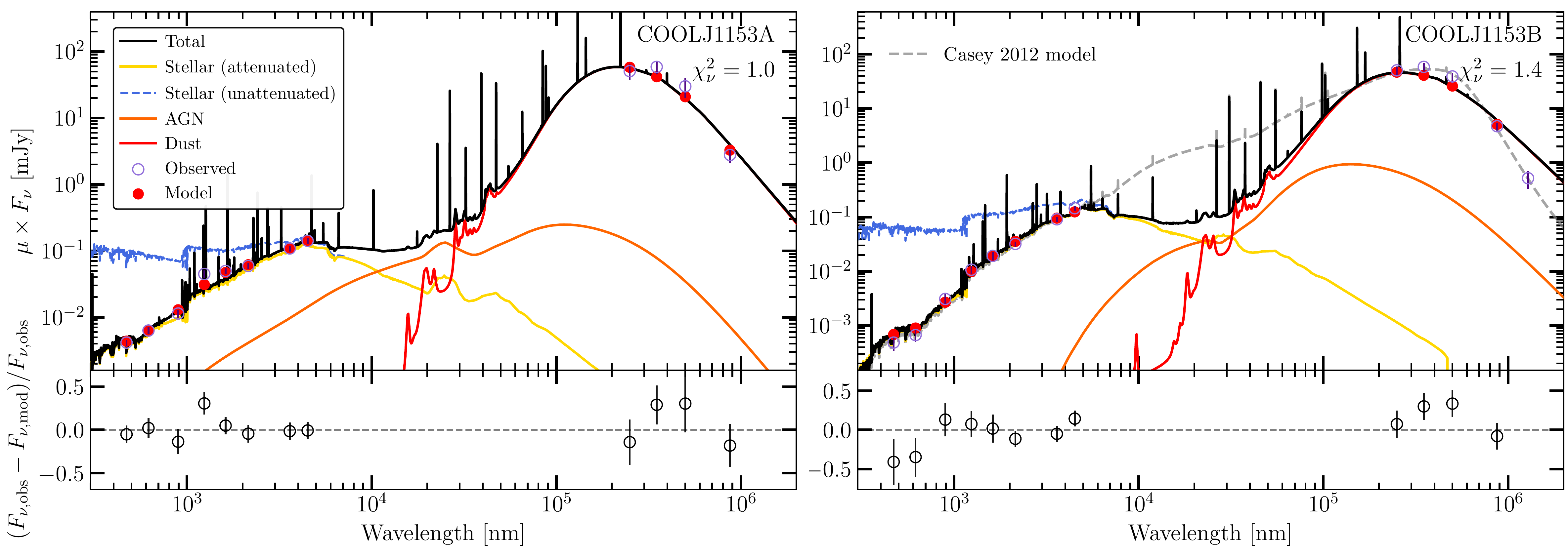}
    \caption{The best-fit SEDs of \targA\ (left panel) and \targB\ (right panel) are consistent with Type I and Type II quasars, respectively. The SED components (stellar, dust, and AGN) are shown in different colored lines. For \targB\, the ALMA Band 6 detection is excluded from the primary fit, as it would require a steeper dust emission curve than the Rayleigh-Jeans law allows, possibly calling for a more complex dust emission model than applied here or an unaccounted-for systematic. An alternative fit using a two-component \citet{Casey:2012} dust emission model with an exceptionally high $\beta = 4.6$ is shown as a gray dashed line.}
\label{fig:cigale_COOLJ1153AB}
\end{figure*}

\begin{table}
\caption{CIGALE Best-Fit SED Parameters}
\label{tab:cigale_params}
\centering
\begin{tabular}{lcc}
\hline
\hline
Parameter & COOLJ1153A & COOLJ1153B \\
\hline
\hline
\multicolumn{3}{c}{Star Formation History (\texttt{sfhdelayed})} \\
\hline
$\tau_{\rm main}$ [Myr] & $1050 \pm 400$ & $8600 \pm 3600$ \\
Age$_{\rm main}$ [Myr] & $690 \pm 300$ & $2520 \pm 420$ \\
\hline
\multicolumn{3}{c}{Dust Attenuation (Calzetti 2000)} \\
\hline
$E(B{-}V)_{\rm lines}$ [mag] & $0.6 \pm 0.1$ & $0.7 \pm 0.05$ \\
$E(B{-}V)_{\rm factor}$ [mag] & $0.7 \pm 0.1$ & $0.8 \pm 0.09$ \\
$\delta_{\rm dust}$ & $-0.23 \pm 0.35$ & $0.0 \pm 0.0$ \\
\hline
\multicolumn{3}{c}{Dust Emission (\texttt{DL2014})} \\
\hline
$q_{\rm PAH}$ & $1.1 \pm 0.63$ & $1.1 \pm 0.62$ \\
$U_{\rm min}$ & $16 \pm 3.6$ & $10 \pm 2.8$ \\
$\alpha$ & $2.4 \pm 0.79$ & $2.9 \pm 0.08$ \\
$\gamma$ & $<0.1$ & $0.9 \pm 0.05$ \\
\hline
\multicolumn{3}{c}{AGN Torus (\texttt{SKIRTOR2016})} \\
\hline
$\tau$ & $9.0 \pm 1.6$ & $5.0 \pm 1.6$ \\
$\delta_{\rm AGN}$ & $-0.1 \pm 0.6$ & $-0.5 \pm 0.7$ \\
$\mathrm{frac}_{\rm AGN}$ & $0.2 \pm 0.07$ & $<0.02$ \\
$E(B{-}V)_{\rm AGN}$ [mag] & $0.16 \pm 0.12$ & $0.14 \pm 0.11$ \\
$T_d$ [K] & $82 \pm 34$ & $52 \pm 42$ \\
\hline
\end{tabular}
\raggedright
\textbf{Notes.} Parameters represent Bayesian estimates with $1\sigma$ uncertainties. We caution the reader that systematic uncertainties may be larger.
\end{table}

We employ an iterative approach to constrain the multidimensional parameter space by doing a module-by-module search and a grid-based parameter optimization. First, we isolate and optimize the parameter space for each SED component module independently, identifying physically reasonable regions that produce viable fits to the observed photometry. We then perform a grid search over the combined parameter space, focusing on regions identified in the previous step, refining our search until convergence to the global best-fit solution. The grid parameter spaces used are listed in Appendix~\ref{app:cigale_grid}. To account for systematic uncertainties in the photometry and avoid over-constraining the model parameters, we include a 10\% additional uncertainty on our reported photometry (see Appendix~\ref{app:photometry}). Due to the uncertainties in the magnifications, we choose to perform the fit for the observed fluxes without demagnifying them.

The ALMA Band 6 detection in \targB\ cannot be accommodated by the \texttt{dl2014} dust-emission model and requires a dust tail that is steeper than Rayleigh-Jeans. We therefore exclude it from the nominal fit. To investigate this tension, we construct an alternative model using a \citet{Casey:2012} dust template with an exceptionally high dust opacity index $\beta = 4.6$ (shown as a gray dashed line in Fig.~\ref{fig:cigale_COOLJ1153AB}). This value is notably uncommon -- typical AGN host galaxies show $\beta \sim 1.6$–$2.0$ -- and would correspond to a dust temperature of only $\sim 10 \, \mathrm{K}$. While this solution appears physically implausible, it may be an intriguing direction for further study and may suggest either an unusual dust composition in \targB\ or the need for more sophisticated dust models to capture this behavior.

To identify the best-fit SED parameters, \texttt{CIGALE} minimizes $\chi^2$ across the parameter grid and then performs a local Bayesian optimization around the global minimum. This yields likelihood-weighted means and standard deviations, which we report in Table~\ref{tab:cigale_params}. To test the robustness of our results against potentially underreported telescope systematic uncertainties, we additionally ran fits with a 5\% error floor in addition to the default 10\% uncertainty on fluxes. The fitted parameters remained consistent with the nominal model, confirming that our main conclusions are insensitive to reasonable variations in the systematic uncertainty floor.

The best-fit SEDs are presented in Fig.~\ref{fig:cigale_COOLJ1153AB}. By integrating the SEDs, we find that both sources exhibit high bolometric luminosities: \targA\ has $L_{\rm bol} = 2.2 \times 10^{46}$ erg s$^{-1}$ and \targB\ has $L_{\rm bol} = 2.7 \times 10^{46}$ erg s$^{-1}$.
These are not yet corrected for magnification, which we will do in the next section.

\subsection{Lens Model} \label{sec:slmodel}

\begin{figure*}
    \centering
    \includegraphics[width=\linewidth]{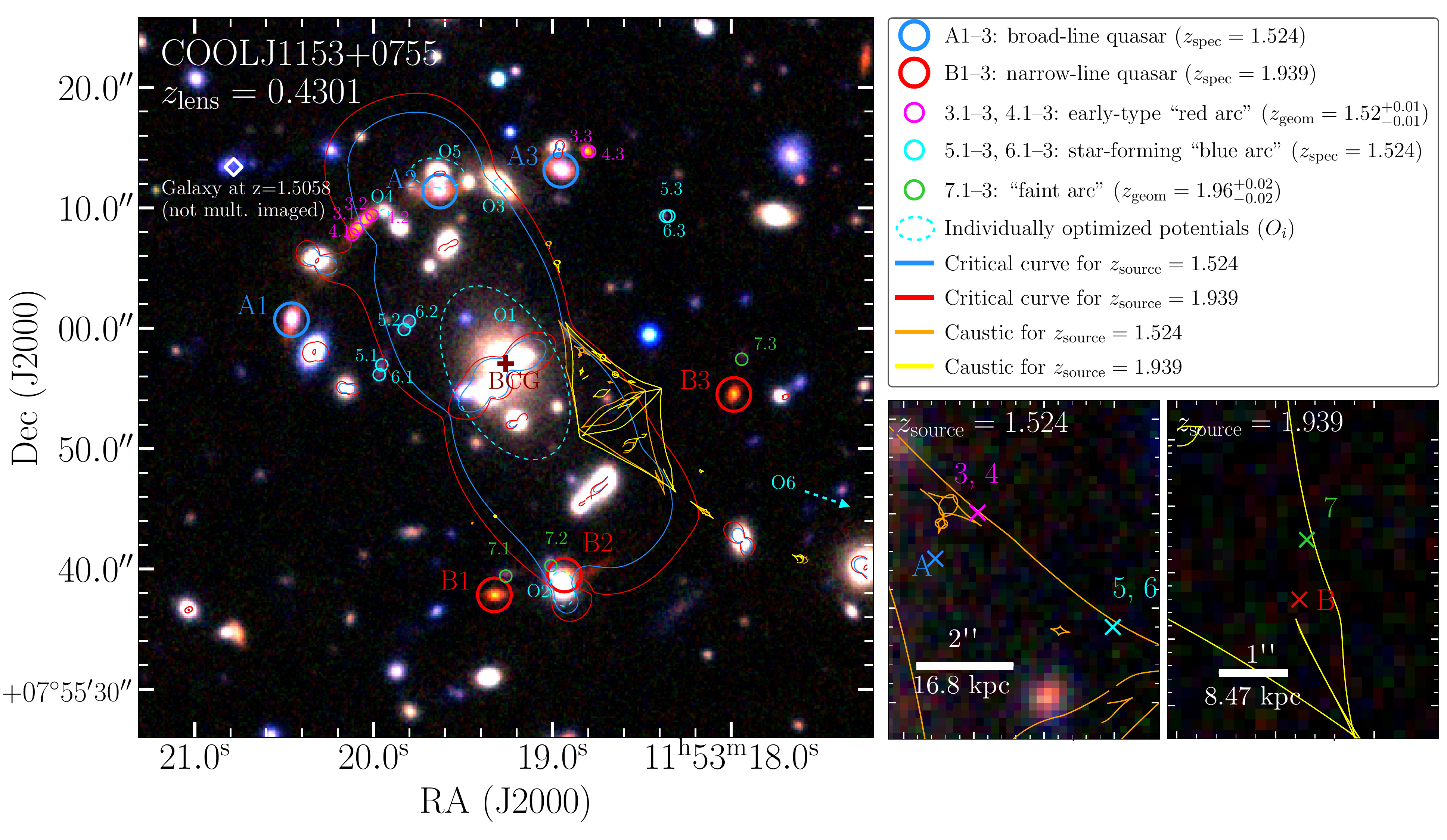}
    \caption{The lens model successfully reproduces the observed image configurations. \textit{Left:} RGB color image (combining $g+r$ in blue, $z+J$ in green, $H+K_s$ in red from LDSS3 and IMACS) overlaid with the critical curves and caustics of the strong-lensing model of \targ. The multiply-imaged sources are marked: \targA\ in blue, \targB\ in red, and the multiply-imaged ``blue arc'' at the same redshift as \targA\ in cyan circles, with additional constraints in other colors. A galaxy northeast of \targA\ at $z=1.5058$ and lying outside the cluster's critical radius (and therefore not multiply imaged) is marked with a white diamond. \textit{Bottom right:} Source planes for each of the quasar redshifts showing their local group environment. Each point is calculated from the mean of the delensed image position from the best-fit lens model.}
    \label{fig:rgbimage}
\end{figure*}

We use the lens modeling software \texttt{lenstool} \citep{Jullo:2007} to model the mass distribution of the foreground lens potential. Throughout this work, we configure \texttt{lenstool} to optimize models in the image plane. 
Capturing the detailed substructure of the lens potential is crucial to produce a good lens model. We therefore build our lens model iteratively from several individual components, and for each halo mass, we adopt a pseudo-isothermal elliptical mass distribution \citep[PIEMD, described in][]{Eliasdottir:2007}. A PIEMD is constrained by its position (x, y), ellipticity ($\varepsilon$), position angle ($\theta$), core radius ($r_{\mathrm{core}}$), cut radius ($r_{\mathrm{cut}}$), and velocity dispersion ($\sigma_v$). We identify seven background galaxies that are multiply imaged and use their image positions as constraints in the lens model. For each galaxy, we assign a unique source ID (\textit{X}), and for each image of that source, we assign a corresponding image ID (\textit{Y}), resulting in identifiers of the form ``X.Y.'' To distinguish the quasars from the other galaxy images, we assign the quasar IDs of the format ``ZY'' where \textit{Z} is a letter designating the quasar source, in contrast to the ``X.Y'' format used for the other galaxies. When we identify individual clumps across multiple images of the same source, we treat them as separate constraints. The redshifts of the sources for our constraints are set to $z=1.524$ for \targA\ (A1-3), $z=1.525$ for the blue arc (5.X and 6.X), and $z=1.939$ for \targB\ (B1-3). The redshifts, $z_{\rm geom}$, of sources 3, 4 and 7 are free to vary within broad priors around the redshifts of \targA\ and \targB. In Fig.~\ref{fig:rgbimage}, we represent each source with a distinct color for clarity.

We begin by modeling a single large-scale halo centered on the BCG, constrained within a $10''\times10''$ box around its measured position. This component reproduces the approximate Einstein radius required to explain the separation of the quasar images. We fix $r_{\mathrm{cut}}$ to 1500 kpc since the mass density beyond this radius is outside the area that can be constrained by the image constraints. All other parameters are free to vary within broad priors.

After including the large cluster-scale halo, we also need to optimize the individual masses of cluster member galaxies as these, too, significantly influence the detailed gravitational potential of the lensing cluster. However, directly optimizing the mass of each member galaxy is computationally expensive and introduces more free parameters than there are constraints. A common approach, then, is to scale their masses based on their observed brightnesses, allowing for a more efficient parameterization of their contribution to the cluster potential that is physically grounded in their relative brightnesses. To implement this, we use the LS-DR9 TRACTOR $z$-band catalog
to identify foreground cluster members using the red-sequence technique \citep[][]{Gladders:2000} where we use their colors to identify their association with the cluster. Our final sample of 42 cluster galaxies (not including four individually optimized halos, noted later) are collectively optimized in \texttt{lenstool} by scaling their velocity dispersions, $\sigma_v$, and their cut radii, $r_{\rm cut}$, by their measured luminosities, $L$, as $\sigma_v = \sigma^{*}_v \big(\frac{L}{L^{*}}\big)^{\frac{1}{4}}$ and $r_{\rm cut} = r_{\rm cut}^* \big(\frac{L}{L^{*}}\big)^{\frac{1}{2}}$. The pivot parameter, $L^*$, is the magnitude of an $L^{\star}$ galaxy at the lens redshift and is set to $19.53$. The arguments x, y, $\varepsilon(a,b)$, and $\theta$ are fixed to their measured values. In \texttt{lenstool}, the ellipticity of the PIEMD model, $\varepsilon$, is related to the semi-major, $a$, and semi-minor, $b$, axes by $\varepsilon = \frac{a^2 - b^2}{a^2 + b^2}$. We fix their core radii to a small value, $r_{\rm core} =0.15 \, {\rm kpc}$.

We observe that this does not quite remove the scatter in the predicted image positions, so to capture local mass concentrations within the cluster we iteratively ``free'' four galaxy-scale halos by individually optimizing their parameters. For these halos, we allow $\sigma_v$, $\varepsilon$, $\theta$, $r_{\mathrm{core}}$, and $r_{\rm cut}$ to vary within broad priors with one exception: we fix $\theta$ for two of the halos given their high ellipticity which makes it easier to constrain their position angles.

To capture additional shear from the second cluster southwest of the \targ\ cluster, we include a second cluster-scale spherical halo, centered $180''$ west and $80''$ south of the BCG, which we allow to vary within a $20''\times20''$ box. This halo is necessary to reproduce the three \targB\ images seen in the ground-based imaging. We fix this halo's ellipticity and core radius to $\varepsilon = 0$ and $r_{\rm core} = 0$ kpc (effectively making it a singular isothermal ellipsoid, SIE), respectively, and set $r_{\rm cut} = 1500$ kpc. Its velocity dispersion, $\sigma_v$, is allowed to vary within broad priors. Although this second halo lies at a different redshift ($z=0.5766 \pm 0.0007$), \texttt{lenstool} does not allow for optimizing multiple mass planes; we therefore assign it the same redshift as the main cluster. 

We then proceed to optimize using \texttt{lenstool}'s built in Bayesian framework using $N_{\rm chain}=10$ Markov chains for $N_{\rm iter}=200$ iterations with 200 burn-in steps.

Our best-fit lens model optimizes the predicted images in the image plane and provides a well-constrained mass distribution across the lensing cluster (30 constraints, 29 free parameters, image-plane RMS$=0.23''$, $\chi_{\nu}^2 \sim 0.95$). See Table~\ref{tab:lensmodelparams} in Appendix~\ref{app:cornerplot} for the resulting optimized parameters. We show the critical curves (caustics) for sources at the quasar redshifts $z=1.524$ in blue (orange) and $z=1.939$ in red (yellow) in Fig.~\ref{fig:rgbimage}. Each individually optimized halo, $O_i$, is shown with cyan ellipses with position angle and ellipticities of the best-fit halos and relative sizes set by their velocity dispersions.

\begin{figure}[h!]
    \centering
    \includegraphics[width=\columnwidth]{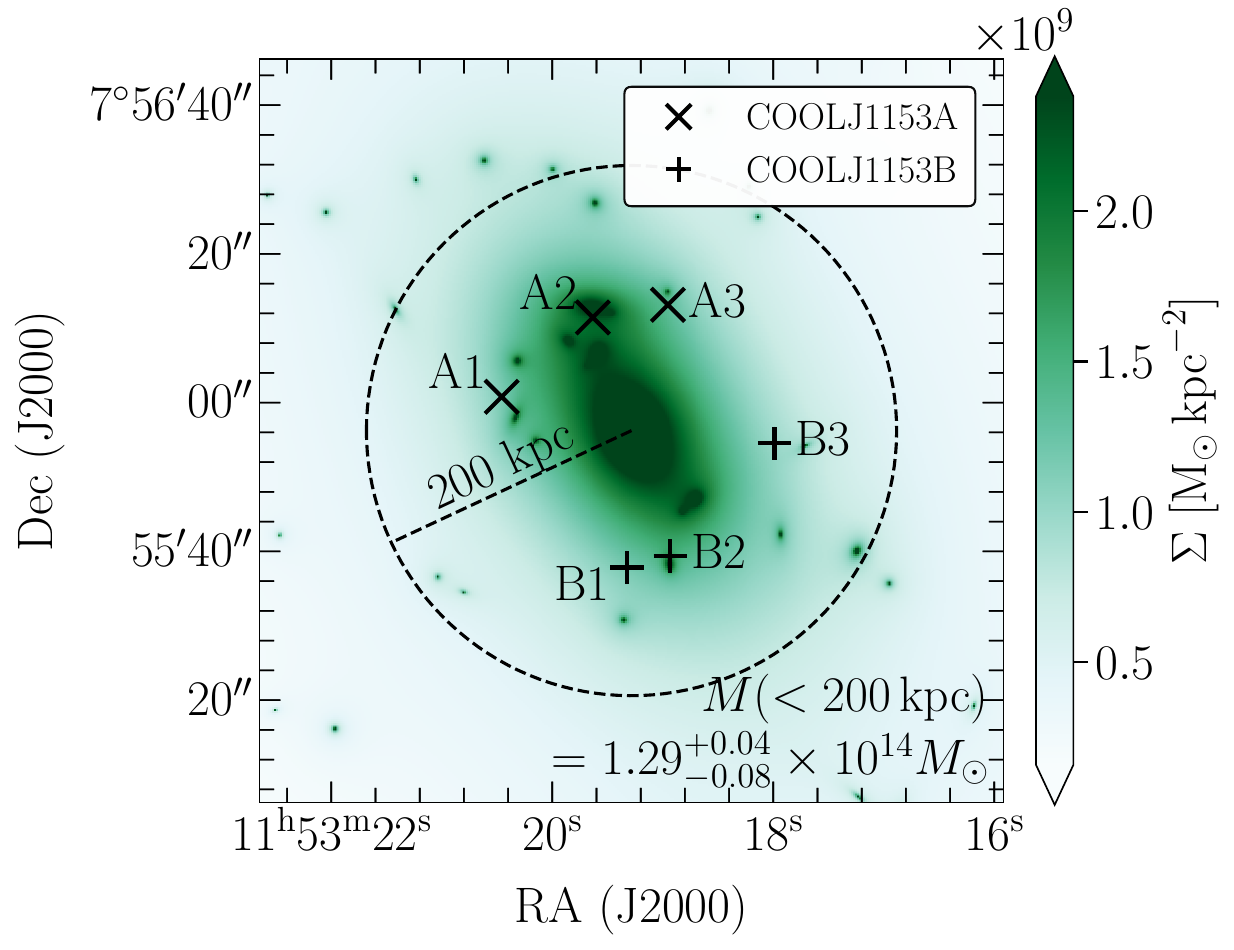}
    \caption{Projected mass density, $\Sigma$, of our best-fit lens model in a $80''\times80''$ box centered on the BCG. We report the integrated mass density of the lens within 200 kpc, $M(<200 \, {\rm kpc})$, and show the image locations with crosses for \targA\ and \targB\ to guide the eye.}
    \label{fig:mass_lens}
\end{figure}

\begin{figure}[h!]
    \centering
    \includegraphics[width=\columnwidth]{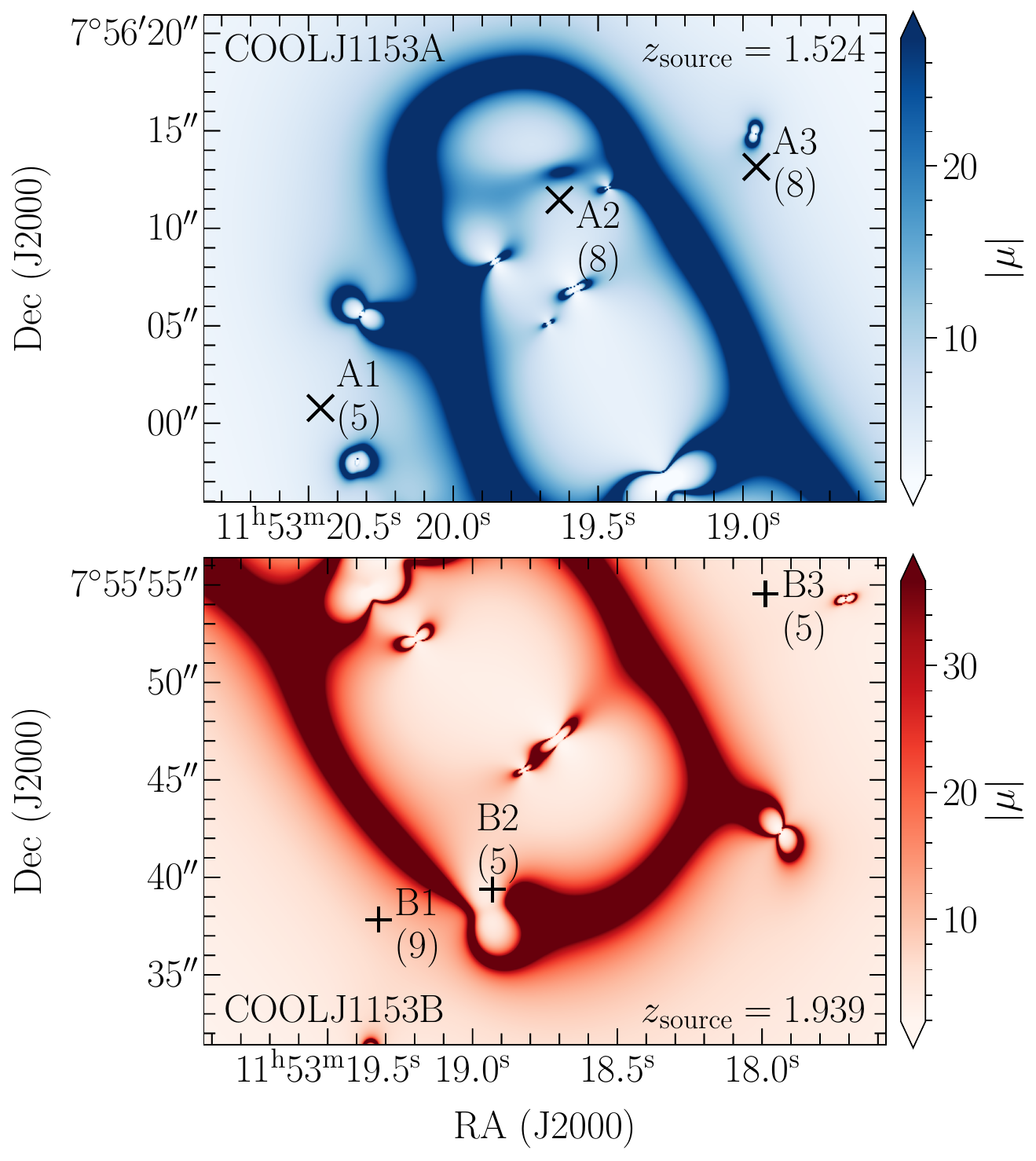}
    \caption{Absolute value magnification, $|\mu|$, in $35''\times25''$ cutouts for \targA\ \textit{(top)} and \targB\ \textit{(bottom)}. The color scaling is clipped
    for visual clarity; the pointed ends of the colorbars indicate that magnifications exceed the displayed limits, e.g. near critical curves. The magnification at each quasar image position is shown in parentheses next to its label.}
    \label{fig:magnification_lens}
\end{figure}

The parameter uncertainties are derived from the full MCMC posterior distribution of \texttt{lenstool}; we derive the 1$\sigma$ confidence intervals for each parameter, as reported in Appendix \ref{app:cornerplot}. To estimate uncertainties on derived quantities like magnifications, time delays, and mass, we randomly sample $100$ of the $N_{\rm chain} \times N_{\rm iter}=2000$ lens model realizations and use these $100$ models to reconstruct the distributions of these quantities. From the lens model mass distribution of the primary cluster, we infer a mean projected mass within 200 kpc of the BCG center of $M (<200 {\rm kpc}) = 1.33^{+0.04}_{-0.02}
\times 10^{14} {\rm M}_{\odot}$ (see Fig.~\ref{fig:mass_lens}) and within 500 kpc of $M (<500 {\rm kpc}) = 3.30^{+0.15}_{-0.07}
\times 10^{14} {\rm M}_{\odot}$. This is consistent with the expected mass of a massive galaxy cluster. We present magnification maps for $z_{\rm source} = [1.524, 1.939]$ from our best-fit lens model in Fig.~\ref{fig:magnification_lens}. In this paper, we produce magnifications using \texttt{lenstool}'s \texttt{ampli} runmode. This approach provides accurate magnification calculations for unresolved compact sources like quasars. The lens model magnifications of the \targA\ quasar images are $\mu_{A1} = 5.5^{+1.5}_{-1.0}$, $\mu_{A2} = 5.4^{+2.5}_{-1.6}$, and $\mu_{A3} = 8.4^{+4.2}_{-2.2}$. For the \targB\ quasar, the magnifications are $\mu_{B1} = 6.8^{+1.9}_{-1.3}$, $\mu_{B2} = 5.2^{+2.7}_{-1.7}$, and $\mu_{B3} = 4.1^{+1.0}_{-0.7}$. Correcting the bolometric luminosity measurements from Section~\ref{sec:sed_modeling} for the magnification factors of images A1 and B3 we calculate intrinsic bolometric luminosities of $L_{\rm bol} = 4.0^{+0.9}_{-0.9} \times 10^{45}$ erg s$^{-1}$ for \targA\ and $L_{\rm bol} = 6.6^{+1.3}_{-1.3} \times 10^{45}$ erg s$^{-1}$ for \targB. 

Our lens model for the mass distribution of the foreground galaxy cluster predicts the expected arrival time delays of the quasar images. Using the equation below, we can estimate the absolute time delay across the full field of our best-fit lens model from the deflection fields for a source at the redshifts of the two quasars ($z_{\rm source}=[1.524, \, 1.939]$).
The expected arrival-time delays, the \textit{Fermat potential}, can be derived from the lensing potential, $\psi(\boldsymbol{\theta})$ \citep[see reviews by][]{Schneider:1985, Narayan:1996}. For an image-plane position $\boldsymbol{\theta}$ and source position $\boldsymbol{\beta}$, the absolute time delay surface is given by

\begin{equation}
\label{eqn:tau}
\tau(\boldsymbol{\theta};\boldsymbol{\beta})
= \frac{1+z_{\rm l}}{c}\,
\frac{D_{\rm l}\,D_{\rm s}}{D_{\rm ls}}\,
\Bigg[
\frac{1}{2}\,\bigl\lvert \boldsymbol{\theta}-\boldsymbol{\beta} \bigr\rvert^{2}
\;-\;
\psi(\boldsymbol{\theta})
\Bigg] ,
\end{equation}
where $z_{\rm l}$ is the lens redshift and $D_{\rm l}$, $D_{\rm s}$, and $D_{\rm ls}$ are angular-diameter distances to the lens, the source and from the lens to the source, respectively. Note that the time-delay surface is sensitive to the source-plane position, and the lens model predicts slightly different source positions for each of the quasar images. We choose to adopt the mean position of these predicted source positions of each source as $\boldsymbol{\beta}$.

\begin{figure}[ht!]
    \centering
    \includegraphics[width=\columnwidth]{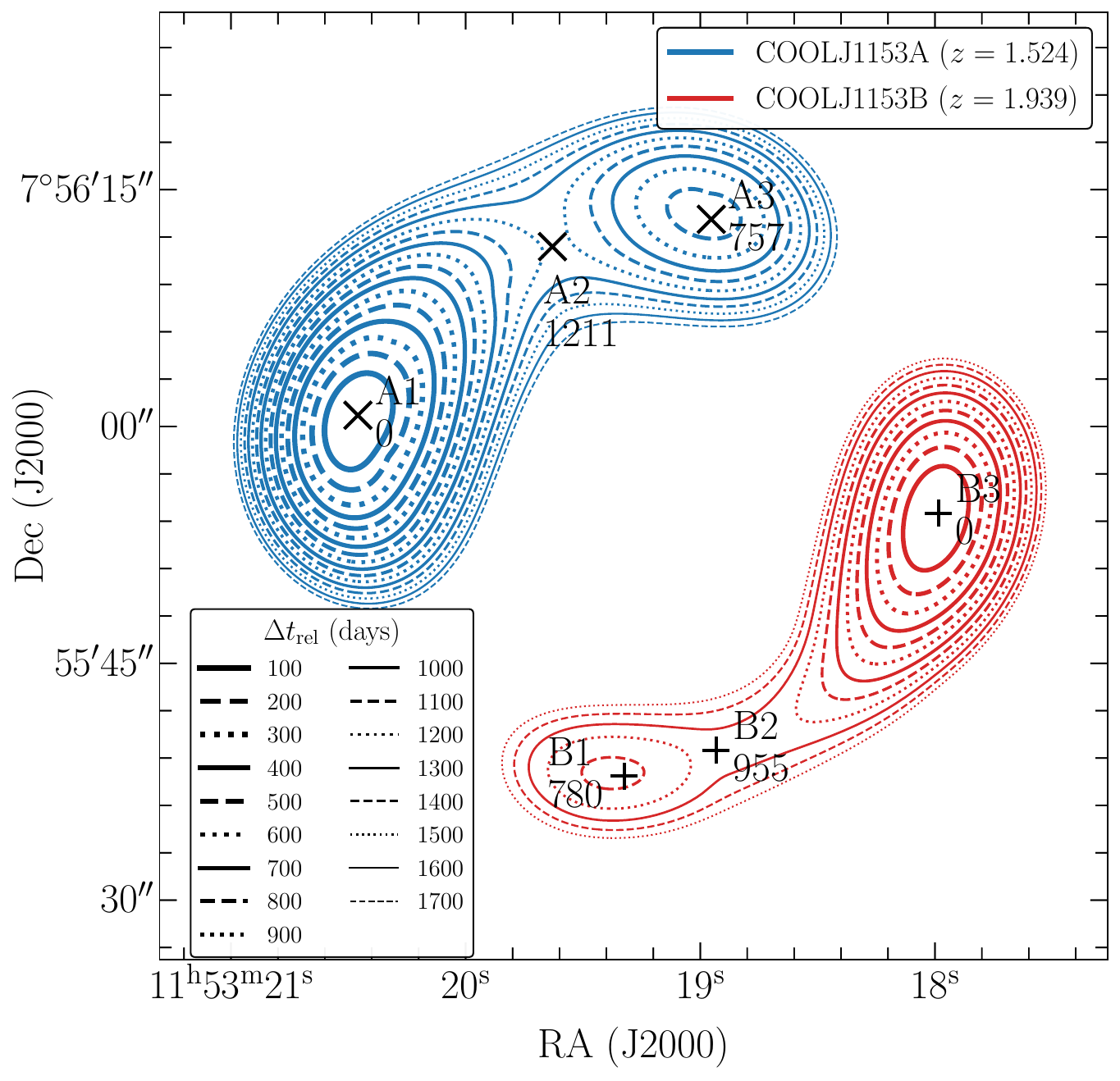}
    \caption{Time delay surfaces relative to the leading images of the two quasars, $\Delta t_{\rm rel, \, A}$ and $\Delta t_{\rm rel, \, B}$ (see Eqns. \ref{eqn:tau} and \ref{eqn:tau_rel}), in \targ\ from our best-fit lens model in units of days. Each set of contours corresponds to $\Delta t_{\rm rel}$ for a source at source redshift $z_{\rm source}=1.524$ (\targA, $\Delta t_{\rm rel, \, A}$) and $z_{\rm source}=1.939$ (\targB, $\Delta t_{\rm rel, \, B}$). Observed images of the quasars, appearing at critical points on the $\Delta t_{\rm rel}$ surfaces, are marked with crosses (A1-3 and B1-3) and annotated with their best-fit time delays. The box is $60''\times60''$ centered on the BCG.}
    \label{fig:fermat}
\end{figure}

The relative time-delay surface is then obtained by subtracting the time delay of the leading image:

\begin{equation}
\label{eqn:tau_rel}
\Delta t(\boldsymbol{\theta}\,|\,\boldsymbol{\theta}_{\rm lead})
\equiv
\tau(\boldsymbol{\theta};\boldsymbol{\beta})
-
\tau(\boldsymbol{\theta}_{\rm lead};\boldsymbol{\beta}) ,
\end{equation}
where $\boldsymbol{\theta}_{\rm lead}$ is the image-plane position of the leading image. Fig.~\ref{fig:fermat} shows $\Delta t$ in days, relative to the leading image for each quasar.

According to our lens model, the leading image in \targA\ is image A1, and in \targB\ it is image B3. The time delays relative to these images are then, $\Delta t_{\rm A2, A1} = 1253^{+42}_{-50}$, $\Delta t_{\rm A3,A1} = 789^{+55}_{-59}$, $\Delta t_{\rm B1,B3} = 733^{+104}_{-139}$, and $\Delta t_{\rm B2,B3} = 915^{+97}_{-150}$. An updated lens model incorporating space-based observations presented later in this work provides refined time-delay estimates, discussed in Section~\ref{sec:space}.

\subsection{Compact Structure in the Quasar Environments}
\label{sec:quasar_environments}

Our observations suggest that \targA\ resides in a densely populated environment. We find two other galaxies at the same redshift: a triply imaged blue star-forming galaxy (nicknamed the ``blue arc,'' 5.1-3 and 6.1-3 in Fig.~\ref{fig:rgbimage}), and an early-type galaxy (``red arc,'' 3.1-3 and 4.1-3). The multiply-imaged ``blue arc'' is spectroscopically confirmed to lie at the same redshift as the quasar. FIRE spectroscopy places the blue arc at $z = 1.525$ based on a single line (H$\alpha$). The early-type galaxy is not spectroscopically confirmed, as we see no clear emission lines in attempted FIRE and LDSS3 multislit spectra, but the lens model predicts a redshift of $z = 1.52^{+0.01}_{-0.01}$. Deeper spectroscopy may be needed to resolve the continuum and sample any absorption lines we would expect if it indeed is an early-type galaxy. We also identify a bright blue galaxy with LDSS3 longslit spectroscopy at $z = 1.5058$, placing it ``nearby'' but outside the group. It is located outside the critical radius of the lens and thereby not multiply imaged but magnified by $|\mu|\sim 3 - 4$ (marked by a white diamond in Fig.~\ref{fig:rgbimage}). We do not observe any lines in attempted FIRE spectroscopy of the ``faint arc'' (source 7). However, the lens model favors a redshift, $z=1.96^{+0.02}_{-0.02}$, consistent (within the error bars) with placing it in the neighborhood of \targB.

We delens the quasar images and companion galaxies to the redshifts $z_{\rm source} = 1.524$ and $z_{\rm source} = 1.939$ and reconstruct their source-plane configurations (see right panels in Fig.~\ref{fig:rgbimage}). We find compact projected source-plane separations: \targA\ is separated from the blue arc (source 3, 4) by $\sim 11$ kpc and the red arc (source 5, 6) by $\sim 34$ kpc. Assuming the same redshifts, \targB\ is separated from source 7 by $\sim 8$ kpc. This strongly suggests that both quasars exist in overdense environments.
Close associations among galaxies are not merely incidental; rather, dense environments are thought to influence AGN activity through interactions, tidal encounters, and mergers, which both trigger and regulate black-hole growth \citep[e.g.,][]{Hopkins:2010}. 

\section{Space Telescope Imaging} \label{sec:space}

\begin{figure*}[ht!]
    \centering
    \includegraphics[width=\linewidth]{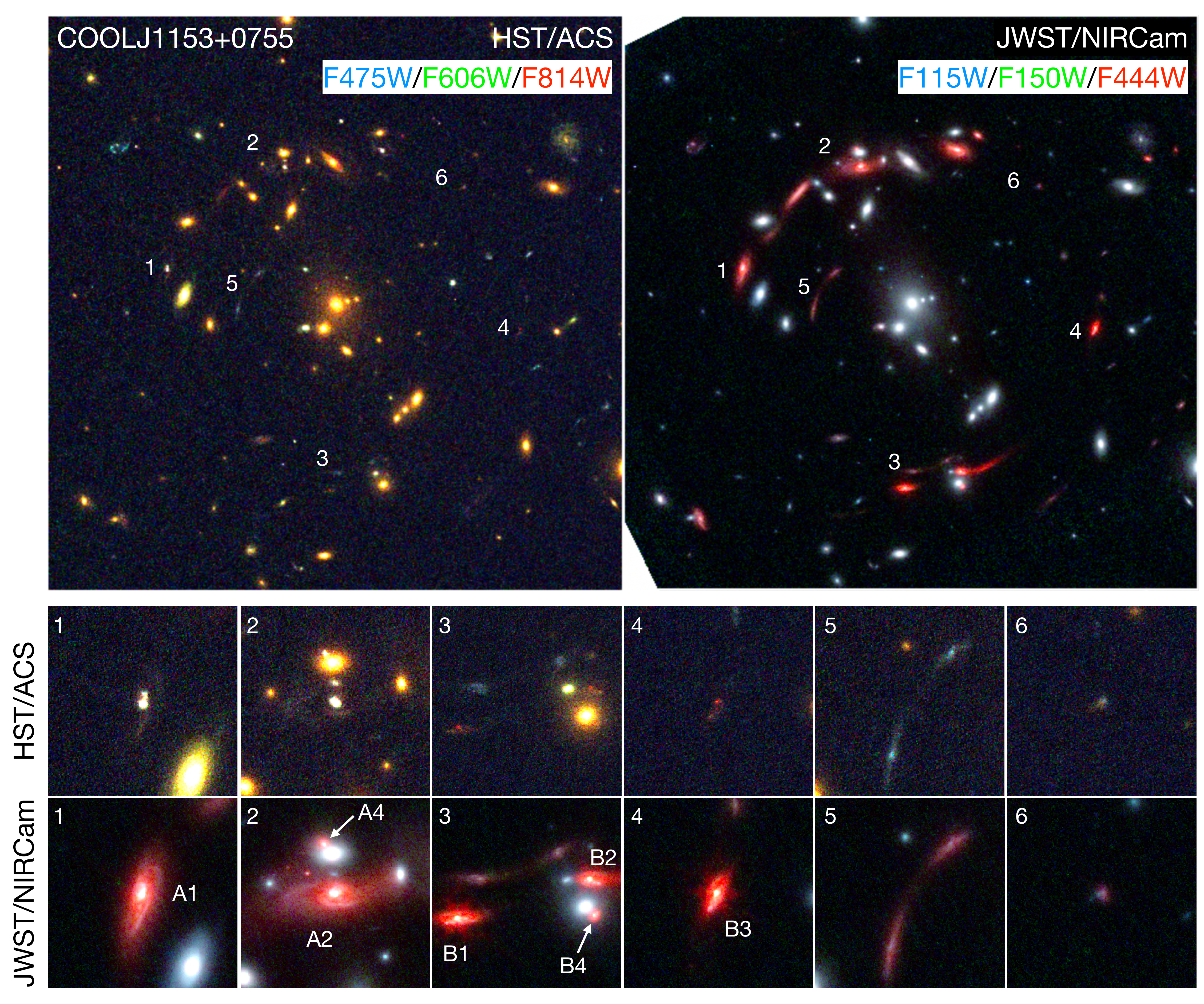}
    \caption{$1'\times1'$ images from HST/ACS (\textit{top left}) and JWST/NIRCam (\textit{top right}) centered on the BCG in \targ.
    \textit{Top left:} The blue, green, and red channels correspond to the F475W, F606W, and F814W bands, respectively. 
    \textit{Top right:} The blue, green, and red channels correspond to the F115W, F150W, and F444W bands, respectively.
    Enlarged panels highlight selected regions of \targ: (1) northeastern image of \targA, (2) northern images of \targA, (3) \targB\ southeastern and southern images, (4) \targB\ western image, (5) blue arc, and (6) blue arc counterimage.
    The point-source morphology of \targB\ (see enlarged panels 3 and 4) is especially prominent in the infrared JWST/NIRCam bands where dust attenuation is minimal, as evidenced by the characteristic diffraction spikes of the JWST point-spread function, which further confirms our hypothesis from ground-based data that \targB\ is a quasar. A fourth image of each quasar appears in close line-of-sight proximity to cluster members (panels 2 and 3), while two conspicuous point sources are visible in the outskirts of the northern image of \targA\ (panel 2).}
    \label{fig:hstjwstimages}
\end{figure*}

Late in the development of this manuscript, space-based imaging from HST (GO 17431, PI: Michael Gladders) and JWST (GO 6675, PI: H\r{a}kon Dahle) became available (see Fig. \ref{fig:hstjwstimages}). Here, we present an initial analysis of \targ\ based on these data, building upon the conclusions from the ground-based data. In doing so, we develop an updated version of our lens model.

High-resolution HST/ACS and JWST/NIRCam imaging provide direct confirmation that \targB\ is a quasar. The NIRCam data were reduced as described in \citet[][]{Florian:2025}. In the HST and JWST images, the unresolved central point source is clearly isolated from surrounding extended emission, and in the NIRCam imaging, it exhibits the characteristic diffraction spikes of JWST observations of compact nuclear sources \citep[see e.g.,][]{Ortiz:2024}. Additionally, the point source exhibits distinctly redder colors compared to the surrounding diffuse host galaxy -- a signature of dust obscuration. The HST F814W and near-infrared JWST observations penetrate the dust layers that obscure the nucleus at the wavelengths of HST F475W and F606W, directly revealing the central quasar. This observed optical extinction combined with the relatively low opacity in the infrared confirms that \targB\ is a heavily dust-obscured quasar.

Despite the imaging being relatively shallow ($1021 {\rm s}$ in the deepest filter, F444W) compared to other JWST imaging, it reveals a \textit{fourth image} of both quasars in close line-of-sight separation from foreground cluster members (Fig. \ref{fig:hstjwstimages}, panels 2 and 3). This demonstrates the utility of space-based data for WSLQ systems; the superior resolution and sensitivity of JWST and HST are necessary to isolate the light of these fainter images from the glare of foreground galaxies \citep[see also][]{Inada:2005, Sharon:2017}.

\begin{figure}[ht!]
    \centering
    \includegraphics[width=\linewidth]{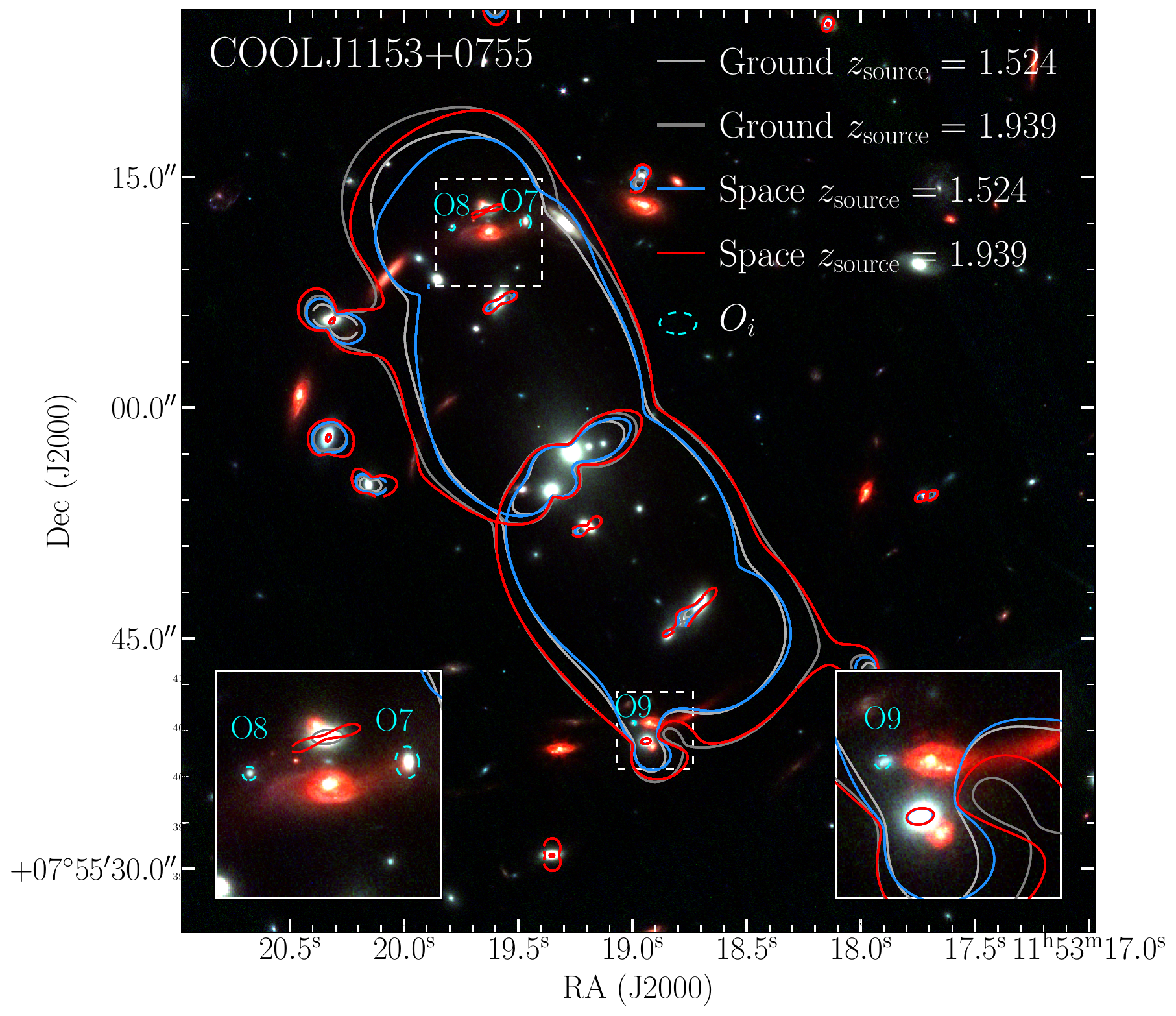}
    \caption{Progression of lens model by adding space-based model constraints, like improved centroid measurements of quasar images and nearby halo potentials. Here, we show an updated lens model based on improved constraints from the space-based data (blue and red) compared to the critical curves from Fig.~\ref{fig:rgbimage} (light gray and gray) overlaid on the JWST/NIRCam (F115W, F150W, F444W) imaging. Cyan dashed ellipses mark the three halos that were individually optimized in the space-based model (and not in the ground-based models). Enlarged panels show regions around the fourth images identified from space-based imaging.}
    \label{fig:jwst_lensmodel}
\end{figure}

These lensed images provide additional constraints for the lens model.
We therefore refine our lens model by incorporating the newly identified fourth images, A4 and B4, as constraints. We measure the precise centroid positions of images A1-4 and B1-4 and the positions, position angle and ellipticity of their closest foreground cluster galaxies in the JWST/F150W image. 
The ground-based model does not reproduce the positions of the fourth images, even when they are included as constraints and the model re-optimized. Reproducing the position of A4 requires freeing the halo west of A2/A4 (O7 in Fig.~\ref{fig:jwst_lensmodel}) from the cluster-member scaling relation and optimizing it as an independent mass component.
We note that a spiral galaxy attributed to this halo appears slightly redder than the cluster members in the space-based imaging which could be from a variety of factors, such as dust, age and metallicity. With our LDSS3 spectrum, we derive a redshift for this galaxy of $z=0.57253$, indicating it is located behind the cluster, but we treat it as a cluster member in our lens model due to the limitations of \texttt{lenstool} (see \citet{Raney:2020} for a discussion of the validity of this approach). We additionally include a new individually optimized halo at the location of a foreground galaxy southeast of A4 (O8 in Fig.~\ref{fig:jwst_lensmodel}). Both potentials are allowed to vary in $r_{\rm core}$, $r_{\rm cut}$, and $\sigma_v$. We add an individually optimized halo northeast of B4 (O9 in Fig.~\ref{fig:jwst_lensmodel}) with $r_{\rm cut} = 50 \, {\rm kpc}$ fixed and $r_{\rm core}$, $\sigma_v$ and $\epsilon$ as free parameters.
The close image splitting of these images, A2/A4 and B2/B4, by cluster members raises the intriguing possibility of constraining the central SMBH mass in these foreground galaxies. While we have experimented with adding point sources to the two cluster galaxies that produce the image splitting, we find that this preliminary model does not require such a component to reproduce the fourth quasar images.

\begin{table*}
\begin{center}
\caption{Magnification and Relative Time Delays}
\label{tab:lensmodelupdate}
\begin{tabular}{l|cc|cc|cc|cc}
\hline
\hline
& \multicolumn{4}{c|}{\textbf{3-image model (ground-based constraints)}} & \multicolumn{4}{c}{\textbf{4-image model (space-based constraints)}} \\
\hline
& \multicolumn{2}{c|}{$|\mu|$} & \multicolumn{2}{c|}{$\Delta t_{\rm rel}$} & \multicolumn{2}{c|}{$|\mu|$} & \multicolumn{2}{c}{$\Delta t_{\rm rel}$} \\
\hline
Object ID & Best & [16, 50, 84]\% & Best & [16, 50, 84]\% & Best & [16, 50, 84]\% & Best & [16, 50, 84]\%\\
\hline
\hline
A1 & 5.1 & $5.5^{+1.5}_{-1.0}$ & 0 & 0 & 5.4 & $6.5^{+1.6}_{-0.8}$ & 0 & 0 \\
A2 & 7.7 & $5.4^{+2.5}_{-1.6}$ & 1211 & $1253^{+42}_{-50}$ & 4.7 & $5.3^{+1.6}_{-1.1}$ & 1077 & $1146^{+107}_{-54}$ \\
A3 & 8.1 & $8.4^{+4.2}_{-2.2}$ & 757 & $789^{+55}_{-59}$ & 7.4 & $10.0^{+2.9}_{-1.8}$ & 749 & $788^{+48}_{-55}$ \\
A4 & ... & ... & ... & ... & 1.4 & $17.9^{+8.1}_{-2.9}$ \tnote{a}$^{\rm *}$ & 1110 & $1172^{+99}_{-59}$ \\
\hline
B1 & 8.8 & $6.8^{+1.9}_{-1.3}$ & 780 & $733^{+104}_{-139}$ & 9.3 & $7.3^{+2.0}_{-1.1}$ & 668 & $590^{+77}_{-88}$ \\
B2 & 4.9 & $5.2^{+2.7}_{-1.7}$ & 955 & $915^{+97}_{-150}$ & 4.0 & $3.9^{+0.9}_{-0.7}$ & 860 & $805^{+81}_{-106}$ \\
B3 & 4.9 & $4.1^{+1.0}_{-0.7}$ & 0 & 0 & 5.1 & $5.0^{+0.9}_{-0.7}$ & 0 & 0 \\
B4 & ... & ... & ... & ... & 0.8 & $2.5^{+1.6}_{-1.0}$ & 898 & $834^{+74}_{-114}$ \\
\hline
\end{tabular}
\end{center}

\raggedright

\textbf{Notes.} Values are reported for lensed images for our two lens models using 3 versus 4 images of each quasar (ground- vs. space-based constraints). We report the best-fit values from the maximum likelihood model and the 16th, 50th, and 84th percentile values from the posterior distribution of the MCMC from \texttt{lenstool}.
\raggedright

$^{\rm *}$ Image A4's proximity to the critical curve makes the distribution of its derived quantities sensitive to small changes in the potential and hence the quoted statistical uncertainty is large.
\end{table*}

Precision cluster mass lens modeling requires high-resolution space-based imaging. In Fig.~\ref{fig:rgbimage}, we compare the critical curves for our model based on ground-based data alone with the model based on updated constraints from space-based imaging, which is for the purpose of this paper preliminary and subject to further analysis. We also highlight the locations of the halo that needed to be optimized as an independent mass component (O7) and the halos that were added (O8 and O9) in order to reproduce the fourth images of the quasars. Our updated (space-based) lens model reproduces the fourth images, A4 and B4, introduced as new constraints from the space-based imaging and has a model quality comparable to the ground-based lens model from Section~\ref{sec:slmodel} (34 constraints, 33 free parameters, image-plane RMS$=0.25''$, $\chi_{\nu}^2 \sim 1.10$). The integrated mass of the primary cluster in this model is $M (<200 {\rm kpc}) = 1.34^{+0.04}_{-0.03}
\times 10^{14} {\rm M}_{\odot}$ and $M (<500 {\rm kpc}) = 3.31^{+0.09}_{-0.09}
\times 10^{14} {\rm M}_{\odot}$ which is consistent within the error bars of the ground-based model. We compare $|\mu|$ and $\Delta t_{\rm rel}$ for the lens model from earlier in the paper (ground-based) and the updated (space-based) lens model in Table~\ref{tab:lensmodelupdate}. All magnification factors and time delays of the three brightest images of each quasar are consistent within the error bars between the two lens models. Adding these space-based constraints (i.e., fourth images of the quasars and improved measurements of the halos near them) leads to a $\sim 28\%$ reduction in the average uncertainties on the magnifications and a $\sim 11\%$ reduction in the average uncertainties on the time delays. We note that additional improvements in magnifications and time delays are expected with a more complex modeling using the space-based imaging. Applying these refined magnification factors to the measured bolometric luminosities (Section~\ref{sec:sed_modeling}), \targA\ has $L_{\rm bol} = 3.4^{+0.5}_{-0.7} \times 10^{45}$ erg s$^{-1}$ and \targB\ has $L_{\rm bol} = 5.4^{+0.9}_{-0.8} \times 10^{45}$ erg s$^{-1}$. The space telescope data corroborate our identification of other image families used as constraints in our ground-based lens model. The arc near \targB\ is unambiguously an image family (panel 3), and the blue arc's counterimage, which we teased out in ground-based FIRE spectroscopy, is securely detected (panel 6). The space-based imaging data thus underscores the importance of deep, high-resolution observations in the study of cluster-scale lensed quasars. These data will feed a much more complex analysis in a future paper (Solhaug et al. in prep.).

\section{Discussion} \label{sec:discussion}

The discovery of \targ\ provides a valuable case study for understanding the diversity of WSLQs and the environments that produce them. In this section, we discuss the nature of the two multiply-imaged quasars, the implications for identifying similar systems in upcoming surveys, and the potential of \targ\ as a system of interest for time-delay cosmography.

\subsection{A Cluster Lensing a Quasar Duo}

\targ\ represents the first known case of a galaxy cluster lensing two distinct background quasars, and in fact, this is the first lens ever discovered lensing two individual quasars into multiple images. All previously identified WSLQs exhibit multiple images of a single quasar, whereas \targ\ uniquely produces two sets of multiply-imaged quasars: \targA, a broad-line (Type I) quasar, and \targB, a narrow-line (Type II) quasar.

While \targA\ exhibits unmistakable broad lines and high luminosity which unambiguously make it a quasar, \targB\ requires more classification effort due to the lack of broad lines. The elevated nebular emission-line ratios [\NII] $\lambda 6584$ / \Halpha\ and [\OIII] $\lambda 5007$ / \Hbeta, clearly place \targB\ within the AGN regime of the BPT diagram (Fig.~\ref{fig:bpt}). These ratios are characteristic of hard ionizing spectra and consistent with an AGN-like ionization source.

Additionally, high-resolution HST- and JWST-imaging provides further confirmation that \targB\ is a quasar by revealing a compact, red point source at the center of \targB\ in the F475W-F606W-F814W color composite (Fig.~\ref{fig:hstjwstimages}). In ground-based imaging, the host galaxy of \targB\ is marginally resolved and well described by a single S\'ersic component surrounding a central PSF, but without visible substructure. The HST imaging, by contrast, resolves the central point source and separates it cleanly from the host. A full morphological and SED analysis of these space-based data will be presented in future work. Together, the spectroscopic diagnostics and resolved morphology establish \targ\ as the first confirmed case of a cluster-scale lens producing two sets of multiply-imaged quasars.

\subsection{Implications for Future WSLQ Searches}

The discovery of \targ\ underscores both the potential and the current limitations of WSLQ searches in large imaging surveys. Notably, none of \targA's quasar images were classified as point sources in the DECaLS catalog, and \targB\ does not exist in the catalog. Focusing exclusively on point-source morphologies risks the exclusion of blended or extended systems, particularly in crowded cluster fields that distort and contaminate the observed quasar surface brightness profiles.

Future surveys -- particularly LSST (Rubin), Euclid, and Roman -- are expected to significantly enhance WSLQ discovery rates through deeper imaging and improved sensitivity at redder wavelengths \citep[][]{Napier:2023b}. 
However, maximizing completeness of the WSLQ sample will require search techniques that can recognize quasar candidates in crowded sightlines and with extended, non-point-source-like morphologies. Understanding the biases that lead to false non-detections -- whether due to extended host galaxies, blending in crowded cluster fields, or survey-dependent systematics (even the survey classifications are not optimized for all types of galaxies) -- will be key to building a statistically robust census of WSLQs.

\subsection{Prospects for Time-Delay Cosmography}

Luminosity variability in quasars provides a natural clock for measuring time delays between lensed images, thereby constraining the Fermat potential and, ultimately, cosmological parameters such as the Hubble constant, $H_0$. Cluster-scale lenses like \targ\ are particularly valuable in this context: their large separations produce long, well-measurable time delays.

We are conducting an ongoing $g$-band monitoring campaign of \targA\ with NOT/ALFOSC, building upon the broader WSLQ monitoring program at the Nordic Optical Telescope (PI: H\r{a}kon Dahle).
As shown by \citet{Napier:2023b}, a sample of $\sim100$ lensed image pairs could yield percent-level constraints on $H_0$.

\begin{figure}[h!]
    \centering
    \includegraphics[width=\columnwidth]{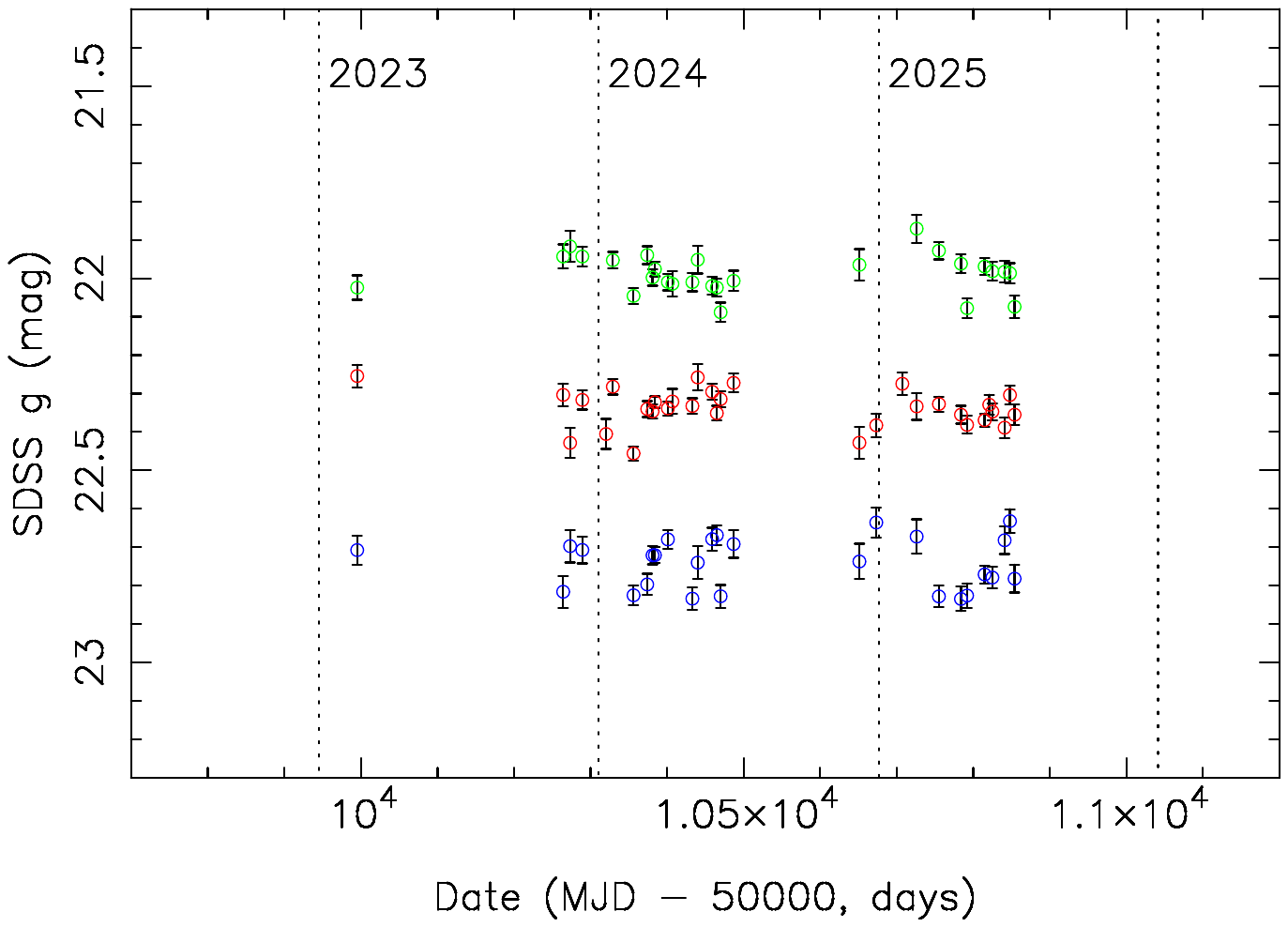}
    \caption{Nordic Optical Telescope/ALFOSC monitoring of the three brightest images of \targA\ (blue circles: A1; green: A2, offset by -0.7 mag for clarity; red: A3). As the monitoring continues, the time delay of the quasar images will become measurable. \targB\ lacks sufficient blue flux for monitoring in the $g$-band.}
    \label{fig:alfosc_monitoring}
\end{figure}

So far, the \targA\ $g$-band light curves (Fig.~\ref{fig:alfosc_monitoring}) show hints of variability, but the current sampling and monitoring time span are insufficient to measure a reliable time delay. The ongoing monitoring -- and incoming LSST data from the Rubin Observatory -- will continue to test for correlated brightness fluctuations among the three images.  Our lens model predicts relative time delays of approximately $\Delta t_{\rm A1, \, A3} \sim 800$ and $\Delta t_{\rm A1, \, A2} \sim 1200$ days (Fig.~\ref{fig:fermat}).

Because \targB\ is heavily reddened, it is useless for $g$-band monitoring; future campaigns will require redder filters to detect any variability. An initial \texttt{GALFIT} analysis of the JWST imaging, in which we fit the light from the central point source and the host galaxy separately, indicates that the quasar contributes a significant fraction of the total flux in the near-infrared. The measured quasar-to-host flux ratios are 0.27, 0.41 and 0.74 in the F115W, F150W, and F444W bands, respectively. These results indicate that the quasar remains luminous at red wavelengths despite heavy dust obscuration, suggesting time-delays for \targB\ can be measured with ground-based monitoring in red or near-infrared bands, even if the system is only partially resolved.

\section{Conclusions} \label{sec:conclusions}

In this paper, we present the discovery of \targ, a new wide-separation lensed quasar (WSLQ) system and the first known gravitational lens to produce multiple images of two background quasars, \targA\ and \targB. The system was identified by the COOL-LAMPS collaboration through photometric selection in DECaLS survey data and confirmed with follow-up imaging and spectroscopy from the Nordic Optical Telescope and the Magellan Telescopes.

\targ\ is the \textit{eighth} known WSLQ system (or the \textit{eighth} and \textit{ninth} if counting individual quasars), comprising spectroscopically confirmed quasars at $z_{\rm QSO} = 1.524$ (\targA) and $z_{\rm QSO} = 1.939$ (\targB), lensed by a primary cluster at redshift $z_{\rm lens} = 0.4301$ with additional influence from a secondary line-of-sight cluster at $z_{\rm lens} = 0.5766$, offset by $\sim 1.8$ Mpc in projection.

High-resolution HST and JWST imaging further confirm the lensing configuration and reveal, in remarkable detail, the morphologies of their host galaxies. These space-based observations corroborate our interpretation that \targB\ at $z=1.939$ is a heavily dust-obscured quasar, lacking the broad emission lines characteristic of unobscured quasars, while \targA\ at $z=1.524$ displays clear broad lines. Both sources exhibit high bolometric luminosities after correcting for magnification factors: \targA\ has $L_{\rm bol} = 3.4^{+0.5}_{-0.7} \times 10^{45}$ erg s$^{-1}$ and \targB\ has $L_{\rm bol} = 5.4^{+0.9}_{-0.8} \times 10^{45}$ erg s$^{-1}$. 
Notably, \targA\ resides in a dense galaxy group, raising the intriguing possibility that environmental interactions have triggered and continue to sustain its AGN activity.

Our two lens models (one based on ground-based data only and one updated with space-based constraints) 
imply a projected mass for the primary cluster of $M(<200 \,{\rm kpc}) = 1.2-1.3 \times 10^{14} M_{\odot}$.
We predict time delays that, with continued monitoring, would yield at least 4 additional image pairs suitable for time-delay cosmography, bringing us closer to an independent, high-precision measurement of the Hubble constant, $H_0$.

\targ\ adds a rich field to the growing family of WSLQs, offering two new laboratories for studying quasar physics and their environments at Cosmic Noon ($z \sim 2$). All lensed images of both quasars are cataloged as extended sources in the ground-based DECaLS image catalog, making it the third and fourth known WSLQ that would be impossible to identify in a search selecting for point sources in a DECaLS-like survey. A search strategy that does not solely focus on classification of point-source-like objects -- i.e., as was used to find \targ\ -- may be needed in order to identify WSLQs that have been overlooked in survey data.
With the advent of large surveys like LSST (Rubin), Euclid and Roman, careful selection strategies that remain sensitive to complex cluster-quasar constellations -- particularly given the increased depth and, in the case of Euclid and Roman, enhanced infrared sensitivity -- will be crucial. In doing so, we move closer to a larger sample of WSLQs and a more complete understanding of quasar-galaxy co-evolution during the era of peak AGN and star formation activity, Cosmic Noon.

\section{Acknowledgments}

This paper uses data obtained with the 6.5-m Magellan Telescopes located at Las Campanas Observatory, Chile. Magellan observing time for this project was granted by the time allocation committees of the University of Chicago and the University of Michigan. We would like to express our gratitude to the staff of Las Campanas Observatory in Chile whose support was instrumental in facilitating data collection. Furthermore, the data presented here were obtained in part with ALFOSC, which is provided by the Instituto de Astrofisica de Andalucia (IAA) under a joint agreement with the University of Copenhagen and the Nordic Optical Telescope (NOT). This research uses services or data provided by the SPectra Analysis and Retrievable Catalog Lab (SPARCL) and the Astro Data Lab, which are both part of the Community Science and Data Center (CSDC) program at NSF National Optical-Infrared Astronomy Research Laboratory. NOIRLab is operated by the Association of Universities for Research in Astronomy (AURA), Inc. under a cooperative agreement with the National Science Foundation. This research is based on observations made with the NASA/ESA Hubble Space Telescope obtained from the Space Telescope Science Institute, which is operated by the Association of Universities for Research in Astronomy, Inc., under NASA contract NAS 5–26555. These observations are associated with program: GO 17431. This work is based in part on observations made with the NASA/ESA/CSA James Webb Space Telescope. The data were obtained from the Mikulski Archive for Space Telescopes at the Space Telescope Science Institute, which is operated by the Association of Universities for Research in Astronomy, Inc., under NASA contract NAS 5-03127 for JWST. These observations are associated with program \#6675. Support for program \#6675 was provided by NASA through a grant from the Space Telescope Science Institute, which is operated by the Association of Universities for Research in Astronomy, Inc., under NASA contract NAS 5-03127. This work has made use of data from the European Space Agency (ESA) mission
{\it Gaia} (\url{https://www.cosmos.esa.int/gaia}), processed by the {\it Gaia}
Data Processing and Analysis Consortium (DPAC,
\url{https://www.cosmos.esa.int/web/gaia/dpac/consortium}). Funding for the DPAC
has been provided by national institutions, in particular the institutions
participating in the {\it Gaia} Multilateral Agreement. This publication makes use of data products from the Wide-field Infrared Survey Explorer (WISE), which is a joint project of the University of California, Los Angeles, and the Jet Propulsion Laboratory/California Institute of Technology, funded by the National Aeronautics and Space Administration. This research has made use of the NASA/IPAC Infrared Science Archive, which is funded by the National Aeronautics and Space Administration and operated by the California Institute of Technology. This work is supported by The College Undergraduate program at the University of Chicago, and the Department of Astronomy and Astrophysics at the University of Chicago. S.D.M. would like to thank Rafael Ortiz III for useful discussions and advice regarding parameter inference with CIGALE. We acknowledge the use of Claude Haiku 4.5 (Anthropic) for editorial and code assistance. Responsibility for the accuracy and contents of this manuscript remains entirely with the authors.

%

\vspace{5mm}
\facilities{Spitzer (IRAC), WISE, IRSA, Herschel (SPIRE), Magellan: Baade (IMACS, FIRE, FOURSTAR), Magellan: Clay (LDSS3-C), HST (ACS), JWST (NIRCam), Astro Data Lab}


\vspace{5mm}
\software{
    \texttt{SPARCL} \citep[][]{Juneau:2021},
    \texttt{astropy} \citep{Astropy:2013, Astropy:2018},
    \texttt{DS9} \citep{Smithsonian:2000},
    \texttt{IRAF} \citep{Tody:1986, Tody:1993},
    \texttt{matplotlib} \citep{Hunter:2007},
    \texttt{numpy} \citep{Harris:2020},
    \texttt{scipy} \citep{Virtanen:2020},
    \texttt{Source Extractor} \citep[][]{Bertin:1996},
    \texttt{lenstool} \citep[][]{Jullo:2007},
    \texttt{GALFIT} \citep[][]{Peng:2002, Peng:2010},
    \texttt{CIGALE} \citep[][]{Boquien:2019},
    \texttt{pyspeckit} \citep[][]{Ginsburg:2022,Ginsburg:2011},
    \texttt{CARTA} \citep[][]{Comrie:2021, Comrie:2021_zndo},
    \texttt{Claude} \citep[][]{Claude:2025}
    }



\bibliography{CJ1153}{}

@ARTICLE{Jullo:2007,
       author = {{Jullo}, E. and {Kneib}, J. -P. and {Limousin}, M. and {El{\'\i}asd{\'o}ttir}, {\'A}. and {Marshall}, P.~J. and {Verdugo}, T.},
        title = "{A Bayesian approach to strong lensing modelling of galaxy clusters}",
      journal = {New Journal of Physics},
         year = 2007,
        month = dec,
       volume = {9},
       number = {12},
        pages = {447},
          doi = {10.1088/1367-2630/9/12/447},
archivePrefix = {arXiv},
       eprint = {0706.0048},
 primaryClass = {astro-ph},
       adsurl = {https://ui.adsabs.harvard.edu/abs/2007NJPh....9..447J}
}

@ARTICLE{Gladders:2000,
       author = {{Gladders}, Michael D. and {Yee}, H.~K.~C.},
        title = "{A New Method For Galaxy Cluster Detection. I. The Algorithm}",
      journal = {\aj},
         year = 2000,
        month = oct,
       volume = {120},
       number = {4},
        pages = {2148-2162},
          doi = {10.1086/301557},
archivePrefix = {arXiv},
       eprint = {astro-ph/0004092},
 primaryClass = {astro-ph},
       adsurl = {https://ui.adsabs.harvard.edu/abs/2000AJ....120.2148G}
}

@ARTICLE{Boquien:2019,
       author = {{Boquien}, M. and {Burgarella}, D. and {Roehlly}, Y. and {Buat}, V. and {Ciesla}, L. and {Corre}, D. and {Inoue}, A.~K. and {Salas}, H.},
        title = "{CIGALE: a python Code Investigating GALaxy Emission}",
      journal = {\aap},
         year = 2019,
        month = feb,
       volume = {622},
          eid = {A103},
        pages = {A103},
          doi = {10.1051/0004-6361/201834156},
archivePrefix = {arXiv},
       eprint = {1811.03094},
 primaryClass = {astro-ph.GA},
       adsurl = {https://ui.adsabs.harvard.edu/abs/2019A&A...622A.103B}
}

@INPROCEEDINGS{Tody:1986,
       author = {{Tody}, Doug},
        title = "{The IRAF Data Reduction and Analysis System}",
    booktitle = {Instrumentation in astronomy VI},
         year = 1986,
       editor = {{Crawford}, David L.},
       series = {Society of Photo-Optical Instrumentation Engineers (SPIE) Conference Series},
       volume = {627},
        month = jan,
        pages = {733},
          doi = {10.1117/12.968154},
       adsurl = {https://ui.adsabs.harvard.edu/abs/1986SPIE..627..733T}
}

@INPROCEEDINGS{Tody:1993,
       author = {{Tody}, Doug},
        title = "{IRAF in the Nineties}",
    booktitle = {Astronomical Data Analysis Software and Systems II},
         year = 1993,
       editor = {{Hanisch}, R.~J. and {Brissenden}, R.~J.~V. and {Barnes}, J.},
       series = {Astronomical Society of the Pacific Conference Series},
       volume = {52},
        month = jan,
        pages = {173},
       adsurl = {https://ui.adsabs.harvard.edu/abs/1993ASPC...52..173T}
}

@ARTICLE{Stalevski:2012,
       author = {{Stalevski}, Marko and {Fritz}, Jacopo and {Baes}, Maarten and {Nakos}, Theodoros and {Popovi{\'c}}, Luka {\v{C}}.},
        title = "{3D radiative transfer modelling of the dusty tori around active galactic nuclei as a clumpy two-phase medium}",
      journal = {\mnras},
         year = 2012,
        month = mar,
       volume = {420},
       number = {4},
        pages = {2756-2772},
          doi = {10.1111/j.1365-2966.2011.19775.x},
archivePrefix = {arXiv},
       eprint = {1109.1286},
 primaryClass = {astro-ph.CO},
       adsurl = {https://ui.adsabs.harvard.edu/abs/2012MNRAS.420.2756S}
}

@ARTICLE{Stalevski:2016,
       author = {{Stalevski}, Marko and {Ricci}, Claudio and {Ueda}, Yoshihiro and {Lira}, Paulina and {Fritz}, Jacopo and {Baes}, Maarten},
        title = "{The dust covering factor in active galactic nuclei}",
      journal = {\mnras},
         year = 2016,
        month = may,
       volume = {458},
       number = {3},
        pages = {2288-2302},
          doi = {10.1093/mnras/stw444},
archivePrefix = {arXiv},
       eprint = {1602.06954},
 primaryClass = {astro-ph.GA},
       adsurl = {https://ui.adsabs.harvard.edu/abs/2016MNRAS.458.2288S}
}

@ARTICLE{Kauffman:2003,
       author = {{Kauffmann}, Guinevere and {Heckman}, Timothy M. and {Tremonti}, Christy and {Brinchmann}, Jarle and {Charlot}, St{\'e}phane and {White}, Simon D.~M. and {Ridgway}, Susan E. and {Brinkmann}, Jon and {Fukugita}, Masataka and {Hall}, Patrick B. and {Ivezi{\'c}}, {\v{Z}}eljko and {Richards}, Gordon T. and {Schneider}, Donald P.},
        title = "{The host galaxies of active galactic nuclei}",
      journal = {\mnras},
         year = 2003,
        month = dec,
       volume = {346},
       number = {4},
        pages = {1055-1077},
          doi = {10.1111/j.1365-2966.2003.07154.x},
archivePrefix = {arXiv},
       eprint = {astro-ph/0304239},
 primaryClass = {astro-ph},
       adsurl = {https://ui.adsabs.harvard.edu/abs/2003MNRAS.346.1055K}
}

@ARTICLE{Baldwin:1981,
       author = {{Baldwin}, J.~A. and {Phillips}, M.~M. and {Terlevich}, R.},
        title = "{Classification parameters for the emission-line spectra of extragalactic objects.}",
      journal = {\pasp},
         year = 1981,
        month = feb,
       volume = {93},
        pages = {5-19},
          doi = {10.1086/130766},
       adsurl = {https://ui.adsabs.harvard.edu/abs/1981PASP...93....5B}
}

@ARTICLE{Napier:2023a,
       author = {{Napier}, Kate and {Gladders}, Michael D. and {Sharon}, Keren and {Dahle}, H{\r{a}}kon and {Cloonan}, Aidan P. and {Mahler}, Guillaume and {Escapa}, Isaiah and {Garza}, Josh and {Kisare}, Andrew and {Malagon}, Natalie and {Mork}, Simon and {Niu}, Kunwanhui and {Rosener}, Riley and {Sullivan}, Jamar and {Tagliavia}, Marie and {Tamargo-Arizmendi}, Marcos and {Teixeira}, Raul and {Tsiane}, Kabelo and {Wagner}, Grace and {Zhang}, Yunchong and {Zhao}, Megan},
        title = "{COOL-LAMPS. Discovery of COOL J0335-1927, a Gravitationally Lensed Quasar at z = 3.27 with an Image Separation of 23.″3}",
      journal = {\apjl},
         year = {2023a},
        month = sep,
       volume = {954},
       number = {2},
          eid = {L38},
        pages = {L38},
          doi = {10.3847/2041-8213/acf132},
archivePrefix = {arXiv},
       eprint = {2305.14317},
 primaryClass = {astro-ph.GA},
       adsurl = {https://ui.adsabs.harvard.edu/abs/2023ApJ...954L..38N}
}

@ARTICLE{Inada:2006,
       author = {{Inada}, Naohisa and {Oguri}, Masamune and {Morokuma}, Tomoki and {Doi}, Mamoru and {Yasuda}, Naoki and {Becker}, Robert H. and {Richards}, Gordon T. and {Kochanek}, Christopher S. and {Kayo}, Issha and {Konishi}, Kohki and {Utsunomiya}, Hiroyuki and {Shin}, Min-Su and {Strauss}, Michael A. and {Sheldon}, Erin S. and {York}, Donald G. and {Hennawi}, Joseph F. and {Schneider}, Donald P. and {Dai}, Xinyu and {Fukugita}, Masataka},
        title = "{SDSS J1029+2623: A Gravitationally Lensed Quasar with an Image Separation of 22.5''}",
      journal = {\apjl},
         year = 2006,
        month = dec,
       volume = {653},
       number = {2},
        pages = {L97-L100},
          doi = {10.1086/510671},
archivePrefix = {arXiv},
       eprint = {astro-ph/0611275},
 primaryClass = {astro-ph},
       adsurl = {https://ui.adsabs.harvard.edu/abs/2006ApJ...653L..97I}
}

@ARTICLE{Dahle:2013,
       author = {{Dahle}, H. and {Gladders}, M.~D. and {Sharon}, K. and {Bayliss}, M.~B. and {Wuyts}, E. and {Abramson}, L.~E. and {Koester}, B.~P. and {Groeneboom}, N. and {Brinckmann}, T.~E. and {Kristensen}, M.~T. and {Lindholmer}, M.~O. and {Nielsen}, A. and {Krogager}, J. -K. and {Fynbo}, J.~P.~U.},
        title = "{SDSS J2222+2745: A Gravitationally Lensed Sextuple Quasar with a Maximum Image Separation of 15.''1 Discovered in the Sloan Giant Arcs Survey}",
      journal = {\apj},
         year = 2013,
        month = aug,
       volume = {773},
       number = {2},
          eid = {146},
        pages = {146},
          doi = {10.1088/0004-637X/773/2/146},
archivePrefix = {arXiv},
       eprint = {1211.1091},
 primaryClass = {astro-ph.CO},
       adsurl = {https://ui.adsabs.harvard.edu/abs/2013ApJ...773..146D}
}

@ARTICLE{Inada:2003,
       author = {{Inada}, Naohisa and {Oguri}, Masamune and {Pindor}, Bartosz and {Hennawi}, Joseph F. and {Chiu}, Kuenley and {Zheng}, Wei and {Ichikawa}, Shin-Ichi and {Gregg}, Michael D. and {Becker}, Robert H. and {Suto}, Yasushi and {Strauss}, Michael A. and {Turner}, Edwin L. and {Keeton}, Charles R. and {Annis}, James and {Castander}, Francisco J. and {Eisenstein}, Daniel J. and {Frieman}, Joshua A. and {Fukugita}, Masataka and {Gunn}, James E. and {Johnston}, David E. and {Kent}, Stephen M. and {Nichol}, Robert C. and {Richards}, Gordon T. and {Rix}, Hans-Walter and {Sheldon}, Erin Scott and {Bahcall}, Neta A. and {Brinkmann}, J. and {Ivezi{\'c}}, {\v{Z}}eljko and {Lamb}, Don Q. and {McKay}, Timothy A. and {Schneider}, Donald P. and {York}, Donald G.},
        title = "{A gravitationally lensed quasar with quadruple images separated by 14.62arcseconds}",
      journal = {\nat},
         year = 2003,
        month = dec,
       volume = {426},
       number = {6968},
        pages = {810-812},
          doi = {10.1038/nature02153},
archivePrefix = {arXiv},
       eprint = {astro-ph/0312427},
 primaryClass = {astro-ph},
       adsurl = {https://ui.adsabs.harvard.edu/abs/2003Natur.426..810I}
}

@ARTICLE{Shu:2018,
       author = {{Shu}, Yiping and {Marques-Chaves}, Rui and {Evans}, N. Wyn and {P{\'e}rez-Fournon}, Ismael},
        title = "{SDSS J0909+4449: A large-separation strongly lensed quasar at z {\ensuremath{\sim}} 2.8 with three images}",
      journal = {\mnras},
         year = 2018,
        month = nov,
       volume = {481},
       number = {1},
        pages = {L136-L140},
          doi = {10.1093/mnrasl/sly174},
archivePrefix = {arXiv},
       eprint = {1809.07337},
 primaryClass = {astro-ph.GA},
       adsurl = {https://ui.adsabs.harvard.edu/abs/2018MNRAS.481L.136S}
}

@ARTICLE{Shu:2019,
       author = {{Shu}, Yiping and {Koposov}, Sergey E. and {Evans}, N. Wyn and {Belokurov}, Vasily and {McMahon}, Richard G. and {Auger}, Matthew W. and {Lemon}, Cameron A.},
        title = "{Catalogues of active galactic nuclei from Gaia and unWISE data}",
      journal = {\mnras},
         year = 2019,
        month = nov,
       volume = {489},
       number = {4},
        pages = {4741-4759},
          doi = {10.1093/mnras/stz2487},
archivePrefix = {arXiv},
       eprint = {1909.02010},
 primaryClass = {astro-ph.GA},
       adsurl = {https://ui.adsabs.harvard.edu/abs/2019MNRAS.489.4741S}
}

@ARTICLE{Martinez:2023,
       author = {{Martinez}, Michael N. and {Napier}, Kate A. and {Cloonan}, Aidan P. and {Sukay}, Ezra and {Gozman}, Katya and {Merz}, Kaiya and {Khullar}, Gourav and {Lin}, Jason J. and {Matthews Acu{\~n}a}, Owen S. and {Medina}, Elisabeth and {Sanchez}, Jorge A. and {Sisco}, Emily E. and {Kavin Stein}, Daniel J. and {Tavangar}, Kiyan and {Gonz{\'a}lez}, Juan Remolina and {Mahler}, Guillaume and {Sharon}, Keren and {Dahle}, H{\r{a}}kon and {Gladders}, Michael D.},
        title = "{COOL-LAMPS. III. Discovery of a 25.″9 Separation Quasar Lensed by a Merging Galaxy Cluster}",
      journal = {\apj},
         year = 2023,
        month = apr,
       volume = {946},
       number = {2},
          eid = {63},
        pages = {63},
          doi = {10.3847/1538-4357/acbe39},
archivePrefix = {arXiv},
       eprint = {2209.03972},
 primaryClass = {astro-ph.GA},
       adsurl = {https://ui.adsabs.harvard.edu/abs/2023ApJ...946...63M}
}

@ARTICLE{Dey:2019,
       author = {{Dey}, Arjun and {Schlegel}, David J. and {Lang}, Dustin and {Blum}, Robert and {Burleigh}, Kaylan and {Fan}, Xiaohui and {Findlay}, Joseph R. and {Finkbeiner}, Doug and {Herrera}, David and {Juneau}, St{\'e}phanie and {Landriau}, Martin and {Levi}, Michael and {McGreer}, Ian and {Meisner}, Aaron and {Myers}, Adam D. and {Moustakas}, John and {Nugent}, Peter and {Patej}, Anna and {Schlafly}, Edward F. and {Walker}, Alistair R. and {Valdes}, Francisco and {Weaver}, Benjamin A. and {Y{\`e}che}, Christophe and {Zou}, Hu and {Zhou}, Xu and {Abareshi}, Behzad and {Abbott}, T.~M.~C. and {Abolfathi}, Bela and {Aguilera}, C. and {Alam}, Shadab and {Allen}, Lori and {Alvarez}, A. and {Annis}, James and {Ansarinejad}, Behzad and {Aubert}, Marie and {Beechert}, Jacqueline and {Bell}, Eric F. and {BenZvi}, Segev Y. and {Beutler}, Florian and {Bielby}, Richard M. and {Bolton}, Adam S. and {Brice{\~n}o}, C{\'e}sar and {Buckley-Geer}, Elizabeth J. and {Butler}, Karen and {Calamida}, Annalisa and {Carlberg}, Raymond G. and {Carter}, Paul and {Casas}, Ricard and {Castander}, Francisco J. and {Choi}, Yumi and {Comparat}, Johan and {Cukanovaite}, Elena and {Delubac}, Timoth{\'e}e and {DeVries}, Kaitlin and {Dey}, Sharmila and {Dhungana}, Govinda and {Dickinson}, Mark and {Ding}, Zhejie and {Donaldson}, John B. and {Duan}, Yutong and {Duckworth}, Christopher J. and {Eftekharzadeh}, Sarah and {Eisenstein}, Daniel J. and {Etourneau}, Thomas and {Fagrelius}, Parker A. and {Farihi}, Jay and {Fitzpatrick}, Mike and {Font-Ribera}, Andreu and {Fulmer}, Leah and {G{\"a}nsicke}, Boris T. and {Gaztanaga}, Enrique and {George}, Koshy and {Gerdes}, David W. and {Gontcho}, Satya Gontcho A. and {Gorgoni}, Claudio and {Green}, Gregory and {Guy}, Julien and {Harmer}, Diane and {Hernandez}, M. and {Honscheid}, Klaus and {Huang}, Lijuan Wendy and {James}, David J. and {Jannuzi}, Buell T. and {Jiang}, Linhua and {Joyce}, Richard and {Karcher}, Armin and {Karkar}, Sonia and {Kehoe}, Robert and {Kneib}, Jean-Paul and {Kueter-Young}, Andrea and {Lan}, Ting-Wen and {Lauer}, Tod R. and {Le Guillou}, Laurent and {Le Van Suu}, Auguste and {Lee}, Jae Hyeon and {Lesser}, Michael and {Perreault Levasseur}, Laurence and {Li}, Ting S. and {Mann}, Justin L. and {Marshall}, Robert and {Mart{\'\i}nez-V{\'a}zquez}, C.~E. and {Martini}, Paul and {du Mas des Bourboux}, H{\'e}lion and {McManus}, Sean and {Meier}, Tobias Gabriel and {M{\'e}nard}, Brice and {Metcalfe}, Nigel and {Mu{\~n}oz-Guti{\'e}rrez}, Andrea and {Najita}, Joan and {Napier}, Kevin and {Narayan}, Gautham and {Newman}, Jeffrey A. and {Nie}, Jundan and {Nord}, Brian and {Norman}, Dara J. and {Olsen}, Knut A.~G. and {Paat}, Anthony and {Palanque-Delabrouille}, Nathalie and {Peng}, Xiyan and {Poppett}, Claire L. and {Poremba}, Megan R. and {Prakash}, Abhishek and {Rabinowitz}, David and {Raichoor}, Anand and {Rezaie}, Mehdi and {Robertson}, A.~N. and {Roe}, Natalie A. and {Ross}, Ashley J. and {Ross}, Nicholas P. and {Rudnick}, Gregory and {Safonova}, Sasha and {Saha}, Abhijit and {S{\'a}nchez}, F. Javier and {Savary}, Elodie and {Schweiker}, Heidi and {Scott}, Adam and {Seo}, Hee-Jong and {Shan}, Huanyuan and {Silva}, David R. and {Slepian}, Zachary and {Soto}, Christian and {Sprayberry}, David and {Staten}, Ryan and {Stillman}, Coley M. and {Stupak}, Robert J. and {Summers}, David L. and {Sien Tie}, Suk and {Tirado}, H. and {Vargas-Maga{\~n}a}, Mariana and {Vivas}, A. Katherina and {Wechsler}, Risa H. and {Williams}, Doug and {Yang}, Jinyi and {Yang}, Qian and {Yapici}, Tolga and {Zaritsky}, Dennis and {Zenteno}, A. and {Zhang}, Kai and {Zhang}, Tianmeng and {Zhou}, Rongpu and {Zhou}, Zhimin},
        title = "{Overview of the DESI Legacy Imaging Surveys}",
      journal = {\aj},
         year = 2019,
        month = may,
       volume = {157},
       number = {5},
          eid = {168},
        pages = {168},
          doi = {10.3847/1538-3881/ab089d},
archivePrefix = {arXiv},
       eprint = {1804.08657},
 primaryClass = {astro-ph.IM},
       adsurl = {https://ui.adsabs.harvard.edu/abs/2019AJ....157..168D}
}

@ARTICLE{Oguri:2006,
       author = {{Oguri}, Masamune and {Inada}, Naohisa and {Pindor}, Bartosz and {Strauss}, Michael A. and {Richards}, Gordon T. and {Hennawi}, Joseph F. and {Turner}, Edwin L. and {Lupton}, Robert H. and {Schneider}, Donald P. and {Fukugita}, Masataka and {Brinkmann}, Jon},
        title = "{The Sloan Digital Sky Survey Quasar Lens Search. I. Candidate Selection Algorithm}",
      journal = {\aj},
         year = 2006,
        month = sep,
       volume = {132},
       number = {3},
        pages = {999-1013},
          doi = {10.1086/506019},
archivePrefix = {arXiv},
       eprint = {astro-ph/0605571},
 primaryClass = {astro-ph},
       adsurl = {https://ui.adsabs.harvard.edu/abs/2006AJ....132..999O}
}

@ARTICLE{Flesch:2023,
       author = {{Flesch}, Eric Wim},
        title = "{The Million Quasars (Milliquas) Catalogue, v8}",
      journal = {The Open Journal of Astrophysics},
         year = 2023,
        month = dec,
       volume = {6},
          eid = {49},
        pages = {49},
          doi = {10.21105/astro.2308.01505},
archivePrefix = {arXiv},
       eprint = {2308.01505},
 primaryClass = {astro-ph.GA},
       adsurl = {https://ui.adsabs.harvard.edu/abs/2023OJAp....6E..49F}
}

@ARTICLE{Dux:2024,
       author = {{Dux}, Fr{\'e}d{\'e}ric and {Lemon}, Cameron and {Courbin}, Fr{\'e}d{\'e}ric and {Neira}, Favio and {Anguita}, Timo and {Galan}, Aymeric and {Kim}, Sam and {Hempel}, Maren and {Hempel}, Angela and {Lachaume}, R{\'e}gis},
        title = "{Nine lensed quasars and quasar pairs discovered through spatially extended variability in Pan-STARRS}",
      journal = {\aap},
         year = 2024,
        month = feb,
       volume = {682},
          eid = {A47},
        pages = {A47},
          doi = {10.1051/0004-6361/202347598},
archivePrefix = {arXiv},
       eprint = {2307.13729},
 primaryClass = {astro-ph.GA},
       adsurl = {https://ui.adsabs.harvard.edu/abs/2024A&A...682A..47D}
}

@ARTICLE{Anguita:2018,
       author = {{Anguita}, T. and {Schechter}, P.~L. and {Kuropatkin}, N. and {Morgan}, N.~D. and {Ostrovski}, F. and {Abramson}, L.~E. and {Agnello}, A. and {Apostolovski}, Y. and {Fassnacht}, C.~D. and {Hsueh}, J.~W. and {Motta}, V. and {Rojas}, K. and {Rusu}, C.~E. and {Treu}, T. and {Williams}, P. and {Auger}, M. and {Buckley-Geer}, E. and {Lin}, H. and {McMahon}, R. and {Abbott}, T.~M.~C. and {Allam}, S. and {Annis}, J. and {Bernstein}, R.~A. and {Bertin}, E. and {Brooks}, D. and {Burke}, D.~L. and {Carnero Rosell}, A. and {Carrasco-Kind}, M. and {Carretero}, J. and {Cunha}, C.~E. and {D'Andrea}, C.~B. and {De Vicente}, J. and {DePoy}, D.~L. and {Desai}, S. and {Diehl}, H.~T. and {Doel}, P. and {Flaugher}, B. and {Garc{\'\i}a-Bellido}, J. and {Gerdes}, D.~W. and {Gruen}, D. and {Gruendl}, R.~A. and {Gschwend}, J. and {Hartley}, W.~G. and {Hollowood}, D.~L. and {Honscheid}, K. and {James}, D.~J. and {Kuehn}, K. and {Lima}, M. and {Maia}, M.~A.~G. and {Miquel}, R. and {Plazas}, A.~A. and {Sanchez}, E. and {Scarpine}, V. and {Smith}, M. and {Soares-Santos}, M. and {Sobreira}, F. and {Suchyta}, E. and {Tarle}, G. and {Walker}, A.~R.},
        title = "{The STRong lensing Insights into the Dark Energy Survey (STRIDES) 2016 follow-up campaign - II. New quasar lenses from double component fitting}",
      journal = {\mnras},
         year = 2018,
        month = nov,
       volume = {480},
       number = {4},
        pages = {5017-5028},
          doi = {10.1093/mnras/sty2172},
archivePrefix = {arXiv},
       eprint = {1805.12151},
 primaryClass = {astro-ph.GA},
       adsurl = {https://ui.adsabs.harvard.edu/abs/2018MNRAS.480.5017A}
}

@ARTICLE{Lemon:2023,
       author = {{Lemon}, C. and {Anguita}, T. and {Auger-Williams}, M.~W. and {Courbin}, F. and {Galan}, A. and {McMahon}, R. and {Neira}, F. and {Oguri}, M. and {Schechter}, P. and {Shajib}, A. and {Treu}, T. and {Agnello}, A. and {Spiniello}, C.},
        title = "{Gravitationally lensed quasars in Gaia - IV. 150 new lenses, quasar pairs, and projected quasars}",
      journal = {\mnras},
         year = 2023,
        month = apr,
       volume = {520},
       number = {3},
        pages = {3305-3328},
          doi = {10.1093/mnras/stac3721},
archivePrefix = {arXiv},
       eprint = {2206.07714},
 primaryClass = {astro-ph.GA},
       adsurl = {https://ui.adsabs.harvard.edu/abs/2023MNRAS.520.3305L}
}

@ARTICLE{Lemon:2019,
       author = {{Lemon}, Cameron A. and {Auger}, Matthew W. and {McMahon}, Richard G.},
        title = "{Gravitationally lensed quasars in Gaia - III. 22 new lensed quasars from Gaia data release 2}",
      journal = {\mnras},
         year = 2019,
        month = mar,
       volume = {483},
       number = {3},
        pages = {4242-4258},
          doi = {10.1093/mnras/sty3366},
archivePrefix = {arXiv},
       eprint = {1810.04480},
 primaryClass = {astro-ph.GA},
       adsurl = {https://ui.adsabs.harvard.edu/abs/2019MNRAS.483.4242L}
}

@ARTICLE{Scaramella:2022,
       author = {{Euclid Collaboration} and {Scaramella}, R. and {Amiaux}, J. and {Mellier}, Y. and {Burigana}, C. and {Carvalho}, C.~S. and {Cuillandre}, J. -C. and {Da Silva}, A. and {Derosa}, A. and {Dinis}, J. and {Maiorano}, E. and {Maris}, M. and {Tereno}, I. and {Laureijs}, R. and {Boenke}, T. and {Buenadicha}, G. and {Dupac}, X. and {Gaspar Venancio}, L.~M. and {G{\'o}mez-{\'A}lvarez}, P. and {Hoar}, J. and {Lorenzo Alvarez}, J. and {Racca}, G.~D. and {Saavedra-Criado}, G. and {Schwartz}, J. and {Vavrek}, R. and {Schirmer}, M. and {Aussel}, H. and {Azzollini}, R. and {Cardone}, V.~F. and {Cropper}, M. and {Ealet}, A. and {Garilli}, B. and {Gillard}, W. and {Granett}, B.~R. and {Guzzo}, L. and {Hoekstra}, H. and {Jahnke}, K. and {Kitching}, T. and {Maciaszek}, T. and {Meneghetti}, M. and {Miller}, L. and {Nakajima}, R. and {Niemi}, S.~M. and {Pasian}, F. and {Percival}, W.~J. and {Pottinger}, S. and {Sauvage}, M. and {Scodeggio}, M. and {Wachter}, S. and {Zacchei}, A. and {Aghanim}, N. and {Amara}, A. and {Auphan}, T. and {Auricchio}, N. and {Awan}, S. and {Balestra}, A. and {Bender}, R. and {Bodendorf}, C. and {Bonino}, D. and {Branchini}, E. and {Brau-Nogue}, S. and {Brescia}, M. and {Candini}, G.~P. and {Capobianco}, V. and {Carbone}, C. and {Carlberg}, R.~G. and {Carretero}, J. and {Casas}, R. and {Castander}, F.~J. and {Castellano}, M. and {Cavuoti}, S. and {Cimatti}, A. and {Cledassou}, R. and {Congedo}, G. and {Conselice}, C.~J. and {Conversi}, L. and {Copin}, Y. and {Corcione}, L. and {Costille}, A. and {Courbin}, F. and {Degaudenzi}, H. and {Douspis}, M. and {Dubath}, F. and {Duncan}, C.~A.~J. and {Dusini}, S. and {Farrens}, S. and {Ferriol}, S. and {Fosalba}, P. and {Fourmanoit}, N. and {Frailis}, M. and {Franceschi}, E. and {Franzetti}, P. and {Fumana}, M. and {Gillis}, B. and {Giocoli}, C. and {Grazian}, A. and {Grupp}, F. and {Haugan}, S.~V.~H. and {Holmes}, W. and {Hormuth}, F. and {Hudelot}, P. and {Kermiche}, S. and {Kiessling}, A. and {Kilbinger}, M. and {Kohley}, R. and {Kubik}, B. and {K{\"u}mmel}, M. and {Kunz}, M. and {Kurki-Suonio}, H. and {Lahav}, O. and {Ligori}, S. and {Lilje}, P.~B. and {Lloro}, I. and {Mansutti}, O. and {Marggraf}, O. and {Markovic}, K. and {Marulli}, F. and {Massey}, R. and {Maurogordato}, S. and {Melchior}, M. and {Merlin}, E. and {Meylan}, G. and {Mohr}, J.~J. and {Moresco}, M. and {Morin}, B. and {Moscardini}, L. and {Munari}, E. and {Nichol}, R.~C. and {Padilla}, C. and {Paltani}, S. and {Peacock}, J. and {Pedersen}, K. and {Pettorino}, V. and {Pires}, S. and {Poncet}, M. and {Popa}, L. and {Pozzetti}, L. and {Raison}, F. and {Rebolo}, R. and {Rhodes}, J. and {Rix}, H. -W. and {Roncarelli}, M. and {Rossetti}, E. and {Saglia}, R. and {Schneider}, P. and {Schrabback}, T. and {Secroun}, A. and {Seidel}, G. and {Serrano}, S. and {Sirignano}, C. and {Sirri}, G. and {Skottfelt}, J. and {Stanco}, L. and {Starck}, J.~L. and {Tallada-Cresp{\'\i}}, P. and {Tavagnacco}, D. and {Taylor}, A.~N. and {Teplitz}, H.~I. and {Toledo-Moreo}, R. and {Torradeflot}, F. and {Trifoglio}, M. and {Valentijn}, E.~A. and {Valenziano}, L. and {Verdoes Kleijn}, G.~A. and {Wang}, Y. and {Welikala}, N. and {Weller}, J. and {Wetzstein}, M. and {Zamorani}, G. and {Zoubian}, J. and {Andreon}, S. and {Baldi}, M. and {Bardelli}, S. and {Boucaud}, A. and {Camera}, S. and {Di Ferdinando}, D. and {Fabbian}, G. and {Farinelli}, R. and {Galeotta}, S. and {Graci{\'a}-Carpio}, J. and {Maino}, D. and {Medinaceli}, E. and {Mei}, S. and {Neissner}, C. and {Polenta}, G. and {Renzi}, A. and {Romelli}, E. and {Rosset}, C. and {Sureau}, F. and {Tenti}, M. and {Vassallo}, T. and {Zucca}, E. and {Baccigalupi}, C. and {Balaguera-Antol{\'\i}nez}, A. and {Battaglia}, P. and {Biviano}, A. and {Borgani}, S. and {Bozzo}, E. and {Cabanac}, R. and {Cappi}, A.},
        title = "{Euclid preparation. I. The Euclid Wide Survey}",
      journal = {\aap},
         year = 2022,
        month = jun,
       volume = {662},
          eid = {A112},
        pages = {A112},
          doi = {10.1051/0004-6361/202141938},
archivePrefix = {arXiv},
       eprint = {2108.01201},
 primaryClass = {astro-ph.CO},
       adsurl = {https://ui.adsabs.harvard.edu/abs/2022A&A...662A.112E}
}

@ARTICLE{Robertson:2020,
       author = {{Robertson}, Andrew and {Smith}, Graham P. and {Massey}, Richard and {Eke}, Vincent and {Jauzac}, Mathilde and {Bianconi}, Matteo and {Ryczanowski}, Dan},
        title = "{What does strong gravitational lensing? The mass and redshift distribution of high-magnification lenses}",
      journal = {\mnras},
         year = 2020,
        month = jul,
       volume = {495},
       number = {4},
        pages = {3727-3739},
          doi = {10.1093/mnras/staa1429},
archivePrefix = {arXiv},
       eprint = {2002.01479},
 primaryClass = {astro-ph.CO},
       adsurl = {https://ui.adsabs.harvard.edu/abs/2020MNRAS.495.3727R}
}

@ARTICLE{Abe:2025,
       author = {{Abe}, Katsuya T. and {Oguri}, Masamune and {Birrer}, Simon and {Khadka}, Narayan and {Marshall}, Philip J. and {Lemon}, Cameron and {More}, Anupreeta and {Collaboration}, the LSST Dark Energy Science},
        title = "{A halo model approach for mock catalogs of time-variable strong gravitational lenses}",
      journal = {The Open Journal of Astrophysics},
         year = 2025,
        month = jan,
       volume = {8},
          eid = {8},
        pages = {8},
          doi = {10.33232/001c.128482},
archivePrefix = {arXiv},
       eprint = {2411.07509},
 primaryClass = {astro-ph.CO},
       adsurl = {https://ui.adsabs.harvard.edu/abs/2025OJAp....8E...8A}
}

@ARTICLE{Dutta:2024,
       author = {{Dutta}, Rajeshwari and {Acebron}, Ana and {Fumagalli}, Michele and {Grillo}, Claudio and {Caminha}, Gabriel B. and {Fossati}, Matteo},
        title = "{Probing coherence in metal absorption towards multiple images of strong gravitationally lensed quasars}",
      journal = {\mnras},
         year = 2024,
        month = feb,
       volume = {528},
       number = {2},
        pages = {1895-1905},
          doi = {10.1093/mnras/stae048},
archivePrefix = {arXiv},
       eprint = {2401.03024},
 primaryClass = {astro-ph.GA},
       adsurl = {https://ui.adsabs.harvard.edu/abs/2024MNRAS.528.1895D}
}

@ARTICLE{Refsdal:1964,
       author = {{Refsdal}, S.},
        title = "{On the possibility of determining Hubble's parameter and the masses of galaxies from the gravitational lens effect}",
      journal = {\mnras},
         year = 1964,
        month = jan,
       volume = {128},
        pages = {307},
          doi = {10.1093/mnras/128.4.307},
       adsurl = {https://ui.adsabs.harvard.edu/abs/1964MNRAS.128..307R}
}

@ARTICLE{Napier:2023b,
       author = {{Napier}, Kate and {Sharon}, Keren and {Dahle}, H{\r{a}}kon and {Bayliss}, Matthew and {Gladders}, Michael D. and {Mahler}, Guillaume and {Rigby}, Jane R. and {Florian}, Michael},
        title = "{Hubble Constant Measurement from Three Large-separation Quasars Strongly Lensed by Galaxy Clusters}",
      journal = {\apj},
         year = {2023b},
        month = dec,
       volume = {959},
       number = {2},
          eid = {134},
        pages = {134},
          doi = {10.3847/1538-4357/ad045a},
archivePrefix = {arXiv},
       eprint = {2301.11240},
 primaryClass = {astro-ph.CO},
       adsurl = {https://ui.adsabs.harvard.edu/abs/2023ApJ...959..134N}
}

@ARTICLE{Wong:2020,
       author = {{Wong}, Kenneth C. and {Suyu}, Sherry H. and {Chen}, Geoff C. -F. and {Rusu}, Cristian E. and {Millon}, Martin and {Sluse}, Dominique and {Bonvin}, Vivien and {Fassnacht}, Christopher D. and {Taubenberger}, Stefan and {Auger}, Matthew W. and {Birrer}, Simon and {Chan}, James H.~H. and {Courbin}, Frederic and {Hilbert}, Stefan and {Tihhonova}, Olga and {Treu}, Tommaso and {Agnello}, Adriano and {Ding}, Xuheng and {Jee}, Inh and {Komatsu}, Eiichiro and {Shajib}, Anowar J. and {Sonnenfeld}, Alessandro and {Blandford}, Roger D. and {Koopmans}, L{\'e}on V.~E. and {Marshall}, Philip J. and {Meylan}, Georges},
        title = "{H0LiCOW - XIII. A 2.4 per cent measurement of H$_{0}$ from lensed quasars: 5.3{\ensuremath{\sigma}} tension between early- and late-Universe probes}",
      journal = {\mnras},
         year = 2020,
        month = oct,
       volume = {498},
       number = {1},
        pages = {1420-1439},
          doi = {10.1093/mnras/stz3094},
archivePrefix = {arXiv},
       eprint = {1907.04869},
 primaryClass = {astro-ph.CO},
       adsurl = {https://ui.adsabs.harvard.edu/abs/2020MNRAS.498.1420W}
}

@ARTICLE{Cloonan:2024,
       author = {{Cloonan}, Aidan P. and {Khullar}, Gourav and {Napier}, Kate A. and {Gladders}, Michael D. and {Dahle}, H{\r{a}}kon and {Rosener}, Riley and {Sullivan}, Jamar and {Bayliss}, Matthew B. and {Chicoine}, Nathalie and {Escapa}, Isaiah and {Garza}, Diego and {Garza}, Josh and {Glusman}, Rowen and {Gozman}, Katya and {Horwath}, Gabriela and {Kisare}, Andi and {Levine}, Benjamin C. and {Liang}, Olina and {Malagon}, Natalie and {Martinez}, Michael N. and {Masegian}, Alexandra and {Matthews Acu{\~n}a}, Owen S. and {Mork}, Simon D. and {Niu}, Kunwanhui and {Owens}, M. Riley and {Pan}, Yue and {Rigby}, Jane R. and {Sharon}, Keren and {Sierra}, Isaac and {Stark}, Antony A. and {Sukay}, Ezra and {Tagliavia}, Marie and {Tamargo-Arizmendi}, Marcos and {Tavangar}, Kiyan and {Teixeira}, Raul and {Tsiane}, Kabelo and {Tu}, Ruoyang and {Wagner}, Grace and {Zaborowski}, Erik A. and {Zhang}, Yunchong and {Zhao}, Yifan ''Megan''},
        title = "{COOL-LAMPS. VIII. Known Wide-separation Lensed Quasars and Their Host Galaxies Reveal a Lack of Evolution in M$_{BH}$/M$_{{\ensuremath{\star}}}$ Since z {\ensuremath{\sim}} 3}",
      journal = {\apj},
         year = 2025,
        month = jul,
       volume = {987},
       number = {2},
          eid = {194},
        pages = {194},
          doi = {10.3847/1538-4357/addabf},
archivePrefix = {arXiv},
       eprint = {2408.03379},
 primaryClass = {astro-ph.GA},
       adsurl = {https://ui.adsabs.harvard.edu/abs/2025ApJ...987..194C}
}

@ARTICLE{Ross:2009,
       author = {{Ross}, N.~R. and {Assef}, R.~J. and {Kochanek}, C.~S. and {Falco}, E. and {Poindexter}, S.~D.},
        title = "{The UV-Mid-IR Spectral Energy Distribution of a z = 1.7 Quasar Host Galaxy}",
      journal = {\apj},
         year = 2009,
        month = sep,
       volume = {702},
       number = {1},
        pages = {472-479},
          doi = {10.1088/0004-637X/702/1/472},
archivePrefix = {arXiv},
       eprint = {0904.1708},
 primaryClass = {astro-ph.CO},
       adsurl = {https://ui.adsabs.harvard.edu/abs/2009ApJ...702..472R}
}

@ARTICLE{Williams:2021,
       author = {{Williams}, Peter R. and {Treu}, Tommaso and {Dahle}, H{\r{a}}kon and {Valenti}, Stefano and {Abramson}, Louis and {Barth}, Aaron J. and {Dyrland}, Karianne and {Gladders}, Michael and {Horne}, Keith and {Sharon}, Keren},
        title = "{The Black Hole Mass of the z = 2.805 Multiply Imaged Quasar SDSS J2222+2745 from Velocity-resolved Time Lags of the C IV Emission Line}",
      journal = {\apj},
         year = 2021,
        month = apr,
       volume = {911},
       number = {1},
          eid = {64},
        pages = {64},
          doi = {10.3847/1538-4357/abe943},
archivePrefix = {arXiv},
       eprint = {2011.02007},
 primaryClass = {astro-ph.GA},
       adsurl = {https://ui.adsabs.harvard.edu/abs/2021ApJ...911...64W}
}

@ARTICLE{Williams:2021b,
       author = {{Williams}, Peter R. and {Treu}, Tommaso and {Dahle}, H{\r{a}}kon and {Valenti}, Stefano and {Abramson}, Louis and {Barth}, Aaron J. and {Brewer}, Brendon J. and {Dyrland}, Karianne and {Gladders}, Michael and {Horne}, Keith and {Sharon}, Keren},
        title = "{Dynamical Modeling of the C IV Broad Line Region of the z = 2.805 Multiply Imaged Quasar SDSS J2222+2745}",
      journal = {\apjl},
         year = 2021,
        month = jul,
       volume = {915},
       number = {1},
          eid = {L9},
        pages = {L9},
          doi = {10.3847/2041-8213/ac081b},
archivePrefix = {arXiv},
       eprint = {2103.10961},
 primaryClass = {astro-ph.GA},
       adsurl = {https://ui.adsabs.harvard.edu/abs/2021ApJ...915L...9W}
}

@ARTICLE{Bayliss:2017,
       author = {{Bayliss}, Matthew B. and {Sharon}, Keren and {Acharyya}, Ayan and {Gladders}, Michael D. and {Rigby}, Jane R. and {Bian}, Fuyan and {Bordoloi}, Rongmon and {Runnoe}, Jessie and {Dahle}, Hakon and {Kewley}, Lisa and {Florian}, Michael and {Johnson}, Traci and {Paterno-Mahler}, Rachel},
        title = "{Spatially Resolved Patchy Ly{\ensuremath{\alpha}} Emission within the Central Kiloparsec of a Strongly Lensed Quasar Host Galaxy at z = 2.8}",
      journal = {\apjl},
         year = 2017,
        month = aug,
       volume = {845},
       number = {2},
          eid = {L14},
        pages = {L14},
          doi = {10.3847/2041-8213/aa831a},
archivePrefix = {arXiv},
       eprint = {1708.00453},
 primaryClass = {astro-ph.GA},
       adsurl = {https://ui.adsabs.harvard.edu/abs/2017ApJ...845L..14B}
}

@ARTICLE{Bertin:1996,
       author = {{Bertin}, E. and {Arnouts}, S.},
        title = "{SExtractor: Software for source extraction.}",
      journal = {\aaps},
         year = 1996,
        month = jun,
       volume = {117},
        pages = {393-404},
          doi = {10.1051/aas:1996164},
       adsurl = {https://ui.adsabs.harvard.edu/abs/1996A&AS..117..393B}
}

@article{Harris:2020,
	author = {Harris, Charles R. and Millman, K. Jarrod and van der Walt, St{\'e}fan J. and Gommers, Ralf and Virtanen, Pauli and Cournapeau, David and Wieser, Eric and Taylor, Julian and Berg, Sebastian and Smith, Nathaniel J. and Kern, Robert and Picus, Matti and Hoyer, Stephan and van Kerkwijk, Marten H. and Brett, Matthew and Haldane, Allan and del R{\'\i}o, Jaime Fern{\'a}ndez and Wiebe, Mark and Peterson, Pearu and G{\'e}rard-Marchant, Pierre and Sheppard, Kevin and Reddy, Tyler and Weckesser, Warren and Abbasi, Hameer and Gohlke, Christoph and Oliphant, Travis E.},
	date = {2020/09/01},
	doi = {10.1038/s41586-020-2649-2},
	id = {Harris2020},
	isbn = {1476-4687},
	journal = {Nature},
	number = {7825},
	pages = {357--362},
	title = {Array programming with NumPy},
	url = {https://doi.org/10.1038/s41586-020-2649-2},
	volume = {585},
	year = {2020}}

@ARTICLE{Hunter:2007,
       author = {{Hunter}, John D.},
        title = "{Matplotlib: A 2D Graphics Environment}",
      journal = {Computing in Science and Engineering},
         year = 2007,
        month = may,
       volume = {9},
       number = {3},
        pages = {90-95},
          doi = {10.1109/MCSE.2007.55},
       adsurl = {https://ui.adsabs.harvard.edu/abs/2007CSE.....9...90H}
}

@ARTICLE{Virtanen:2020,
       author = {{Virtanen}, Pauli and {Gommers}, Ralf and {Oliphant}, Travis E. and {Haberland}, Matt and {Reddy}, Tyler and {Cournapeau}, David and {Burovski}, Evgeni and {Peterson}, Pearu and {Weckesser}, Warren and {Bright}, Jonathan and {van der Walt}, St{\'e}fan J. and {Brett}, Matthew and {Wilson}, Joshua and {Millman}, K. Jarrod and {Mayorov}, Nikolay and {Nelson}, Andrew R.~J. and {Jones}, Eric and {Kern}, Robert and {Larson}, Eric and {Carey}, C.~J. and {Polat}, {\.I}lhan and {Feng}, Yu and {Moore}, Eric W. and {VanderPlas}, Jake and {Laxalde}, Denis and {Perktold}, Josef and {Cimrman}, Robert and {Henriksen}, Ian and {Quintero}, E.~A. and {Harris}, Charles R. and {Archibald}, Anne M. and {Ribeiro}, Ant{\^o}nio H. and {Pedregosa}, Fabian and {van Mulbregt}, Paul and {SciPy 1. 0 Contributors}},
        title = "{SciPy 1.0: fundamental algorithms for scientific computing in Python}",
      journal = {Nature Methods},
         year = 2020,
        month = feb,
       volume = {17},
        pages = {261-272},
          doi = {10.1038/s41592-019-0686-2},
archivePrefix = {arXiv},
       eprint = {1907.10121},
 primaryClass = {cs.MS},
       adsurl = {https://ui.adsabs.harvard.edu/abs/2020NatMe..17..261V}
}

@ARTICLE{Astropy:2013,
       author = {{Astropy Collaboration} and {Robitaille}, Thomas P. and {Tollerud}, Erik J. and {Greenfield}, Perry and {Droettboom}, Michael and {Bray}, Erik and {Aldcroft}, Tom and {Davis}, Matt and {Ginsburg}, Adam and {Price-Whelan}, Adrian M. and {Kerzendorf}, Wolfgang E. and {Conley}, Alexander and {Crighton}, Neil and {Barbary}, Kyle and {Muna}, Demitri and {Ferguson}, Henry and {Grollier}, Fr{\'e}d{\'e}ric and {Parikh}, Madhura M. and {Nair}, Prasanth H. and {Unther}, Hans M. and {Deil}, Christoph and {Woillez}, Julien and {Conseil}, Simon and {Kramer}, Roban and {Turner}, James E.~H. and {Singer}, Leo and {Fox}, Ryan and {Weaver}, Benjamin A. and {Zabalza}, Victor and {Edwards}, Zachary I. and {Azalee Bostroem}, K. and {Burke}, D.~J. and {Casey}, Andrew R. and {Crawford}, Steven M. and {Dencheva}, Nadia and {Ely}, Justin and {Jenness}, Tim and {Labrie}, Kathleen and {Lim}, Pey Lian and {Pierfederici}, Francesco and {Pontzen}, Andrew and {Ptak}, Andy and {Refsdal}, Brian and {Servillat}, Mathieu and {Streicher}, Ole},
        title = "{Astropy: A community Python package for astronomy}",
      journal = {\aap},
         year = 2013,
        month = oct,
       volume = {558},
          eid = {A33},
        pages = {A33},
          doi = {10.1051/0004-6361/201322068},
archivePrefix = {arXiv},
       eprint = {1307.6212},
 primaryClass = {astro-ph.IM},
       adsurl = {https://ui.adsabs.harvard.edu/abs/2013A&A...558A..33A}
}

@ARTICLE{Astropy:2018,
       author = {{Astropy Collaboration} and {Price-Whelan}, A.~M. and {Sip{\H{o}}cz}, B.~M. and {G{\"u}nther}, H.~M. and {Lim}, P.~L. and {Crawford}, S.~M. and {Conseil}, S. and {Shupe}, D.~L. and {Craig}, M.~W. and {Dencheva}, N. and {Ginsburg}, A. and {VanderPlas}, J.~T. and {Bradley}, L.~D. and {P{\'e}rez-Su{\'a}rez}, D. and {de Val-Borro}, M. and {Aldcroft}, T.~L. and {Cruz}, K.~L. and {Robitaille}, T.~P. and {Tollerud}, E.~J. and {Ardelean}, C. and {Babej}, T. and {Bach}, Y.~P. and {Bachetti}, M. and {Bakanov}, A.~V. and {Bamford}, S.~P. and {Barentsen}, G. and {Barmby}, P. and {Baumbach}, A. and {Berry}, K.~L. and {Biscani}, F. and {Boquien}, M. and {Bostroem}, K.~A. and {Bouma}, L.~G. and {Brammer}, G.~B. and {Bray}, E.~M. and {Breytenbach}, H. and {Buddelmeijer}, H. and {Burke}, D.~J. and {Calderone}, G. and {Cano Rodr{\'\i}guez}, J.~L. and {Cara}, M. and {Cardoso}, J.~V.~M. and {Cheedella}, S. and {Copin}, Y. and {Corrales}, L. and {Crichton}, D. and {D'Avella}, D. and {Deil}, C. and {Depagne}, {\'E}. and {Dietrich}, J.~P. and {Donath}, A. and {Droettboom}, M. and {Earl}, N. and {Erben}, T. and {Fabbro}, S. and {Ferreira}, L.~A. and {Finethy}, T. and {Fox}, R.~T. and {Garrison}, L.~H. and {Gibbons}, S.~L.~J. and {Goldstein}, D.~A. and {Gommers}, R. and {Greco}, J.~P. and {Greenfield}, P. and {Groener}, A.~M. and {Grollier}, F. and {Hagen}, A. and {Hirst}, P. and {Homeier}, D. and {Horton}, A.~J. and {Hosseinzadeh}, G. and {Hu}, L. and {Hunkeler}, J.~S. and {Ivezi{\'c}}, {\v{Z}}. and {Jain}, A. and {Jenness}, T. and {Kanarek}, G. and {Kendrew}, S. and {Kern}, N.~S. and {Kerzendorf}, W.~E. and {Khvalko}, A. and {King}, J. and {Kirkby}, D. and {Kulkarni}, A.~M. and {Kumar}, A. and {Lee}, A. and {Lenz}, D. and {Littlefair}, S.~P. and {Ma}, Z. and {Macleod}, D.~M. and {Mastropietro}, M. and {McCully}, C. and {Montagnac}, S. and {Morris}, B.~M. and {Mueller}, M. and {Mumford}, S.~J. and {Muna}, D. and {Murphy}, N.~A. and {Nelson}, S. and {Nguyen}, G.~H. and {Ninan}, J.~P. and {N{\"o}the}, M. and {Ogaz}, S. and {Oh}, S. and {Parejko}, J.~K. and {Parley}, N. and {Pascual}, S. and {Patil}, R. and {Patil}, A.~A. and {Plunkett}, A.~L. and {Prochaska}, J.~X. and {Rastogi}, T. and {Reddy Janga}, V. and {Sabater}, J. and {Sakurikar}, P. and {Seifert}, M. and {Sherbert}, L.~E. and {Sherwood-Taylor}, H. and {Shih}, A.~Y. and {Sick}, J. and {Silbiger}, M.~T. and {Singanamalla}, S. and {Singer}, L.~P. and {Sladen}, P.~H. and {Sooley}, K.~A. and {Sornarajah}, S. and {Streicher}, O. and {Teuben}, P. and {Thomas}, S.~W. and {Tremblay}, G.~R. and {Turner}, J.~E.~H. and {Terr{\'o}n}, V. and {van Kerkwijk}, M.~H. and {de la Vega}, A. and {Watkins}, L.~L. and {Weaver}, B.~A. and {Whitmore}, J.~B. and {Woillez}, J. and {Zabalza}, V. and {Astropy Contributors}},
        title = "{The Astropy Project: Building an Open-science Project and Status of the v2.0 Core Package}",
      journal = {\aj},
         year = 2018,
        month = sep,
       volume = {156},
       number = {3},
          eid = {123},
        pages = {123},
          doi = {10.3847/1538-3881/aabc4f},
archivePrefix = {arXiv},
       eprint = {1801.02634},
 primaryClass = {astro-ph.IM},
       adsurl = {https://ui.adsabs.harvard.edu/abs/2018AJ....156..123A}
}

@software{Smithsonian:2000,
       author = {{Smithsonian Astrophysical Observatory}},
        title = "{SAOImage DS9: A utility for displaying astronomical images in the X11 window environment}",
 howpublished = {Astrophysics Source Code Library, record ascl:0003.002},
         year = 2000,
        month = mar,
          eid = {ascl:0003.002},
       adsurl = {https://ui.adsabs.harvard.edu/abs/2000ascl.soft03002S}
}

@ARTICLE{York:2000,
       author = {{York}, Donald G. and {Adelman}, J. and {Anderson}, Jr., John E. and {Anderson}, Scott F. and {Annis}, James and {Bahcall}, Neta A. and {Bakken}, J.~A. and {Barkhouser}, Robert and {Bastian}, Steven and {Berman}, Eileen and {Boroski}, William N. and {Bracker}, Steve and {Briegel}, Charlie and {Briggs}, John W. and {Brinkmann}, J. and {Brunner}, Robert and {Burles}, Scott and {Carey}, Larry and {Carr}, Michael A. and {Castander}, Francisco J. and {Chen}, Bing and {Colestock}, Patrick L. and {Connolly}, A.~J. and {Crocker}, J.~H. and {Csabai}, Istv{\'a}n and {Czarapata}, Paul C. and {Davis}, John Eric and {Doi}, Mamoru and {Dombeck}, Tom and {Eisenstein}, Daniel and {Ellman}, Nancy and {Elms}, Brian R. and {Evans}, Michael L. and {Fan}, Xiaohui and {Federwitz}, Glenn R. and {Fiscelli}, Larry and {Friedman}, Scott and {Frieman}, Joshua A. and {Fukugita}, Masataka and {Gillespie}, Bruce and {Gunn}, James E. and {Gurbani}, Vijay K. and {de Haas}, Ernst and {Haldeman}, Merle and {Harris}, Frederick H. and {Hayes}, J. and {Heckman}, Timothy M. and {Hennessy}, G.~S. and {Hindsley}, Robert B. and {Holm}, Scott and {Holmgren}, Donald J. and {Huang}, Chi-hao and {Hull}, Charles and {Husby}, Don and {Ichikawa}, Shin-Ichi and {Ichikawa}, Takashi and {Ivezi{\'c}}, {\v{Z}}eljko and {Kent}, Stephen and {Kim}, Rita S.~J. and {Kinney}, E. and {Klaene}, Mark and {Kleinman}, A.~N. and {Kleinman}, S. and {Knapp}, G.~R. and {Korienek}, John and {Kron}, Richard G. and {Kunszt}, Peter Z. and {Lamb}, D.~Q. and {Lee}, B. and {Leger}, R. French and {Limmongkol}, Siriluk and {Lindenmeyer}, Carl and {Long}, Daniel C. and {Loomis}, Craig and {Loveday}, Jon and {Lucinio}, Rich and {Lupton}, Robert H. and {MacKinnon}, Bryan and {Mannery}, Edward J. and {Mantsch}, P.~M. and {Margon}, Bruce and {McGehee}, Peregrine and {McKay}, Timothy A. and {Meiksin}, Avery and {Merelli}, Aronne and {Monet}, David G. and {Munn}, Jeffrey A. and {Narayanan}, Vijay K. and {Nash}, Thomas and {Neilsen}, Eric and {Neswold}, Rich and {Newberg}, Heidi Jo and {Nichol}, R.~C. and {Nicinski}, Tom and {Nonino}, Mario and {Okada}, Norio and {Okamura}, Sadanori and {Ostriker}, Jeremiah P. and {Owen}, Russell and {Pauls}, A. George and {Peoples}, John and {Peterson}, R.~L. and {Petravick}, Donald and {Pier}, Jeffrey R. and {Pope}, Adrian and {Pordes}, Ruth and {Prosapio}, Angela and {Rechenmacher}, Ron and {Quinn}, Thomas R. and {Richards}, Gordon T. and {Richmond}, Michael W. and {Rivetta}, Claudio H. and {Rockosi}, Constance M. and {Ruthmansdorfer}, Kurt and {Sandford}, Dale and {Schlegel}, David J. and {Schneider}, Donald P. and {Sekiguchi}, Maki and {Sergey}, Gary and {Shimasaku}, Kazuhiro and {Siegmund}, Walter A. and {Smee}, Stephen and {Smith}, J. Allyn and {Snedden}, S. and {Stone}, R. and {Stoughton}, Chris and {Strauss}, Michael A. and {Stubbs}, Christopher and {SubbaRao}, Mark and {Szalay}, Alexander S. and {Szapudi}, Istvan and {Szokoly}, Gyula P. and {Thakar}, Anirudda R. and {Tremonti}, Christy and {Tucker}, Douglas L. and {Uomoto}, Alan and {Vanden Berk}, Dan and {Vogeley}, Michael S. and {Waddell}, Patrick and {Wang}, Shu-i. and {Watanabe}, Masaru and {Weinberg}, David H. and {Yanny}, Brian and {Yasuda}, Naoki and {SDSS Collaboration}},
        title = "{The Sloan Digital Sky Survey: Technical Summary}",
      journal = {\aj},
         year = 2000,
        month = sep,
       volume = {120},
       number = {3},
        pages = {1579-1587},
          doi = {10.1086/301513},
archivePrefix = {arXiv},
       eprint = {astro-ph/0006396},
 primaryClass = {astro-ph},
       adsurl = {https://ui.adsabs.harvard.edu/abs/2000AJ....120.1579Y}
}

@ARTICLE{Blanton:2017,
       author = {{Blanton}, Michael R. and {Bershady}, Matthew A. and {Abolfathi}, Bela and {Albareti}, Franco D. and {Allende Prieto}, Carlos and {Almeida}, Andres and {Alonso-Garc{\'\i}a}, Javier and {Anders}, Friedrich and {Anderson}, Scott F. and {Andrews}, Brett and {Aquino-Ort{\'\i}z}, Erik and {Arag{\'o}n-Salamanca}, Alfonso and {Argudo-Fern{\'a}ndez}, Maria and {Armengaud}, Eric and {Aubourg}, Eric and {Avila-Reese}, Vladimir and {Badenes}, Carles and {Bailey}, Stephen and {Barger}, Kathleen A. and {Barrera-Ballesteros}, Jorge and {Bartosz}, Curtis and {Bates}, Dominic and {Baumgarten}, Falk and {Bautista}, Julian and {Beaton}, Rachael and {Beers}, Timothy C. and {Belfiore}, Francesco and {Bender}, Chad F. and {Berlind}, Andreas A. and {Bernardi}, Mariangela and {Beutler}, Florian and {Bird}, Jonathan C. and {Bizyaev}, Dmitry and {Blanc}, Guillermo A. and {Blomqvist}, Michael and {Bolton}, Adam S. and {Boquien}, M{\'e}d{\'e}ric and {Borissova}, Jura and {van den Bosch}, Remco and {Bovy}, Jo and {Brandt}, William N. and {Brinkmann}, Jonathan and {Brownstein}, Joel R. and {Bundy}, Kevin and {Burgasser}, Adam J. and {Burtin}, Etienne and {Busca}, Nicol{\'a}s G. and {Cappellari}, Michele and {Delgado Carigi}, Maria Leticia and {Carlberg}, Joleen K. and {Carnero Rosell}, Aurelio and {Carrera}, Ricardo and {Chanover}, Nancy J. and {Cherinka}, Brian and {Cheung}, Edmond and {G{\'o}mez Maqueo Chew}, Yilen and {Chiappini}, Cristina and {Choi}, Peter Doohyun and {Chojnowski}, Drew and {Chuang}, Chia-Hsun and {Chung}, Haeun and {Cirolini}, Rafael Fernando and {Clerc}, Nicolas and {Cohen}, Roger E. and {Comparat}, Johan and {da Costa}, Luiz and {Cousinou}, Marie-Claude and {Covey}, Kevin and {Crane}, Jeffrey D. and {Croft}, Rupert A.~C. and {Cruz-Gonzalez}, Irene and {Garrido Cuadra}, Daniel and {Cunha}, Katia and {Damke}, Guillermo J. and {Darling}, Jeremy and {Davies}, Roger and {Dawson}, Kyle and {de la Macorra}, Axel and {Dell'Agli}, Flavia and {De Lee}, Nathan and {Delubac}, Timoth{\'e}e and {Di Mille}, Francesco and {Diamond-Stanic}, Aleks and {Cano-D{\'\i}az}, Mariana and {Donor}, John and {Downes}, Juan Jos{\'e} and {Drory}, Niv and {du Mas des Bourboux}, H{\'e}lion and {Duckworth}, Christopher J. and {Dwelly}, Tom and {Dyer}, Jamie and {Ebelke}, Garrett and {Eigenbrot}, Arthur D. and {Eisenstein}, Daniel J. and {Emsellem}, Eric and {Eracleous}, Mike and {Escoffier}, Stephanie and {Evans}, Michael L. and {Fan}, Xiaohui and {Fern{\'a}ndez-Alvar}, Emma and {Fernandez-Trincado}, J.~G. and {Feuillet}, Diane K. and {Finoguenov}, Alexis and {Fleming}, Scott W. and {Font-Ribera}, Andreu and {Fredrickson}, Alexander and {Freischlad}, Gordon and {Frinchaboy}, Peter M. and {Fuentes}, Carla E. and {Galbany}, Llu{\'\i}s and {Garcia-Dias}, R. and {Garc{\'\i}a-Hern{\'a}ndez}, D.~A. and {Gaulme}, Patrick and {Geisler}, Doug and {Gelfand}, Joseph D. and {Gil-Mar{\'\i}n}, H{\'e}ctor and {Gillespie}, Bruce A. and {Goddard}, Daniel and {Gonzalez-Perez}, Violeta and {Grabowski}, Kathleen and {Green}, Paul J. and {Grier}, Catherine J. and {Gunn}, James E. and {Guo}, Hong and {Guy}, Julien and {Hagen}, Alex and {Hahn}, ChangHoon and {Hall}, Matthew and {Harding}, Paul and {Hasselquist}, Sten and {Hawley}, Suzanne L. and {Hearty}, Fred and {Gonzalez Hern{\'a}ndez}, Jonay I. and {Ho}, Shirley and {Hogg}, David W. and {Holley-Bockelmann}, Kelly and {Holtzman}, Jon A. and {Holzer}, Parker H. and {Huehnerhoff}, Joseph and {Hutchinson}, Timothy A. and {Hwang}, Ho Seong and {Ibarra-Medel}, H{\'e}ctor J. and {da Silva Ilha}, Gabriele and {Ivans}, Inese I. and {Ivory}, KeShawn and {Jackson}, Kelly and {Jensen}, Trey W. and {Johnson}, Jennifer A. and {Jones}, Amy and {J{\"o}nsson}, Henrik and {Jullo}, Eric and {Kamble}, Vikrant and {Kinemuchi}, Karen and {Kirkby}, David and {Kitaura}, Francisco-Shu and {Klaene}, Mark and {Knapp}, Gillian R. and {Kneib}, Jean-Paul and {Kollmeier}, Juna A. and {Lacerna}, Ivan and {Lane}, Richard R. and {Lang}, Dustin and {Law}, David R. and {Lazarz}, Daniel and {Lee}, Youngbae and {Le Goff}, Jean-Marc and {Liang}, Fu-Heng and {Li}, Cheng and {Li}, Hongyu and {Lian}, Jianhui and {Lima}, Marcos and {Lin}, Lihwai and {Lin}, Yen-Ting and {Bertran de Lis}, Sara and {Liu}, Chao and {de Icaza Lizaola}, Miguel Angel C. and {Long}, Dan and {Lucatello}, Sara and {Lundgren}, Britt and {MacDonald}, Nicholas K. and {Deconto Machado}, Alice and {MacLeod}, Chelsea L. and {Mahadevan}, Suvrath and {Geimba Maia}, Marcio Antonio and {Maiolino}, Roberto and {Majewski}, Steven R. and {Malanushenko}, Elena and {Malanushenko}, Viktor and {Manchado}, Arturo and {Mao}, Shude and {Maraston}, Claudia and {Marques-Chaves}, Rui and {Masseron}, Thomas and {Masters}, Karen L. and {McBride}, Cameron K. and {McDermid}, Richard M. and {McGrath}, Brianne and {McGreer}, Ian D. and {Medina Pe{\~n}a}, Nicol{\'a}s and {Melendez}, Matthew},
        title = "{Sloan Digital Sky Survey IV: Mapping the Milky Way, Nearby Galaxies, and the Distant Universe}",
      journal = {\aj},
         year = 2017,
        month = jul,
       volume = {154},
       number = {1},
          eid = {28},
        pages = {28},
          doi = {10.3847/1538-3881/aa7567},
archivePrefix = {arXiv},
       eprint = {1703.00052},
 primaryClass = {astro-ph.GA},
       adsurl = {https://ui.adsabs.harvard.edu/abs/2017AJ....154...28B}
}

@ARTICLE{Narayan:1996,
       author = {{Narayan}, Ramesh and {Bartelmann}, Matthias},
        title = "{Lectures on Gravitational Lensing}",
      journal = {arXiv e-prints},
         year = 1996,
        month = jun,
          eid = {astro-ph/9606001},
        pages = {astro-ph/9606001},
          doi = {10.48550/arXiv.astro-ph/9606001},
archivePrefix = {arXiv},
       eprint = {astro-ph/9606001},
 primaryClass = {astro-ph},
       adsurl = {https://ui.adsabs.harvard.edu/abs/1996astro.ph..6001N}
}

@ARTICLE{Schneider:1985,
       author = {{Schneider}, P.},
        title = "{A new formulation of gravitational lens theory, time-delay, and Fermat's principle}",
      journal = {\aap},
         year = 1985,
        month = feb,
       volume = {143},
       number = {2},
        pages = {413-420},
       adsurl = {https://ui.adsabs.harvard.edu/abs/1985A&A...143..413S}
}

@ARTICLE{Ivezic:2019,
       author = {{Ivezi{\'c}}, {\v{Z}}eljko and {Kahn}, Steven M. and {Tyson}, J. Anthony and {Abel}, Bob and {Acosta}, Emily and {Allsman}, Robyn and {Alonso}, David and {AlSayyad}, Yusra and {Anderson}, Scott F. and {Andrew}, John and {Angel}, James Roger P. and {Angeli}, George Z. and {Ansari}, Reza and {Antilogus}, Pierre and {Araujo}, Constanza and {Armstrong}, Robert and {Arndt}, Kirk T. and {Astier}, Pierre and {Aubourg}, {\'E}ric and {Auza}, Nicole and {Axelrod}, Tim S. and {Bard}, Deborah J. and {Barr}, Jeff D. and {Barrau}, Aurelian and {Bartlett}, James G. and {Bauer}, Amanda E. and {Bauman}, Brian J. and {Baumont}, Sylvain and {Bechtol}, Ellen and {Bechtol}, Keith and {Becker}, Andrew C. and {Becla}, Jacek and {Beldica}, Cristina and {Bellavia}, Steve and {Bianco}, Federica B. and {Biswas}, Rahul and {Blanc}, Guillaume and {Blazek}, Jonathan and {Blandford}, Roger D. and {Bloom}, Josh S. and {Bogart}, Joanne and {Bond}, Tim W. and {Booth}, Michael T. and {Borgland}, Anders W. and {Borne}, Kirk and {Bosch}, James F. and {Boutigny}, Dominique and {Brackett}, Craig A. and {Bradshaw}, Andrew and {Brandt}, William Nielsen and {Brown}, Michael E. and {Bullock}, James S. and {Burchat}, Patricia and {Burke}, David L. and {Cagnoli}, Gianpietro and {Calabrese}, Daniel and {Callahan}, Shawn and {Callen}, Alice L. and {Carlin}, Jeffrey L. and {Carlson}, Erin L. and {Chandrasekharan}, Srinivasan and {Charles-Emerson}, Glenaver and {Chesley}, Steve and {Cheu}, Elliott C. and {Chiang}, Hsin-Fang and {Chiang}, James and {Chirino}, Carol and {Chow}, Derek and {Ciardi}, David R. and {Claver}, Charles F. and {Cohen-Tanugi}, Johann and {Cockrum}, Joseph J. and {Coles}, Rebecca and {Connolly}, Andrew J. and {Cook}, Kem H. and {Cooray}, Asantha and {Covey}, Kevin R. and {Cribbs}, Chris and {Cui}, Wei and {Cutri}, Roc and {Daly}, Philip N. and {Daniel}, Scott F. and {Daruich}, Felipe and {Daubard}, Guillaume and {Daues}, Greg and {Dawson}, William and {Delgado}, Francisco and {Dellapenna}, Alfred and {de Peyster}, Robert and {de Val-Borro}, Miguel and {Digel}, Seth W. and {Doherty}, Peter and {Dubois}, Richard and {Dubois-Felsmann}, Gregory P. and {Durech}, Josef and {Economou}, Frossie and {Eifler}, Tim and {Eracleous}, Michael and {Emmons}, Benjamin L. and {Fausti Neto}, Angelo and {Ferguson}, Henry and {Figueroa}, Enrique and {Fisher-Levine}, Merlin and {Focke}, Warren and {Foss}, Michael D. and {Frank}, James and {Freemon}, Michael D. and {Gangler}, Emmanuel and {Gawiser}, Eric and {Geary}, John C. and {Gee}, Perry and {Geha}, Marla and {Gessner}, Charles J.~B. and {Gibson}, Robert R. and {Gilmore}, D. Kirk and {Glanzman}, Thomas and {Glick}, William and {Goldina}, Tatiana and {Goldstein}, Daniel A. and {Goodenow}, Iain and {Graham}, Melissa L. and {Gressler}, William J. and {Gris}, Philippe and {Guy}, Leanne P. and {Guyonnet}, Augustin and {Haller}, Gunther and {Harris}, Ron and {Hascall}, Patrick A. and {Haupt}, Justine and {Hernandez}, Fabio and {Herrmann}, Sven and {Hileman}, Edward and {Hoblitt}, Joshua and {Hodgson}, John A. and {Hogan}, Craig and {Howard}, James D. and {Huang}, Dajun and {Huffer}, Michael E. and {Ingraham}, Patrick and {Innes}, Walter R. and {Jacoby}, Suzanne H. and {Jain}, Bhuvnesh and {Jammes}, Fabrice and {Jee}, M. James and {Jenness}, Tim and {Jernigan}, Garrett and {Jevremovi{\'c}}, Darko and {Johns}, Kenneth and {Johnson}, Anthony S. and {Johnson}, Margaret W.~G. and {Jones}, R. Lynne and {Juramy-Gilles}, Claire and {Juri{\'c}}, Mario and {Kalirai}, Jason S. and {Kallivayalil}, Nitya J. and {Kalmbach}, Bryce and {Kantor}, Jeffrey P. and {Karst}, Pierre and {Kasliwal}, Mansi M. and {Kelly}, Heather and {Kessler}, Richard and {Kinnison}, Veronica and {Kirkby}, David and {Knox}, Lloyd and {Kotov}, Ivan V. and {Krabbendam}, Victor L. and {Krughoff}, K. Simon and {Kub{\'a}nek}, Petr and {Kuczewski}, John and {Kulkarni}, Shri and {Ku}, John and {Kurita}, Nadine R. and {Lage}, Craig S. and {Lambert}, Ron and {Lange}, Travis and {Langton}, J. Brian and {Le Guillou}, Laurent and {Levine}, Deborah and {Liang}, Ming and {Lim}, Kian-Tat and {Lintott}, Chris J. and {Long}, Kevin E. and {Lopez}, Margaux and {Lotz}, Paul J. and {Lupton}, Robert H. and {Lust}, Nate B. and {MacArthur}, Lauren A. and {Mahabal}, Ashish and {Mandelbaum}, Rachel and {Markiewicz}, Thomas W. and {Marsh}, Darren S. and {Marshall}, Philip J. and {Marshall}, Stuart and {May}, Morgan and {McKercher}, Robert and {McQueen}, Michelle and {Meyers}, Joshua and {Migliore}, Myriam and {Miller}, Michelle and {Mills}, David J.},
        title = "{LSST: From Science Drivers to Reference Design and Anticipated Data Products}",
      journal = {\apj},
         year = 2019,
        month = mar,
       volume = {873},
       number = {2},
          eid = {111},
        pages = {111},
          doi = {10.3847/1538-4357/ab042c},
archivePrefix = {arXiv},
       eprint = {0805.2366},
 primaryClass = {astro-ph},
       adsurl = {https://ui.adsabs.harvard.edu/abs/2019ApJ...873..111I}
}

@ARTICLE{pypeit:joss_arXiv,
       author = {{Prochaska}, J. Xavier and {Hennawi}, Joseph F. and {Westfall}, Kyle B. and
         {Cooke}, Ryan J. and {Wang}, Feige and {Hsyu}, Tiffany and
         {Davies}, Frederick B. and {Farina}, Emanuele Paolo},
        title = "{PypeIt: The Python Spectroscopic Data Reduction Pipeline}",
      journal = {arXiv e-prints},
         year = {2020a},
        month = may,
          eid = {arXiv:2005.06505},
        pages = {arXiv:2005.06505},
archivePrefix = {arXiv},
       eprint = {2005.06505},
 primaryClass = {astro-ph.IM},
       adsurl = {https://ui.adsabs.harvard.edu/abs/2020arXiv200506505P}
}

@article{pypeit:joss_pub,
    doi = {10.21105/joss.02308},
    url = {https://doi.org/10.21105/joss.02308},
    year = {2020b},
    publisher = {The Open Journal},
    volume = {5},
    number = {56},
    pages = {2308},
    author = {J. Xavier Prochaska and Joseph F. Hennawi and Kyle B. Westfall and Ryan J. Cooke and Feige Wang and Tiffany Hsyu and Frederick B. Davies and Emanuele Paolo Farina and Debora Pelliccia},
    title = {PypeIt: The Python Spectroscopic Data Reduction Pipeline},
    journal = {Journal of Open Source Software}
}

@MISC{pypeit:zenodo,
       author = {{Prochaska}, J. Xavier and {Hennawi}, Joseph and {Cooke}, Ryan and
         {Westfall}, Kyle and {Wang}, Feige and {EmAstro} and {Tiffanyhsyu} and
         {Wasserman}, Asher and {Villaume}, Alexa and {Marijana777} and
         {Schindler}, JT and {Young}, David and {Simha}, Sunil and
         {Wilde}, Matt and {Tejos}, Nicolas and {Isbell}, Jacob and
         {Fl{\"o}rs}, Andreas and {Sandford}, Nathan and {Vasovi{\'c}}, Zlatan and
         {Betts}, Edward and {Holden}, Brad},
        title = "{pypeit/PypeIt: Release 1.0.0}",
         year = {2020c},
        month = apr,
          eid = {10.5281/zenodo.3743493},
          doi = {10.5281/zenodo.3743493},
      version = {v1.0.0},
    publisher = {Zenodo},
       adsurl = {https://ui.adsabs.harvard.edu/abs/2020zndo...3743493P}
}

@ARTICLE{Veilleux:1987,
       author = {{Veilleux}, Sylvain and {Osterbrock}, Donald E.},
        title = "{Spectral Classification of Emission-Line Galaxies}",
      journal = {\apjs},
         year = 1987,
        month = feb,
       volume = {63},
        pages = {295},
          doi = {10.1086/191166},
       adsurl = {https://ui.adsabs.harvard.edu/abs/1987ApJS...63..295V}
}

@software{Ginsburg:2011,
       author = {{Ginsburg}, Adam and {Mirocha}, Jordan},
        title = "{PySpecKit: Python Spectroscopic Toolkit}",
 howpublished = {Astrophysics Source Code Library, record ascl:1109.001},
         year = 2011,
        month = sep,
          eid = {ascl:1109.001},
       adsurl = {https://ui.adsabs.harvard.edu/abs/2011ascl.soft09001G}
}

@ARTICLE{Ginsburg:2022,
       author = {{Ginsburg}, Adam and {Sokolov}, Vlas and {de Val-Borro}, Miguel and {Rosolowsky}, Erik and {Pineda}, Jaime E. and {Sip{\H{o}}cz}, Brigitta M. and {Henshaw}, Jonathan D.},
        title = "{Pyspeckit: A Spectroscopic Analysis and Plotting Package}",
      journal = {\aj},
         year = 2022,
        month = jun,
       volume = {163},
       number = {6},
          eid = {291},
        pages = {291},
          doi = {10.3847/1538-3881/ac695a},
archivePrefix = {arXiv},
       eprint = {2205.04987},
 primaryClass = {astro-ph.IM},
       adsurl = {https://ui.adsabs.harvard.edu/abs/2022AJ....163..291G}
}

@ARTICLE{Sun:2021,
       author = {{Sun}, Fengwu and {Egami}, Eiichi and {Rawle}, Timothy D. and {Walth}, Gregory L. and {Smail}, Ian and {Dessauges-Zavadsky}, Miroslava and {P{\'e}rez-Gonz{\'a}lez}, Pablo G. and {Richard}, Johan and {Combes}, Francoise and {Ebeling}, Harald and {Pell{\'o}}, Roser and {Van der Werf}, Paul and {Altieri}, Bruno and {Boone}, Fr{\'e}d{\'e}ric and {Cava}, Antonio and {Chapman}, Scott C. and {Cl{\'e}ment}, Benjamin and {Finoguenov}, Alexis and {Nakajima}, Kimihiko and {Rujopakarn}, Wiphu and {Schaerer}, Daniel and {Valtchanov}, Ivan},
        title = "{ALMA 1.3 mm Survey of Lensed Submillimeter Galaxies Selected by Herschel: Discovery of Spatially Extended SMGs and Implications}",
      journal = {\apj},
         year = 2021,
        month = feb,
       volume = {908},
       number = {2},
          eid = {192},
        pages = {192},
          doi = {10.3847/1538-4357/abd6e4},
archivePrefix = {arXiv},
       eprint = {2101.03677},
 primaryClass = {astro-ph.GA},
       adsurl = {https://ui.adsabs.harvard.edu/abs/2021ApJ...908..192S}
}

@ARTICLE{Dressler:2011,
       author = {{Dressler}, Alan and {Bigelow}, Bruce and {Hare}, Tyson and {Sutin}, Brian and {Thompson}, Ian and {Burley}, Greg and {Epps}, Harland and {Oemler}, Jr., Augustus and {Bagish}, Alan and {Birk}, Christoph and {Clardy}, Ken and {Gunnels}, Steve and {Kelson}, Daniel and {Shectman}, Stephen and {Osip}, David},
        title = "{IMACS: The Inamori-Magellan Areal Camera and Spectrograph on Magellan-Baade}",
      journal = {\pasp},
         year = 2011,
        month = mar,
       volume = {123},
       number = {901},
        pages = {288},
          doi = {10.1086/658908},
       adsurl = {https://ui.adsabs.harvard.edu/abs/2011PASP..123..288D}
}

@ARTICLE{Juneau:2021,
       author = {{Juneau}, Stephanie and {Olsen}, Knut and {Nikutta}, Robert and {Jacques}, Alice and {Bailey}, Stephen},
        title = "{Jupyter-Enabled Astrophysical Analysis Using Data-Proximate Computing Platforms}",
      journal = {Computing in Science and Engineering},
         year = 2021,
        month = mar,
       volume = {23},
       number = {2},
        pages = {15-25},
          doi = {10.1109/MCSE.2021.3057097},
archivePrefix = {arXiv},
       eprint = {2104.06527},
 primaryClass = {astro-ph.IM},
       adsurl = {https://ui.adsabs.harvard.edu/abs/2021CSE....23b..15J}
}

@ARTICLE{Kewley:2006,
       author = {{Kewley}, Lisa J. and {Groves}, Brent and {Kauffmann}, Guinevere and {Heckman}, Tim},
        title = "{The host galaxies and classification of active galactic nuclei}",
      journal = {\mnras},
         year = 2006,
        month = nov,
       volume = {372},
       number = {3},
        pages = {961-976},
          doi = {10.1111/j.1365-2966.2006.10859.x},
archivePrefix = {arXiv},
       eprint = {astro-ph/0605681},
 primaryClass = {astro-ph},
       adsurl = {https://ui.adsabs.harvard.edu/abs/2006MNRAS.372..961K}
}

@ARTICLE{Kewley:2019,
       author = {{Kewley}, Lisa J. and {Nicholls}, David C. and {Sutherland}, Ralph S.},
        title = "{Understanding Galaxy Evolution Through Emission Lines}",
      journal = {\araa},
         year = 2019,
        month = aug,
       volume = {57},
        pages = {511-570},
          doi = {10.1146/annurev-astro-081817-051832},
archivePrefix = {arXiv},
       eprint = {1910.09730},
 primaryClass = {astro-ph.GA},
       adsurl = {https://ui.adsabs.harvard.edu/abs/2019ARA&A..57..511K}
}

@ARTICLE{Schawinski:2007,
       author = {{Schawinski}, Kevin and {Thomas}, Daniel and {Sarzi}, Marc and {Maraston}, Claudia and {Kaviraj}, Sugata and {Joo}, Seok-Joo and {Yi}, Sukyoung K. and {Silk}, Joseph},
        title = "{Observational evidence for AGN feedback in early-type galaxies}",
      journal = {\mnras},
         year = 2007,
        month = dec,
       volume = {382},
       number = {4},
        pages = {1415-1431},
          doi = {10.1111/j.1365-2966.2007.12487.x},
archivePrefix = {arXiv},
       eprint = {0709.3015},
 primaryClass = {astro-ph},
       adsurl = {https://ui.adsabs.harvard.edu/abs/2007MNRAS.382.1415S}
}

@ARTICLE{Kewley:2001,
       author = {{Kewley}, L.~J. and {Dopita}, M.~A. and {Sutherland}, R.~S. and {Heisler}, C.~A. and {Trevena}, J.},
        title = "{Theoretical Modeling of Starburst Galaxies}",
      journal = {\apj},
         year = 2001,
        month = jul,
       volume = {556},
       number = {1},
        pages = {121-140},
          doi = {10.1086/321545},
archivePrefix = {arXiv},
       eprint = {astro-ph/0106324},
 primaryClass = {astro-ph},
       adsurl = {https://ui.adsabs.harvard.edu/abs/2001ApJ...556..121K}
}

@ARTICLE{Gaia:2016b,
       author = {{Gaia Collaboration} and {Prusti}, T. and {de Bruijne}, J.~H.~J. and {Brown}, A.~G.~A. and {Vallenari}, A. and {Babusiaux}, C. and {Bailer-Jones}, C.~A.~L. and {Bastian}, U. and {Biermann}, M. and {Evans}, D.~W. and {Eyer}, L. and {Jansen}, F. and {Jordi}, C. and {Klioner}, S.~A. and {Lammers}, U. and {Lindegren}, L. and {Luri}, X. and {Mignard}, F. and {Milligan}, D.~J. and {Panem}, C. and {Poinsignon}, V. and {Pourbaix}, D. and {Randich}, S. and {Sarri}, G. and {Sartoretti}, P. and {Siddiqui}, H.~I. and {Soubiran}, C. and {Valette}, V. and {van Leeuwen}, F. and {Walton}, N.~A. and {Aerts}, C. and {Arenou}, F. and {Cropper}, M. and {Drimmel}, R. and {H{\o}g}, E. and {Katz}, D. and {Lattanzi}, M.~G. and {O'Mullane}, W. and {Grebel}, E.~K. and {Holland}, A.~D. and {Huc}, C. and {Passot}, X. and {Bramante}, L. and {Cacciari}, C. and {Casta{\~n}eda}, J. and {Chaoul}, L. and {Cheek}, N. and {De Angeli}, F. and {Fabricius}, C. and {Guerra}, R. and {Hern{\'a}ndez}, J. and {Jean-Antoine-Piccolo}, A. and {Masana}, E. and {Messineo}, R. and {Mowlavi}, N. and {Nienartowicz}, K. and {Ord{\'o}{\~n}ez-Blanco}, D. and {Panuzzo}, P. and {Portell}, J. and {Richards}, P.~J. and {Riello}, M. and {Seabroke}, G.~M. and {Tanga}, P. and {Th{\'e}venin}, F. and {Torra}, J. and {Els}, S.~G. and {Gracia-Abril}, G. and {Comoretto}, G. and {Garcia-Reinaldos}, M. and {Lock}, T. and {Mercier}, E. and {Altmann}, M. and {Andrae}, R. and {Astraatmadja}, T.~L. and {Bellas-Velidis}, I. and {Benson}, K. and {Berthier}, J. and {Blomme}, R. and {Busso}, G. and {Carry}, B. and {Cellino}, A. and {Clementini}, G. and {Cowell}, S. and {Creevey}, O. and {Cuypers}, J. and {Davidson}, M. and {De Ridder}, J. and {de Torres}, A. and {Delchambre}, L. and {Dell'Oro}, A. and {Ducourant}, C. and {Fr{\'e}mat}, Y. and {Garc{\'\i}a-Torres}, M. and {Gosset}, E. and {Halbwachs}, J. -L. and {Hambly}, N.~C. and {Harrison}, D.~L. and {Hauser}, M. and {Hestroffer}, D. and {Hodgkin}, S.~T. and {Huckle}, H.~E. and {Hutton}, A. and {Jasniewicz}, G. and {Jordan}, S. and {Kontizas}, M. and {Korn}, A.~J. and {Lanzafame}, A.~C. and {Manteiga}, M. and {Moitinho}, A. and {Muinonen}, K. and {Osinde}, J. and {Pancino}, E. and {Pauwels}, T. and {Petit}, J. -M. and {Recio-Blanco}, A. and {Robin}, A.~C. and {Sarro}, L.~M. and {Siopis}, C. and {Smith}, M. and {Smith}, K.~W. and {Sozzetti}, A. and {Thuillot}, W. and {van Reeven}, W. and {Viala}, Y. and {Abbas}, U. and {Abreu Aramburu}, A. and {Accart}, S. and {Aguado}, J.~J. and {Allan}, P.~M. and {Allasia}, W. and {Altavilla}, G. and {{\'A}lvarez}, M.~A. and {Alves}, J. and {Anderson}, R.~I. and {Andrei}, A.~H. and {Anglada Varela}, E. and {Antiche}, E. and {Antoja}, T. and {Ant{\'o}n}, S. and {Arcay}, B. and {Atzei}, A. and {Ayache}, L. and {Bach}, N. and {Baker}, S.~G. and {Balaguer-N{\'u}{\~n}ez}, L. and {Barache}, C. and {Barata}, C. and {Barbier}, A. and {Barblan}, F. and {Baroni}, M. and {Barrado y Navascu{\'e}s}, D. and {Barros}, M. and {Barstow}, M.~A. and {Becciani}, U. and {Bellazzini}, M. and {Bellei}, G. and {Bello Garc{\'\i}a}, A. and {Belokurov}, V. and {Bendjoya}, P. and {Berihuete}, A. and {Bianchi}, L. and {Bienaym{\'e}}, O. and {Billebaud}, F. and {Blagorodnova}, N. and {Blanco-Cuaresma}, S. and {Boch}, T. and {Bombrun}, A. and {Borrachero}, R. and {Bouquillon}, S. and {Bourda}, G. and {Bouy}, H. and {Bragaglia}, A. and {Breddels}, M.~A. and {Brouillet}, N. and {Br{\"u}semeister}, T. and {Bucciarelli}, B. and {Budnik}, F. and {Burgess}, P. and {Burgon}, R. and {Burlacu}, A. and {Busonero}, D. and {Buzzi}, R. and {Caffau}, E. and {Cambras}, J. and {Campbell}, H. and {Cancelliere}, R. and {Cantat-Gaudin}, T. and {Carlucci}, T. and {Carrasco}, J.~M. and {Castellani}, M. and {Charlot}, P. and {Charnas}, J. and {Charvet}, P. and {Chassat}, F. and {Chiavassa}, A. and {Clotet}, M. and {Cocozza}, G. and {Collins}, R.~S. and {Collins}, P. and {Costigan}, G.},
        title = "{The Gaia mission}",
      journal = {\aap},
         year = 2016,
        month = nov,
       volume = {595},
          eid = {A1},
        pages = {A1},
          doi = {10.1051/0004-6361/201629272},
archivePrefix = {arXiv},
       eprint = {1609.04153},
 primaryClass = {astro-ph.IM},
       adsurl = {https://ui.adsabs.harvard.edu/abs/2016A&A...595A...1G}
}

@ARTICLE{Gaia:2023j,
       author = {{Gaia Collaboration} and {Vallenari}, A. and {Brown}, A.~G.~A. and {Prusti}, T. and {de Bruijne}, J.~H.~J. and {Arenou}, F. and {Babusiaux}, C. and {Biermann}, M. and {Creevey}, O.~L. and {Ducourant}, C. and {Evans}, D.~W. and {Eyer}, L. and {Guerra}, R. and {Hutton}, A. and {Jordi}, C. and {Klioner}, S.~A. and {Lammers}, U.~L. and {Lindegren}, L. and {Luri}, X. and {Mignard}, F. and {Panem}, C. and {Pourbaix}, D. and {Randich}, S. and {Sartoretti}, P. and {Soubiran}, C. and {Tanga}, P. and {Walton}, N.~A. and {Bailer-Jones}, C.~A.~L. and {Bastian}, U. and {Drimmel}, R. and {Jansen}, F. and {Katz}, D. and {Lattanzi}, M.~G. and {van Leeuwen}, F. and {Bakker}, J. and {Cacciari}, C. and {Casta{\~n}eda}, J. and {De Angeli}, F. and {Fabricius}, C. and {Fouesneau}, M. and {Fr{\'e}mat}, Y. and {Galluccio}, L. and {Guerrier}, A. and {Heiter}, U. and {Masana}, E. and {Messineo}, R. and {Mowlavi}, N. and {Nicolas}, C. and {Nienartowicz}, K. and {Pailler}, F. and {Panuzzo}, P. and {Riclet}, F. and {Roux}, W. and {Seabroke}, G.~M. and {Sordo}, R. and {Th{\'e}venin}, F. and {Gracia-Abril}, G. and {Portell}, J. and {Teyssier}, D. and {Altmann}, M. and {Andrae}, R. and {Audard}, M. and {Bellas-Velidis}, I. and {Benson}, K. and {Berthier}, J. and {Blomme}, R. and {Burgess}, P.~W. and {Busonero}, D. and {Busso}, G. and {C{\'a}novas}, H. and {Carry}, B. and {Cellino}, A. and {Cheek}, N. and {Clementini}, G. and {Damerdji}, Y. and {Davidson}, M. and {de Teodoro}, P. and {Nu{\~n}ez Campos}, M. and {Delchambre}, L. and {Dell'Oro}, A. and {Esquej}, P. and {Fern{\'a}ndez-Hern{\'a}ndez}, J. and {Fraile}, E. and {Garabato}, D. and {Garc{\'\i}a-Lario}, P. and {Gosset}, E. and {Haigron}, R. and {Halbwachs}, J. -L. and {Hambly}, N.~C. and {Harrison}, D.~L. and {Hern{\'a}ndez}, J. and {Hestroffer}, D. and {Hodgkin}, S.~T. and {Holl}, B. and {Jan{\ss}en}, K. and {Jevardat de Fombelle}, G. and {Jordan}, S. and {Krone-Martins}, A. and {Lanzafame}, A.~C. and {L{\"o}ffler}, W. and {Marchal}, O. and {Marrese}, P.~M. and {Moitinho}, A. and {Muinonen}, K. and {Osborne}, P. and {Pancino}, E. and {Pauwels}, T. and {Recio-Blanco}, A. and {Reyl{\'e}}, C. and {Riello}, M. and {Rimoldini}, L. and {Roegiers}, T. and {Rybizki}, J. and {Sarro}, L.~M. and {Siopis}, C. and {Smith}, M. and {Sozzetti}, A. and {Utrilla}, E. and {van Leeuwen}, M. and {Abbas}, U. and {{\'A}brah{\'a}m}, P. and {Abreu Aramburu}, A. and {Aerts}, C. and {Aguado}, J.~J. and {Ajaj}, M. and {Aldea-Montero}, F. and {Altavilla}, G. and {{\'A}lvarez}, M.~A. and {Alves}, J. and {Anders}, F. and {Anderson}, R.~I. and {Anglada Varela}, E. and {Antoja}, T. and {Baines}, D. and {Baker}, S.~G. and {Balaguer-N{\'u}{\~n}ez}, L. and {Balbinot}, E. and {Balog}, Z. and {Barache}, C. and {Barbato}, D. and {Barros}, M. and {Barstow}, M.~A. and {Bartolom{\'e}}, S. and {Bassilana}, J. -L. and {Bauchet}, N. and {Becciani}, U. and {Bellazzini}, M. and {Berihuete}, A. and {Bernet}, M. and {Bertone}, S. and {Bianchi}, L. and {Binnenfeld}, A. and {Blanco-Cuaresma}, S. and {Blazere}, A. and {Boch}, T. and {Bombrun}, A. and {Bossini}, D. and {Bouquillon}, S. and {Bragaglia}, A. and {Bramante}, L. and {Breedt}, E. and {Bressan}, A. and {Brouillet}, N. and {Brugaletta}, E. and {Bucciarelli}, B. and {Burlacu}, A. and {Butkevich}, A.~G. and {Buzzi}, R. and {Caffau}, E. and {Cancelliere}, R. and {Cantat-Gaudin}, T. and {Carballo}, R. and {Carlucci}, T. and {Carnerero}, M.~I. and {Carrasco}, J.~M. and {Casamiquela}, L. and {Castellani}, M. and {Castro-Ginard}, A. and {Chaoul}, L. and {Charlot}, P. and {Chemin}, L. and {Chiaramida}, V. and {Chiavassa}, A. and {Chornay}, N. and {Comoretto}, G. and {Contursi}, G. and {Cooper}, W.~J. and {Cornez}, T. and {Cowell}, S. and {Crifo}, F. and {Cropper}, M. and {Crosta}, M. and {Crowley}, C. and {Dafonte}, C. and {Dapergolas}, A. and {David}, M. and {David}, P. and {de Laverny}, P. and {De Luise}, F. and {De March}, R.},
        title = "{Gaia Data Release 3. Summary of the content and survey properties}",
      journal = {\aap},
         year = 2023,
        month = jun,
       volume = {674},
          eid = {A1},
        pages = {A1},
          doi = {10.1051/0004-6361/202243940},
archivePrefix = {arXiv},
       eprint = {2208.00211},
 primaryClass = {astro-ph.GA},
       adsurl = {https://ui.adsabs.harvard.edu/abs/2023A&A...674A...1G}
}

@ARTICLE{2MASS:2006,
       author = {{Skrutskie}, M.~F. and {Cutri}, R.~M. and {Stiening}, R. and {Weinberg}, M.~D. and {Schneider}, S. and {Carpenter}, J.~M. and {Beichman}, C. and {Capps}, R. and {Chester}, T. and {Elias}, J. and {Huchra}, J. and {Liebert}, J. and {Lonsdale}, C. and {Monet}, D.~G. and {Price}, S. and {Seitzer}, P. and {Jarrett}, T. and {Kirkpatrick}, J.~D. and {Gizis}, J.~E. and {Howard}, E. and {Evans}, T. and {Fowler}, J. and {Fullmer}, L. and {Hurt}, R. and {Light}, R. and {Kopan}, E.~L. and {Marsh}, K.~A. and {McCallon}, H.~L. and {Tam}, R. and {Van Dyk}, S. and {Wheelock}, S.},
        title = "{The Two Micron All Sky Survey (2MASS)}",
      journal = {\aj},
         year = 2006,
        month = feb,
       volume = {131},
       number = {2},
        pages = {1163-1183},
          doi = {10.1086/498708},
       adsurl = {https://ui.adsabs.harvard.edu/abs/2006AJ....131.1163S}
}

@ARTICLE{Persson:2013,
       author = {{Persson}, S.~E. and {Murphy}, D.~C. and {Smee}, S. and {Birk}, C. and {Monson}, A.~J. and {Uomoto}, A. and {Koch}, E. and {Shectman}, S. and {Barkhouser}, R. and {Orndorff}, J. and {Hammond}, R. and {Harding}, A. and {Scharfstein}, G. and {Kelson}, D. and {Marshall}, J. and {McCarthy}, P.~J.},
        title = "{FourStar: The Near-Infrared Imager for the 6.5 m Baade Telescope at Las Campanas Observatory}",
      journal = {\pasp},
         year = 2013,
        month = jun,
       volume = {125},
       number = {928},
        pages = {654},
          doi = {10.1086/671164},
       adsurl = {https://ui.adsabs.harvard.edu/abs/2013PASP..125..654P}
}

@ARTICLE{Ahumada:2020,
       author = {{Ahumada}, Romina and {Allende Prieto}, Carlos and {Almeida}, Andr{\'e}s and {Anders}, Friedrich and {Anderson}, Scott F. and {Andrews}, Brett H. and {Anguiano}, Borja and {Arcodia}, Riccardo and {Armengaud}, Eric and {Aubert}, Marie and {Avila}, Santiago and {Avila-Reese}, Vladimir and {Badenes}, Carles and {Balland}, Christophe and {Barger}, Kat and {Barrera-Ballesteros}, Jorge K. and {Basu}, Sarbani and {Bautista}, Julian and {Beaton}, Rachael L. and {Beers}, Timothy C. and {Benavides}, B. Izamar T. and {Bender}, Chad F. and {Bernardi}, Mariangela and {Bershady}, Matthew and {Beutler}, Florian and {Bidin}, Christian Moni and {Bird}, Jonathan and {Bizyaev}, Dmitry and {Blanc}, Guillermo A. and {Blanton}, Michael R. and {Boquien}, M{\'e}d{\'e}ric and {Borissova}, Jura and {Bovy}, Jo and {Brandt}, W.~N. and {Brinkmann}, Jonathan and {Brownstein}, Joel R. and {Bundy}, Kevin and {Bureau}, Martin and {Burgasser}, Adam and {Burtin}, Etienne and {Cano-D{\'\i}az}, Mariana and {Capasso}, Raffaella and {Cappellari}, Michele and {Carrera}, Ricardo and {Chabanier}, Sol{\`e}ne and {Chaplin}, William and {Chapman}, Michael and {Cherinka}, Brian and {Chiappini}, Cristina and {Doohyun Choi}, Peter and {Chojnowski}, S. Drew and {Chung}, Haeun and {Clerc}, Nicolas and {Coffey}, Damien and {Comerford}, Julia M. and {Comparat}, Johan and {da Costa}, Luiz and {Cousinou}, Marie-Claude and {Covey}, Kevin and {Crane}, Jeffrey D. and {Cunha}, Katia and {Ilha}, Gabriele da Silva and {Dai}, Yu Sophia and {Damsted}, Sanna B. and {Darling}, Jeremy and {Davidson}, Jr., James W. and {Davies}, Roger and {Dawson}, Kyle and {De}, Nikhil and {de la Macorra}, Axel and {De Lee}, Nathan and {Queiroz}, Anna B{\'a}rbara de Andrade and {Deconto Machado}, Alice and {de la Torre}, Sylvain and {Dell'Agli}, Flavia and {du Mas des Bourboux}, H{\'e}lion and {Diamond-Stanic}, Aleksandar M. and {Dillon}, Sean and {Donor}, John and {Drory}, Niv and {Duckworth}, Chris and {Dwelly}, Tom and {Ebelke}, Garrett and {Eftekharzadeh}, Sarah and {Davis Eigenbrot}, Arthur and {Elsworth}, Yvonne P. and {Eracleous}, Mike and {Erfanianfar}, Ghazaleh and {Escoffier}, Stephanie and {Fan}, Xiaohui and {Farr}, Emily and {Fern{\'a}ndez-Trincado}, Jos{\'e} G. and {Feuillet}, Diane and {Finoguenov}, Alexis and {Fofie}, Patricia and {Fraser-McKelvie}, Amelia and {Frinchaboy}, Peter M. and {Fromenteau}, Sebastien and {Fu}, Hai and {Galbany}, Llu{\'\i}s and {Garcia}, Rafael A. and {Garc{\'\i}a-Hern{\'a}ndez}, D.~A. and {Garma Oehmichen}, Luis Alberto and {Ge}, Junqiang and {Geimba Maia}, Marcio Antonio and {Geisler}, Doug and {Gelfand}, Joseph and {Goddy}, Julian and {Gonzalez-Perez}, Violeta and {Grabowski}, Kathleen and {Green}, Paul and {Grier}, Catherine J. and {Guo}, Hong and {Guy}, Julien and {Harding}, Paul and {Hasselquist}, Sten and {Hawken}, Adam James and {Hayes}, Christian R. and {Hearty}, Fred and {Hekker}, S. and {Hogg}, David W. and {Holtzman}, Jon A. and {Horta}, Danny and {Hou}, Jiamin and {Hsieh}, Bau-Ching and {Huber}, Daniel and {Hunt}, Jason A.~S. and {Ider Chitham}, J. and {Imig}, Julie and {Jaber}, Mariana and {Jimenez Angel}, Camilo Eduardo and {Johnson}, Jennifer A. and {Jones}, Amy M. and {J{\"o}nsson}, Henrik and {Jullo}, Eric and {Kim}, Yerim and {Kinemuchi}, Karen and {Kirkpatrick}, IV, Charles C. and {Kite}, George W. and {Klaene}, Mark and {Kneib}, Jean-Paul and {Kollmeier}, Juna A. and {Kong}, Hui and {Kounkel}, Marina and {Krishnarao}, Dhanesh and {Lacerna}, Ivan and {Lan}, Ting-Wen and {Lane}, Richard R. and {Law}, David R. and {Le Goff}, Jean-Marc and {Leung}, Henry W. and {Lewis}, Hannah and {Li}, Cheng and {Lian}, Jianhui and {Lin}, Lihwai and {Long}, Dan and {Longa-Pe{\~n}a}, Pen{\'e}lope and {Lundgren}, Britt and {Lyke}, Brad W. and {Mackereth}, J. Ted and {MacLeod}, Chelsea L. and {Majewski}, Steven R. and {Manchado}, Arturo and {Maraston}, Claudia and {Martini}, Paul and {Masseron}, Thomas and {Masters}, Karen L. and {Mathur}, Savita and {McDermid}, Richard M. and {Merloni}, Andrea and {Merrifield}, Michael and {M{\'e}sz{\'a}ros}, Szabolcs and {Miglio}, Andrea and {Minniti}, Dante and {Minsley}, Rebecca and {Miyaji}, Takamitsu and {Mohammad}, Faizan Gohar and {Mosser}, Benoit and {Mueller}, Eva-Maria and {Muna}, Demitri and {Mu{\~n}oz-Guti{\'e}rrez}, Andrea and {Myers}, Adam D. and {Nadathur}, Seshadri and {Nair}, Preethi and {Nandra}, Kirpal and {Correa do Nascimento}, Janaina and {Nevin}, Rebecca Jean and {Newman}, Jeffrey A. and {Nidever}, David L. and {Nitschelm}, Christian and {Noterdaeme}, Pasquier and {O'Connell}, Julia E. and {Olmstead}, Matthew D. and {Oravetz}, Daniel and {Oravetz}, Audrey and {Osorio}, Yeisson and {Pace}, Zachary J. and {Padilla}, Nelson and {Palanque-Delabrouille}, Nathalie and {Palicio}, Pedro A.},
        title = "{The 16th Data Release of the Sloan Digital Sky Surveys: First Release from the APOGEE-2 Southern Survey and Full Release of eBOSS Spectra}",
      journal = {\apjs},
         year = 2020,
        month = jul,
       volume = {249},
       number = {1},
          eid = {3},
        pages = {3},
          doi = {10.3847/1538-4365/ab929e},
archivePrefix = {arXiv},
       eprint = {1912.02905},
 primaryClass = {astro-ph.GA},
       adsurl = {https://ui.adsabs.harvard.edu/abs/2020ApJS..249....3A}
}

@ARTICLE{Eliasdottir:2007,
       author = {{El{\'\i}asd{\'o}ttir}, {\'A}rd{\'\i}s and {Limousin}, Marceau and {Richard}, Johan and {Hjorth}, Jens and {Kneib}, Jean-Paul and {Natarajan}, Priya and {Pedersen}, Kristian and {Jullo}, Eric and {Paraficz}, Danuta},
        title = "{Where is the matter in the Merging Cluster Abell 2218?}",
      journal = {arXiv e-prints},
         year = 2007,
        month = oct,
          eid = {arXiv:0710.5636},
        pages = {arXiv:0710.5636},
          doi = {10.48550/arXiv.0710.5636},
archivePrefix = {arXiv},
       eprint = {0710.5636},
 primaryClass = {astro-ph},
       adsurl = {https://ui.adsabs.harvard.edu/abs/2007arXiv0710.5636E}
}

@ARTICLE{Misawa:2016,
       author = {{Misawa}, Toru and {Saez}, Cristian and {Charlton}, Jane C. and {Eracleous}, Michael and {Chartas}, George and {Bauer}, Franz E. and {Inada}, Naohisa and {Uchiyama}, Hisakazu},
        title = "{Multi-Sightline Observation of Narrow Absorption Lines in Lensed Quasar SDSS J1029+2623}",
      journal = {\apj},
         year = 2016,
        month = jul,
       volume = {825},
       number = {1},
          eid = {25},
        pages = {25},
          doi = {10.3847/0004-637X/825/1/25},
archivePrefix = {arXiv},
       eprint = {1605.04775},
 primaryClass = {astro-ph.GA},
       adsurl = {https://ui.adsabs.harvard.edu/abs/2016ApJ...825...25M}
}

@ARTICLE{Misawa:2014,
       author = {{Misawa}, Toru and {Inada}, Naohisa and {Oguri}, Masamune and {Gandhi}, Poshak and {Horiuchi}, Takashi and {Koyamada}, Suzuka and {Okamoto}, Rina},
        title = "{Resolving the Clumpy Structure of the Outflow Winds in the Gravitationally Lensed Quasar SDSS J1029+2623}",
      journal = {\apjl},
         year = 2014,
        month = oct,
       volume = {794},
       number = {2},
          eid = {L20},
        pages = {L20},
          doi = {10.1088/2041-8205/794/2/L20},
archivePrefix = {arXiv},
       eprint = {1410.0791},
 primaryClass = {astro-ph.GA},
       adsurl = {https://ui.adsabs.harvard.edu/abs/2014ApJ...794L..20M}
}

@ARTICLE{Misawa:2013,
       author = {{Misawa}, Toru and {Inada}, Naohisa and {Ohsuga}, Ken and {Gandhi}, Poshak and {Takahashi}, Rohta and {Oguri}, Masamune},
        title = "{Spectroscopy along Multiple, Lensed Sight Lines through Outflowing Winds in the Quasar SDSS J1029+2623}",
      journal = {\aj},
         year = 2013,
        month = feb,
       volume = {145},
       number = {2},
          eid = {48},
        pages = {48},
          doi = {10.1088/0004-6256/145/2/48},
archivePrefix = {arXiv},
       eprint = {1212.6689},
 primaryClass = {astro-ph.CO},
       adsurl = {https://ui.adsabs.harvard.edu/abs/2013AJ....145...48M}
}

@ARTICLE{Peng:2002,
       author = {{Peng}, Chien Y. and {Ho}, Luis C. and {Impey}, Chris D. and {Rix}, Hans-Walter},
        title = "{Detailed Structural Decomposition of Galaxy Images}",
      journal = {\aj},
         year = 2002,
        month = jul,
       volume = {124},
       number = {1},
        pages = {266-293},
          doi = {10.1086/340952},
archivePrefix = {arXiv},
       eprint = {astro-ph/0204182},
 primaryClass = {astro-ph},
       adsurl = {https://ui.adsabs.harvard.edu/abs/2002AJ....124..266P}
}

@ARTICLE{Peng:2010,
       author = {{Peng}, Chien Y. and {Ho}, Luis C. and {Impey}, Chris D. and {Rix}, Hans-Walter},
        title = "{Detailed Decomposition of Galaxy Images. II. Beyond Axisymmetric Models}",
      journal = {\aj},
         year = 2010,
        month = jun,
       volume = {139},
       number = {6},
        pages = {2097-2129},
          doi = {10.1088/0004-6256/139/6/2097},
archivePrefix = {arXiv},
       eprint = {0912.0731},
 primaryClass = {astro-ph.CO},
       adsurl = {https://ui.adsabs.harvard.edu/abs/2010AJ....139.2097P}
}

@article{Hopkins:2010,
	title = {Quasar feedback: more bang for your buck},
	volume = {401},
	issn = {00358711, 13652966},
	url = {https://academic.oup.com/mnras/article-lookup/doi/10.1111/j.1365-2966.2009.15643.x},
	doi = {10.1111/j.1365-2966.2009.15643.x},
	shorttitle = {Quasar feedback},
	pages = {7--14},
	number = {1},
	journaltitle = {Monthly Notices of the Royal Astronomical Society},
	author = {Hopkins, Philip F. and Elvis, Martin},
	urldate = {2025-10-15},
	date = {2010-01-01},
    year = {2010},
	langid = {english},
}

@software{Comrie:2021,
       author = {{Comrie}, Angus and {Wang}, Kuo-Song and {Hsu}, Shou-Chieh and {Moraghan}, Anthony and {Harris}, Pamela and {Pang}, Qi and {Pi{\r{A}}ska}, Adrianna and {Chiang}, Cheng-Chin and {Simmonds}, Rob and {Chang}, Tien-Hao and {Jan}, Hengtai and {Lin}, Ming-Yi},
        title = "{CARTA: Cube Analysis and Rendering Tool for Astronomy}",
 howpublished = {Astrophysics Source Code Library, record ascl:2103.031},
         year = 2021,
        month = mar,
          eid = {ascl:2103.031},
archivePrefix = {ascl},
       eprint = {2103.031},
       adsurl = {https://ui.adsabs.harvard.edu/abs/2021ascl.soft03031C}
}

@dataset{Wright:2019,
       author = {{Wright}, Edward L. and {Eisenhardt}, Peter R.~M. and {Mainzer}, Amy K. and {Ressler}, Michael E. and {Cutri}, Roc M. and {Jarrett}, Thomas and {Kirkpatrick}, J. Davy and {Padgett}, Deborah and {McMillan}, Robert S. and {Skrutskie}, Michael and {Stanford}, S.~A. and {Cohen}, Martin and {Walker}, Russell G. and {Mather}, John C. and {Leisawitz}, David and {Gautier}, III, Thomas N. and {McLean}, Ian and {Benford}, Dominic and {Lonsdale}, Carol J. and {Blain}, Andrew and {Mendez}, Bryan and {Irace}, William R. and {Duval}, Valerie and {Liu}, Fengchuan and {Royer}, Don and {Heinrichsen}, Ingolf and {Howard}, Joan and {Shannon}, Mark and {Kendall}, Martha and {Walsh}, Amy L. and {Larsen}, Mark and {Cardon}, Joel G. and {Schick}, Scott and {Schwalm}, Mark and {Abid}, Mohamed and {Fabinsky}, Beth and {Naes}, Larry and {Tsai}, ChaoWei},
        title = "{AllWISE Source Catalog}",
 howpublished = {NASA IPAC DataSet, IRSA1},
         year = 2019,
        month = jan,
          doi = {10.26131/IRSA1},
       adsurl = {https://ui.adsabs.harvard.edu/abs/2019ipac.data...I1W}
}

@ARTICLE{Lang:2014,
       author = {{Lang}, Dustin},
        title = "{unWISE: Unblurred Coadds of the WISE Imaging}",
      journal = {\aj},
         year = 2014,
        month = may,
       volume = {147},
       number = {5},
          eid = {108},
        pages = {108},
          doi = {10.1088/0004-6256/147/5/108},
archivePrefix = {arXiv},
       eprint = {1405.0308},
 primaryClass = {astro-ph.IM},
       adsurl = {https://ui.adsabs.harvard.edu/abs/2014AJ....147..108L}
}

@ARTICLE{Meisner:2017a,
       author = {{Meisner}, Aaron M. and {Lang}, Dustin and {Schlegel}, David J.},
        title = "{Full-depth Coadds of the WISE and First-year NEOWISE-reactivation Images}",
      journal = {\aj},
         year = 2017,
        month = jan,
       volume = {153},
       number = {1},
          eid = {38},
        pages = {38},
          doi = {10.3847/1538-3881/153/1/38},
archivePrefix = {arXiv},
       eprint = {1603.05664},
 primaryClass = {astro-ph.IM},
       adsurl = {https://ui.adsabs.harvard.edu/abs/2017AJ....153...38M}
}

@ARTICLE{Meisner:2017b,
       author = {{Meisner}, A.~M. and {Lang}, D. and {Schlegel}, D.~J.},
        title = "{Deep Full-sky Coadds from Three Years of WISE and NEOWISE Observations}",
      journal = {\aj},
         year = 2017,
        month = oct,
       volume = {154},
       number = {4},
          eid = {161},
        pages = {161},
          doi = {10.3847/1538-3881/aa894e},
archivePrefix = {arXiv},
       eprint = {1705.06746},
 primaryClass = {astro-ph.IM},
       adsurl = {https://ui.adsabs.harvard.edu/abs/2017AJ....154..161M}
}

@software{Comrie:2021_zndo,
       author = {{Comrie}, Angus and {Wang}, Kuo-Song and {Hsu}, Shou-Chieh and {Moraghan}, Anthony and {Harris}, Pamela and {Pang}, Qi and {Pi{\'n}ska}, Adrianna and {Chiang}, Cheng-Chin and {Chang}, Tien-Hao and {Hwang}, Yu-Hsuan and {Jan}, Hengtai and {Lin}, Ming-Yi and {Simmonds}, Rob},
        title = "{CARTA: The Cube Analysis and Rendering Tool for Astronomy}",
         year = 2021,
        month = jun,
          eid = {10.5281/zenodo.3377984},
          doi = {10.5281/zenodo.3377984},
      version = {2.0.0},
    publisher = {Zenodo},
       adsurl = {https://ui.adsabs.harvard.edu/abs/2021zndo...3377984C}
}

@article{Griffin:2010,
	title = {The Herschel-{SPIRE} instrument and its in-flight performance},
	volume = {518},
	issn = {0004-6361, 1432-0746},
	url = {http://arxiv.org/abs/1005.5123},
	doi = {10.1051/0004-6361/201014519},
	pages = {L3},
	journaltitle = {Astronomy and Astrophysics},
	shortjournal = {A\&A},
	author = {Griffin, M. J. and Abergel, A. and Abreu, A. and Ade, P. A. R. and André, P. and Augueres, J.-L. and Babbedge, T. and Bae, Y. and Baillie, T. and Baluteau, J.-P. and Barlow, M. J. and Bendo, G. and Benielli, D. and Bock, J. J. and Bonhomme, P. and Brisbin, D. and Brockley-Blatt, C. and Caldwell, M. and Cara, C. and Castro-Rodriguez, N. and Cerulli, R. and Chanial, P. and Chen, S. and Clark, E. and Clements, D. L. and Clerc, L. and Coker, J. and Communal, D. and Conversi, L. and Cox, P. and Crumb, D. and Cunningham, C. and Daly, F. and Davis, G. R. and Antoni, P. De and Delderfield, J. and Devin, N. and Giorgio, A. Di and Didschuns, I. and Dohlen, K. and Donati, M. and Dowell, A. and Dowell, C. D. and Duband, L. and Dumaye, L. and Emery, R. J. and Ferlet, M. and Ferrand, D. and Fontignie, J. and Fox, M. and Franceschini, A. and Frerking, M. and Fulton, T. and Garcia, J. and Gastaud, R. and Gear, W. K. and Glenn, J. and Goizel, A. and Griffin, D. K. and Grundy, T. and Guest, S. and Guillemet, L. and Hargrave, P. C. and Harwit, M. and Hastings, P. and Hatziminaoglou, E. and Herman, M. and Hinde, B. and Hristov, V. and Huang, M. and Imhof, P. and Isaak, K. J. and Israelsson, U. and Ivison, R. J. and Jennings, D. and Kiernan, B. and King, K. J. and Lange, A. E. and Latter, W. and Laurent, G. and Laurent, P. and Leeks, S. J. and Lellouch, E. and Levenson, L. and Li, B. and Li, J. and Lilienthal, J. and Lim, T. and Liu, J. and Lu, N. and Madden, S. and Mainetti, G. and Marliani, P. and {McKay}, D. and Mercier, K. and Molinari, S. and Morris, H. and Moseley, H. and Mulder, J. and Mur, M. and Naylor, D. A. and Nguyen, H. and O'Halloran, B. and Oliver, S. and Olofsson, G. and Olofsson, H.-G. and Orfei, R. and Page, M. J. and Pain, I. and Panuzzo, P. and Papageorgiou, A. and Parks, G. and Parr-Burman, P. and Pearce, A. and Pearson, C. and Pérez-Fournon, I. and Pinsard, F. and Pisano, G. and Podosek, J. and Pohlen, M. and Polehampton, E. T. and Pouliquen, D. and Rigopoulou, D. and Rizzo, D. and Roseboom, I. G. and Roussel, H. and Rowan-Robinson, M. and Rownd, B. and Saraceno, P. and Sauvage, M. and Savage, R. and Savini, G. and Sawyer, E. and Scharmberg, C. and Schmitt, D. and Schneider, N. and Schulz, B. and Schwartz, A. and Shafer, R. and Shupe, D. L. and Sibthorpe, B. and Sidher, S. and Smith, A. and Smith, A. J. and Smith, D. and Spencer, L. and Stobie, B. and Sudiwala, R. and Sukhatme, K. and Surace, C. and Stevens, J. A. and Swinyard, B. M. and Trichas, M. and Tourette, T. and Triou, H. and Tseng, S. and Tucker, C. and Turner, A. and Vaccari, M. and Valtchanov, I. and Vigroux, L. and Virique, E. and Voellmer, G. and Walker, H. and Ward, R. and Waskett, T. and Weilert, M. and Wesson, R. and White, G. J. and Whitehouse, N. and Wilson, C. D. and Winter, B. and Woodcraft, A. L. and Wright, G. S. and Xu, C. K. and Zavagno, A. and Zemcov, M. and Zhang, L. and Zonca, E.},
	urldate = {2025-11-12},
	date = {2010-07},
    year = {2010},
	eprinttype = {arxiv},
	eprint = {1005.5123 [astro-ph]},
}

@ARTICLE{Yuan:2016,
       author = {{Yuan}, Sihan and {Strauss}, Michael A. and {Zakamska}, Nadia L.},
        title = "{Spectroscopic identification of type 2 quasars at z < 1 in SDSS-III/BOSS}",
      journal = {\mnras},
         year = 2016,
        month = oct,
       volume = {462},
       number = {2},
        pages = {1603-1615},
          doi = {10.1093/mnras/stw1747},
archivePrefix = {arXiv},
       eprint = {1606.04976},
 primaryClass = {astro-ph.GA},
       adsurl = {https://ui.adsabs.harvard.edu/abs/2016MNRAS.462.1603Y}
}

@ARTICLE{Hopkins:2007,
       author = {{Hopkins}, Philip F. and {Richards}, Gordon T. and {Hernquist}, Lars},
        title = "{An Observational Determination of the Bolometric Quasar Luminosity Function}",
      journal = {\apj},
         year = 2007,
        month = jan,
       volume = {654},
       number = {2},
        pages = {731-753},
          doi = {10.1086/509629},
archivePrefix = {arXiv},
       eprint = {astro-ph/0605678},
 primaryClass = {astro-ph},
       adsurl = {https://ui.adsabs.harvard.edu/abs/2007ApJ...654..731H}
}

@ARTICLE{Akeson:2019,
       author = {{Akeson}, Rachel and {Armus}, Lee and {Bachelet}, Etienne and {Bailey}, Vanessa and {Bartusek}, Lisa and {Bellini}, Andrea and {Benford}, Dominic and {Bennett}, David and {Bhattacharya}, Aparna and {Bohlin}, Ralph and {Boyer}, Martha and {Bozza}, Valerio and {Bryden}, Geoffrey and {Calchi Novati}, Sebastiano and {Carpenter}, Kenneth and {Casertano}, Stefano and {Choi}, Ami and {Content}, David and {Dayal}, Pratika and {Dressler}, Alan and {Dor{\'e}}, Olivier and {Fall}, S. Michael and {Fan}, Xiaohui and {Fang}, Xiao and {Filippenko}, Alexei and {Finkelstein}, Steven and {Foley}, Ryan and {Furlanetto}, Steven and {Kalirai}, Jason and {Gaudi}, B. Scott and {Gilbert}, Karoline and {Girard}, Julien and {Grady}, Kevin and {Greene}, Jenny and {Guhathakurta}, Puragra and {Heinrich}, Chen and {Hemmati}, Shoubaneh and {Hendel}, David and {Henderson}, Calen and {Henning}, Thomas and {Hirata}, Christopher and {Ho}, Shirley and {Huff}, Eric and {Hutter}, Anne and {Jansen}, Rolf and {Jha}, Saurabh and {Johnson}, Samson and {Jones}, David and {Kasdin}, Jeremy and {Kelly}, Patrick and {Kirshner}, Robert and {Koekemoer}, Anton and {Kruk}, Jeffrey and {Lewis}, Nikole and {Macintosh}, Bruce and {Madau}, Piero and {Malhotra}, Sangeeta and {Mandel}, Kaisey and {Massara}, Elena and {Masters}, Daniel and {McEnery}, Julie and {McQuinn}, Kristen and {Melchior}, Peter and {Melton}, Mark and {Mennesson}, Bertrand and {Peeples}, Molly and {Penny}, Matthew and {Perlmutter}, Saul and {Pisani}, Alice and {Plazas}, Andr{\'e}s and {Poleski}, Radek and {Postman}, Marc and {Ranc}, Cl{\'e}ment and {Rauscher}, Bernard and {Rest}, Armin and {Roberge}, Aki and {Robertson}, Brant and {Rodney}, Steven and {Rhoads}, James and {Rhodes}, Jason and {Ryan}, Jr., Russell and {Sahu}, Kailash and {Sand}, David and {Scolnic}, Dan and {Seth}, Anil and {Shvartzvald}, Yossi and {Siellez}, Karelle and {Smith}, Arfon and {Spergel}, David and {Stassun}, Keivan and {Street}, Rachel and {Strolger}, Louis-Gregory and {Szalay}, Alexander and {Trauger}, John and {Troxel}, M.~A. and {Turnbull}, Margaret and {van der Marel}, Roeland and {von der Linden}, Anja and {Wang}, Yun and {Weinberg}, David and {Williams}, Benjamin and {Windhorst}, Rogier and {Wollack}, Edward and {Wu}, Hao-Yi and {Yee}, Jennifer and {Zimmerman}, Neil},
        title = "{The Wide Field Infrared Survey Telescope: 100 Hubbles for the 2020s}",
      journal = {arXiv e-prints},
         year = 2019,
        month = feb,
          eid = {arXiv:1902.05569},
        pages = {arXiv:1902.05569},
          doi = {10.48550/arXiv.1902.05569},
archivePrefix = {arXiv},
       eprint = {1902.05569},
 primaryClass = {astro-ph.IM},
       adsurl = {https://ui.adsabs.harvard.edu/abs/2019arXiv190205569A}
}

@INPROCEEDINGS{Simcoe:2008,
       author = {{Simcoe}, Robert A. and {Burgasser}, Adam J. and {Bernstein}, Rebecca A. and {Bigelow}, Bruce C. and {Fishner}, Jason and {Forrest}, William J. and {McMurtry}, Craig and {Pipher}, Judith L. and {Schechter}, Paul L. and {Smith}, Matthew},
        title = "{FIRE: a near-infrared cross-dispersed echellette spectrometer for the Magellan telescopes}",
    booktitle = {Ground-based and Airborne Instrumentation for Astronomy II},
         year = 2008,
       editor = {{McLean}, Ian S. and {Casali}, Mark M.},
       series = {Society of Photo-Optical Instrumentation Engineers (SPIE) Conference Series},
       volume = {7014},
        month = jul,
          eid = {70140U},
        pages = {70140U},
          doi = {10.1117/12.790414},
       adsurl = {https://ui.adsabs.harvard.edu/abs/2008SPIE.7014E..0US}
}

@ARTICLE{Chabrier:2003,
       author = {{Chabrier}, Gilles},
        title = "{Galactic Stellar and Substellar Initial Mass Function}",
      journal = {\pasp},
         year = 2003,
        month = jul,
       volume = {115},
       number = {809},
        pages = {763-795},
          doi = {10.1086/376392},
archivePrefix = {arXiv},
       eprint = {astro-ph/0304382},
 primaryClass = {astro-ph},
       adsurl = {https://ui.adsabs.harvard.edu/abs/2003PASP..115..763C}
}

@ARTICLE{Coelho:2020,
       author = {{Coelho}, Paula R.~T. and {Bruzual}, Gustavo and {Charlot}, St{\'e}phane},
        title = "{To use or not to use synthetic stellar spectra in population synthesis models?}",
      journal = {\mnras},
         year = 2020,
        month = jan,
       volume = {491},
       number = {2},
        pages = {2025-2042},
          doi = {10.1093/mnras/stz3023},
archivePrefix = {arXiv},
       eprint = {1910.11902},
 primaryClass = {astro-ph.GA},
       adsurl = {https://ui.adsabs.harvard.edu/abs/2020MNRAS.491.2025C}
}

@ARTICLE{Calzetti:2000,
       author = {{Calzetti}, Daniela and {Armus}, Lee and {Bohlin}, Ralph C. and {Kinney}, Anne L. and {Koornneef}, Jan and {Storchi-Bergmann}, Thaisa},
        title = "{The Dust Content and Opacity of Actively Star-forming Galaxies}",
      journal = {\apj},
         year = 2000,
        month = apr,
       volume = {533},
       number = {2},
        pages = {682-695},
          doi = {10.1086/308692},
archivePrefix = {arXiv},
       eprint = {astro-ph/9911459},
 primaryClass = {astro-ph},
       adsurl = {https://ui.adsabs.harvard.edu/abs/2000ApJ...533..682C}
}

@ARTICLE{Khullar:2021,
       author = {{Khullar}, Gourav and {Gozman}, Katya and {Lin}, Jason J. and {Martinez}, Michael N. and {Matthews Acu{\~n}a}, Owen S. and {Medina}, Elisabeth and {Merz}, Kaiya and {Sanchez}, Jorge A. and {Sisco}, Emily E. and {Kavin Stein}, Daniel J. and {Sukay}, Ezra O. and {Tavangar}, Kiyan and {Bayliss}, Matthew B. and {Bleem}, Lindsey E. and {Brownsberger}, Sasha and {Dahle}, H{\r{A}}kon and {Florian}, Michael K. and {Gladders}, Michael D. and {Mahler}, Guillaume and {Rigby}, Jane R. and {Sharon}, Keren and {Stark}, Antony A.},
        title = "{COOL-LAMPS. I. An Extraordinarily Bright Lensed Galaxy at Redshift 5.04}",
      journal = {\apj},
         year = 2021,
        month = jan,
       volume = {906},
       number = {2},
          eid = {107},
        pages = {107},
          doi = {10.3847/1538-4357/abcb86},
archivePrefix = {arXiv},
       eprint = {2011.06601},
 primaryClass = {astro-ph.GA},
       adsurl = {https://ui.adsabs.harvard.edu/abs/2021ApJ...906..107K}
}

@ARTICLE{Stark:2013,
       author = {{Stark}, Daniel P. and {Auger}, Matthew and {Belokurov}, Vasily and {Jones}, Tucker and {Robertson}, Brant and {Ellis}, Richard S. and {Sand}, David J. and {Moiseev}, Alexei and {Eagle}, Will and {Myers}, Thomas},
        title = "{The CASSOWARY spectroscopy survey: a new sample of gravitationally lensed galaxies in SDSS}",
      journal = {\mnras},
         year = 2013,
        month = dec,
       volume = {436},
       number = {2},
        pages = {1040-1056},
          doi = {10.1093/mnras/stt1624},
archivePrefix = {arXiv},
       eprint = {1302.2663},
 primaryClass = {astro-ph.CO},
       adsurl = {https://ui.adsabs.harvard.edu/abs/2013MNRAS.436.1040S}
}

@ARTICLE{Sharon:2017,
       author = {{Sharon}, Keren and {Bayliss}, Matthew B. and {Dahle}, H{\r{a}}kon and {Florian}, Michael K. and {Gladders}, Michael D. and {Johnson}, Traci L. and {Paterno-Mahler}, Rachel and {Rigby}, Jane R. and {Whitaker}, Katherine E. and {Wuyts}, Eva},
        title = "{Lens Model and Time Delay Predictions for the Sextuply Lensed Quasar SDSS J2222+2745}",
      journal = {\apj},
         year = 2017,
        month = jan,
       volume = {835},
       number = {1},
          eid = {5},
        pages = {5},
          doi = {10.3847/1538-4357/835/1/5},
archivePrefix = {arXiv},
       eprint = {1609.08848},
 primaryClass = {astro-ph.GA},
       adsurl = {https://ui.adsabs.harvard.edu/abs/2017ApJ...835....5S}
}

@ARTICLE{Lawther:2012,
       author = {{Lawther}, D. and {Paarup}, T. and {Schmidt}, M. and {Vestergaard}, M. and {Hjorth}, J. and {Malesani}, D.},
        title = "{Constraints on the relative sizes of intervening Mg II-absorbing clouds and quasar emitting regions}",
      journal = {\aap},
         year = 2012,
        month = oct,
       volume = {546},
          eid = {A67},
        pages = {A67},
          doi = {10.1051/0004-6361/201219326},
archivePrefix = {arXiv},
       eprint = {1211.4777},
 primaryClass = {astro-ph.CO},
       adsurl = {https://ui.adsabs.harvard.edu/abs/2012A&A...546A..67L}
}

@ARTICLE{Chisholm:2020,
       author = {{Chisholm}, J. and {Prochaska}, J.~X. and {Schaerer}, D. and {Gazagnes}, S. and {Henry}, A.},
        title = "{Optically thin spatially resolved Mg II emission maps the escape of ionizing photons}",
      journal = {\mnras},
         year = 2020,
        month = oct,
       volume = {498},
       number = {2},
        pages = {2554-2574},
          doi = {10.1093/mnras/staa2470},
archivePrefix = {arXiv},
       eprint = {2008.06059},
 primaryClass = {astro-ph.GA},
       adsurl = {https://ui.adsabs.harvard.edu/abs/2020MNRAS.498.2554C}
}

@ARTICLE{Ortiz:2024,
       author = {{Ortiz}, Rafael and {Windhorst}, Rogier A. and {Cohen}, Seth H. and {Willner}, Steven P. and {Jansen}, Rolf A. and {Carleton}, Timothy and {Kamieneski}, Patrick S. and {Rutkowski}, Michael J. and {Smith}, Brent M. and {Summers}, Jake and {Cheng}, Cheng and {Coe}, Dan and {Conselice}, Christopher J. and {Diego}, Jose M. and {Driver}, Simon P. and {D'Silva}, Jordan C.~J. and {Frye}, Brenda L. and {Gim}, Hansung B. and {Grogin}, Norman A. and {Hammel}, Heidi B. and {Hathi}, Nimish P. and {Holwerda}, Benne W. and {Hyun}, Minhee and {Im}, Myungshin and {Keel}, William C. and {Koekemoer}, Anton M. and {Li}, Juno and {Marshall}, Madeline A. and {McCabe}, Tyler J. and {McLeod}, Noah J. and {Milam}, Stefanie N. and {O'Brien}, Rosalia and {Pirzkal}, Nor and {Robotham}, Aaron S.~G. and {Ryan}, Russell E. and {Willmer}, Christopher N.~A. and {Yan}, Haojing and {Yun}, Min S. and {Zitrin}, Adi},
        title = "{PEARLS: Discovery of Point-source Features within Galaxies in the North Ecliptic Pole Time Domain Field}",
      journal = {\apj},
         year = 2024,
        month = oct,
       volume = {974},
       number = {2},
          eid = {258},
        pages = {258},
          doi = {10.3847/1538-4357/ad6d5e},
archivePrefix = {arXiv},
       eprint = {2404.10709},
 primaryClass = {astro-ph.GA},
       adsurl = {https://ui.adsabs.harvard.edu/abs/2024ApJ...974..258O}
}

@ARTICLE{Pillbratt:2010,
       author = {{Pilbratt}, G.~L. and {Riedinger}, J.~R. and {Passvogel}, T. and {Crone}, G. and {Doyle}, D. and {Gageur}, U. and {Heras}, A.~M. and {Jewell}, C. and {Metcalfe}, L. and {Ott}, S. and {Schmidt}, M.},
        title = "{Herschel Space Observatory. An ESA facility for far-infrared and submillimetre astronomy}",
      journal = {\aap},
         year = 2010,
        month = jul,
       volume = {518},
          eid = {L1},
        pages = {L1},
          doi = {10.1051/0004-6361/201014759},
archivePrefix = {arXiv},
       eprint = {1005.5331},
 primaryClass = {astro-ph.IM},
       adsurl = {https://ui.adsabs.harvard.edu/abs/2010A&A...518L...1P}
}

@ARTICLE{Bleem:2015,
       author = {{Bleem}, L.~E. and {Stalder}, B. and {de Haan}, T. and {Aird}, K.~A. and {Allen}, S.~W. and {Applegate}, D.~E. and {Ashby}, M.~L.~N. and {Bautz}, M. and {Bayliss}, M. and {Benson}, B.~A. and {Bocquet}, S. and {Brodwin}, M. and {Carlstrom}, J.~E. and {Chang}, C.~L. and {Chiu}, I. and {Cho}, H.~M. and {Clocchiatti}, A. and {Crawford}, T.~M. and {Crites}, A.~T. and {Desai}, S. and {Dietrich}, J.~P. and {Dobbs}, M.~A. and {Foley}, R.~J. and {Forman}, W.~R. and {George}, E.~M. and {Gladders}, M.~D. and {Gonzalez}, A.~H. and {Halverson}, N.~W. and {Hennig}, C. and {Hoekstra}, H. and {Holder}, G.~P. and {Holzapfel}, W.~L. and {Hrubes}, J.~D. and {Jones}, C. and {Keisler}, R. and {Knox}, L. and {Lee}, A.~T. and {Leitch}, E.~M. and {Liu}, J. and {Lueker}, M. and {Luong-Van}, D. and {Mantz}, A. and {Marrone}, D.~P. and {McDonald}, M. and {McMahon}, J.~J. and {Meyer}, S.~S. and {Mocanu}, L. and {Mohr}, J.~J. and {Murray}, S.~S. and {Padin}, S. and {Pryke}, C. and {Reichardt}, C.~L. and {Rest}, A. and {Ruel}, J. and {Ruhl}, J.~E. and {Saliwanchik}, B.~R. and {Saro}, A. and {Sayre}, J.~T. and {Schaffer}, K.~K. and {Schrabback}, T. and {Shirokoff}, E. and {Song}, J. and {Spieler}, H.~G. and {Stanford}, S.~A. and {Staniszewski}, Z. and {Stark}, A.~A. and {Story}, K.~T. and {Stubbs}, C.~W. and {Vanderlinde}, K. and {Vieira}, J.~D. and {Vikhlinin}, A. and {Williamson}, R. and {Zahn}, O. and {Zenteno}, A.},
        title = "{Galaxy Clusters Discovered via the Sunyaev-Zel'dovich Effect in the 2500-Square-Degree SPT-SZ Survey}",
      journal = {\apjs},
         year = 2015,
        month = feb,
       volume = {216},
       number = {2},
          eid = {27},
        pages = {27},
          doi = {10.1088/0067-0049/216/2/27},
archivePrefix = {arXiv},
       eprint = {1409.0850},
 primaryClass = {astro-ph.CO},
       adsurl = {https://ui.adsabs.harvard.edu/abs/2015ApJS..216...27B}
}

@ARTICLE{Egami:2010,
       author = {{Egami}, E. and {Rex}, M. and {Rawle}, T.~D. and {P{\'e}rez-Gonz{\'a}lez}, P.~G. and {Richard}, J. and {Kneib}, J.-P. and {Schaerer}, D. and {Altieri}, B. and {Valtchanov}, I. and {Blain}, A.~W. and {Fadda}, D. and {Zemcov}, M. and {Bock}, J.~J. and {Boone}, F. and {Bridge}, C.~R. and {Clement}, B. and {Combes}, F. and {Dessauges-Zavadsky}, M. and {Dowell}, C.~D. and {Ilbert}, O. and {Ivison}, R.~J. and {Jauzac}, M. and {Lutz}, D. and {Metcalfe}, L. and {Omont}, A. and {Pell{\'o}}, R. and {Pereira}, M.~J. and {Rieke}, G.~H. and {Rodighiero}, G. and {Smail}, I. and {Smith}, G.~P. and {Tramoy}, G. and {Walth}, G.~L. and {van der Werf}, P. and {Werner}, M.~W.},
        title = "{The Herschel Lensing Survey (HLS): Overview}",
      journal = {\aap},
         year = 2010,
        month = jul,
       volume = {518},
          eid = {L12},
        pages = {L12},
          doi = {10.1051/0004-6361/201014696},
archivePrefix = {arXiv},
       eprint = {1005.3820},
 primaryClass = {astro-ph.CO},
       adsurl = {https://ui.adsabs.harvard.edu/abs/2010A&A...518L..12E}
}

@ARTICLE{Pradhan:2006,
       author = {{Pradhan}, Anil K. and {Montenegro}, Maximiliano and {Nahar}, Sultana N. and {Eissner}, Werner},
        title = "{[OII] line ratios}",
      journal = {\mnras},
         year = 2006,
        month = feb,
       volume = {366},
       number = {1},
        pages = {L6-L9},
          doi = {10.1111/j.1745-3933.2005.00119.x},
archivePrefix = {arXiv},
       eprint = {astro-ph/0510099},
 primaryClass = {astro-ph},
       adsurl = {https://ui.adsabs.harvard.edu/abs/2006MNRAS.366L...6P}
}

@ARTICLE{Predehl:2021,
       author = {{Predehl}, P. and {Andritschke}, R. and {Arefiev}, V. and {Babyshkin}, V. and {Batanov}, O. and {Becker}, W. and {B{\"o}hringer}, H. and {Bogomolov}, A. and {Boller}, T. and {Borm}, K. and {Bornemann}, W. and {Br{\"a}uninger}, H. and {Br{\"u}ggen}, M. and {Brunner}, H. and {Brusa}, M. and {Bulbul}, E. and {Buntov}, M. and {Burwitz}, V. and {Burkert}, W. and {Clerc}, N. and {Churazov}, E. and {Coutinho}, D. and {Dauser}, T. and {Dennerl}, K. and {Doroshenko}, V. and {Eder}, J. and {Emberger}, V. and {Eraerds}, T. and {Finoguenov}, A. and {Freyberg}, M. and {Friedrich}, P. and {Friedrich}, S. and {F{\"u}rmetz}, M. and {Georgakakis}, A. and {Gilfanov}, M. and {Granato}, S. and {Grossberger}, C. and {Gueguen}, A. and {Gureev}, P. and {Haberl}, F. and {H{\"a}lker}, O. and {Hartner}, G. and {Hasinger}, G. and {Huber}, H. and {Ji}, L. and {Kienlin}, A. v. and {Kink}, W. and {Korotkov}, F. and {Kreykenbohm}, I. and {Lamer}, G. and {Lomakin}, I. and {Lapshov}, I. and {Liu}, T. and {Maitra}, C. and {Meidinger}, N. and {Menz}, B. and {Merloni}, A. and {Mernik}, T. and {Mican}, B. and {Mohr}, J. and {M{\"u}ller}, S. and {Nandra}, K. and {Nazarov}, V. and {Pacaud}, F. and {Pavlinsky}, M. and {Perinati}, E. and {Pfeffermann}, E. and {Pietschner}, D. and {Ramos-Ceja}, M.~E. and {Rau}, A. and {Reiffers}, J. and {Reiprich}, T.~H. and {Robrade}, J. and {Salvato}, M. and {Sanders}, J. and {Santangelo}, A. and {Sasaki}, M. and {Scheuerle}, H. and {Schmid}, C. and {Schmitt}, J. and {Schwope}, A. and {Shirshakov}, A. and {Steinmetz}, M. and {Stewart}, I. and {Str{\"u}der}, L. and {Sunyaev}, R. and {Tenzer}, C. and {Tiedemann}, L. and {Tr{\"u}mper}, J. and {Voron}, V. and {Weber}, P. and {Wilms}, J. and {Yaroshenko}, V.},
        title = "{The eROSITA X-ray telescope on SRG}",
      journal = {\aap},
         year = 2021,
        month = mar,
       volume = {647},
          eid = {A1},
        pages = {A1},
          doi = {10.1051/0004-6361/202039313},
archivePrefix = {arXiv},
       eprint = {2010.03477},
 primaryClass = {astro-ph.HE},
       adsurl = {https://ui.adsabs.harvard.edu/abs/2021A&A...647A...1P}
}

@ARTICLE{Merloni:2024,
       author = {{Merloni}, A. and {Lamer}, G. and {Liu}, T. and {Ramos-Ceja}, M.~E. and {Brunner}, H. and {Bulbul}, E. and {Dennerl}, K. and {Doroshenko}, V. and {Freyberg}, M.~J. and {Friedrich}, S. and {Gatuzz}, E. and {Georgakakis}, A. and {Haberl}, F. and {Igo}, Z. and {Kreykenbohm}, I. and {Liu}, A. and {Maitra}, C. and {Malyali}, A. and {Mayer}, M.~G.~F. and {Nandra}, K. and {Predehl}, P. and {Robrade}, J. and {Salvato}, M. and {Sanders}, J.~S. and {Stewart}, I. and {Tub{\'\i}n-Arenas}, D. and {Weber}, P. and {Wilms}, J. and {Arcodia}, R. and {Artis}, E. and {Aschersleben}, J. and {Avakyan}, A. and {Aydar}, C. and {Bahar}, Y.~E. and {Balzer}, F. and {Becker}, W. and {Berger}, K. and {Boller}, T. and {Bornemann}, W. and {Br{\"u}ggen}, M. and {Brusa}, M. and {Buchner}, J. and {Burwitz}, V. and {Camilloni}, F. and {Clerc}, N. and {Comparat}, J. and {Coutinho}, D. and {Czesla}, S. and {Dannhauer}, S.~M. and {Dauner}, L. and {Dauser}, T. and {Dietl}, J. and {Dolag}, K. and {Dwelly}, T. and {Egg}, K. and {Ehl}, E. and {Freund}, S. and {Friedrich}, P. and {Gaida}, R. and {Garrel}, C. and {Ghirardini}, V. and {Gokus}, A. and {Gr{\"u}nwald}, G. and {Grandis}, S. and {Grotova}, I. and {Gruen}, D. and {Gueguen}, A. and {H{\"a}mmerich}, S. and {Hamaus}, N. and {Hasinger}, G. and {Haubner}, K. and {Homan}, D. and {Ider Chitham}, J. and {Joseph}, W.~M. and {Joyce}, A. and {K{\"o}nig}, O. and {Kaltenbrunner}, D.~M. and {Khokhriakova}, A. and {Kink}, W. and {Kirsch}, C. and {Kluge}, M. and {Knies}, J. and {Krippendorf}, S. and {Krumpe}, M. and {Kurpas}, J. and {Li}, P. and {Liu}, Z. and {Locatelli}, N. and {Lorenz}, M. and {M{\"u}ller}, S. and {Magaudda}, E. and {Mannes}, C. and {McCall}, H. and {Meidinger}, N. and {Michailidis}, M. and {Migkas}, K. and {Mu{\~n}oz-Giraldo}, D. and {Musiimenta}, B. and {Nguyen-Dang}, N.~T. and {Ni}, Q. and {Olechowska}, A. and {Ota}, N. and {Pacaud}, F. and {Pasini}, T. and {Perinati}, E. and {Pires}, A.~M. and {Pommranz}, C. and {Ponti}, G. and {Poppenhaeger}, K. and {P{\"u}hlhofer}, G. and {Rau}, A. and {Reh}, M. and {Reiprich}, T.~H. and {Roster}, W. and {Saeedi}, S. and {Santangelo}, A. and {Sasaki}, M. and {Schmitt}, J. and {Schneider}, P.~C. and {Schrabback}, T. and {Schuster}, N. and {Schwope}, A. and {Seppi}, R. and {Serim}, M.~M. and {Shreeram}, S. and {Sokolova-Lapa}, E. and {Starck}, H. and {Stelzer}, B. and {Stierhof}, J. and {Suleimanov}, V. and {Tenzer}, C. and {Traulsen}, I. and {Tr{\"u}mper}, J. and {Tsuge}, K. and {Urrutia}, T. and {Veronica}, A. and {Waddell}, S.~G.~H. and {Willer}, R. and {Wolf}, J. and {Yeung}, M.~C.~H. and {Zainab}, A. and {Zangrandi}, F. and {Zhang}, X. and {Zhang}, Y. and {Zheng}, X.},
        title = "{The SRG/eROSITA all-sky survey. First X-ray catalogues and data release of the western Galactic hemisphere}",
      journal = {\aap},
         year = 2024,
        month = feb,
       volume = {682},
          eid = {A34},
        pages = {A34},
          doi = {10.1051/0004-6361/202347165},
archivePrefix = {arXiv},
       eprint = {2401.17274},
 primaryClass = {astro-ph.HE},
       adsurl = {https://ui.adsabs.harvard.edu/abs/2024A&A...682A..34M}
}

@ARTICLE{Yang:2022,
       author = {{Yang}, Guang and {Boquien}, M{\'e}d{\'e}ric and {Brandt}, W.~N. and {Buat}, V{\'e}ronique and {Burgarella}, Denis and {Ciesla}, Laure and {Lehmer}, Bret D. and {Ma{\l}ek}, Katarzyna and {Mountrichas}, George and {Papovich}, Casey and {Pons}, Estelle and {Stalevski}, Marko and {Theul{\'e}}, Patrice and {Zhu}, Shifu},
        title = "{Fitting AGN/Galaxy X-Ray-to-radio SEDs with CIGALE and Improvement of the Code}",
      journal = {\apj},
         year = 2022,
        month = mar,
       volume = {927},
       number = {2},
          eid = {192},
        pages = {192},
          doi = {10.3847/1538-4357/ac4971},
archivePrefix = {arXiv},
       eprint = {2201.03718},
 primaryClass = {astro-ph.GA},
       adsurl = {https://ui.adsabs.harvard.edu/abs/2022ApJ...927..192Y}
}

@misc{Charlot:2019,
  author = {Bruzual, G. and Charlot, S.},
  year = {2019},
  title = {Stellar population synthesis models},
  url = {https://www.bruzual.org/CB19/},
  note = {Accessed: 2026-02-22}
}

@ARTICLE{Schartmann:2005,
       author = {{Schartmann}, M. and {Meisenheimer}, K. and {Camenzind}, M. and {Wolf}, S. and {Henning}, Th.},
        title = "{Towards a physical model of dust tori in Active Galactic Nuclei. Radiative transfer calculations for a hydrostatic torus model}",
      journal = {\aap},
         year = 2005,
        month = jul,
       volume = {437},
       number = {3},
        pages = {861-881},
          doi = {10.1051/0004-6361:20042363},
archivePrefix = {arXiv},
       eprint = {astro-ph/0504105},
 primaryClass = {astro-ph},
       adsurl = {https://ui.adsabs.harvard.edu/abs/2005A&A...437..861S}
}

@ARTICLE{Draine:2014,
       author = {{Draine}, B.~T. and {Aniano}, G. and {Krause}, Oliver and {Groves}, Brent and {Sandstrom}, Karin and {Braun}, Robert and {Leroy}, Adam and {Klaas}, Ulrich and {Linz}, Hendrik and {Rix}, Hans-Walter and {Schinnerer}, Eva and {Schmiedeke}, Anika and {Walter}, Fabian},
        title = "{Andromeda's Dust}",
      journal = {\apj},
         year = 2014,
        month = jan,
       volume = {780},
       number = {2},
          eid = {172},
        pages = {172},
          doi = {10.1088/0004-637X/780/2/172},
archivePrefix = {arXiv},
       eprint = {1306.2304},
 primaryClass = {astro-ph.CO},
       adsurl = {https://ui.adsabs.harvard.edu/abs/2014ApJ...780..172D}
}

@ARTICLE{Casey:2012,
       author = {{Casey}, Caitlin M.},
        title = "{Far-infrared spectral energy distribution fitting for galaxies near and far}",
      journal = {\mnras},
         year = 2012,
        month = oct,
       volume = {425},
       number = {4},
        pages = {3094-3103},
          doi = {10.1111/j.1365-2966.2012.21455.x},
archivePrefix = {arXiv},
       eprint = {1206.1595},
 primaryClass = {astro-ph.CO},
       adsurl = {https://ui.adsabs.harvard.edu/abs/2012MNRAS.425.3094C}
}

@ARTICLE{Florian:2025,
       author = {{Florian}, Michael K and {Gladders}, Michael D and {Khullar}, Gourav and {Sharon}, Keren and {Cloonan}, Aidan P and {Solhaug}, Eirk and {Welch}, Brian and {Bayliss}, Matthew and {Dahle}, Hakon and {Hutchison}, Taylor A and {Rigby}, Jane R},
        title = "{JWST Catches a Strongly Gravitationally Lensed AGN In Transition from Type II to Type I}",
      journal = {arXiv e-prints},
         year = 2025,
        month = oct,
          eid = {arXiv:2510.10376},
        pages = {arXiv:2510.10376},
          doi = {10.48550/arXiv.2510.10376},
archivePrefix = {arXiv},
       eprint = {2510.10376},
 primaryClass = {astro-ph.GA},
       adsurl = {https://ui.adsabs.harvard.edu/abs/2025arXiv251010376F}
}

@ARTICLE{Greene:2005,
       author = {{Greene}, Jenny E. and {Ho}, Luis C.},
        title = "{Estimating Black Hole Masses in Active Galaxies Using the H{\ensuremath{\alpha}} Emission Line}",
      journal = {\apj},
         year = 2005,
        month = sep,
       volume = {630},
       number = {1},
        pages = {122-129},
          doi = {10.1086/431897},
archivePrefix = {arXiv},
       eprint = {astro-ph/0508335},
 primaryClass = {astro-ph},
       adsurl = {https://ui.adsabs.harvard.edu/abs/2005ApJ...630..122G}
}

@ARTICLE{Raney:2020,
       author = {{Raney}, Catie A. and {Keeton}, Charles R. and {Brennan}, Sean},
        title = "{Exploring Effects on Magnifications due to Line-of-Sight Galaxies in the Hubble Frontier Fields}",
      journal = {\mnras},
         year = 2020,
        month = feb,
       volume = {492},
       number = {1},
        pages = {503-527},
          doi = {10.1093/mnras/stz3116},
archivePrefix = {arXiv},
       eprint = {1911.02101},
 primaryClass = {astro-ph.CO},
       adsurl = {https://ui.adsabs.harvard.edu/abs/2020MNRAS.492..503R}
}

@ARTICLE{Suyu:2017,
       author = {{Suyu}, S.~H. and {Bonvin}, V. and {Courbin}, F. and {Fassnacht}, C.~D. and {Rusu}, C.~E. and {Sluse}, D. and {Treu}, T. and {Wong}, K.~C. and {Auger}, M.~W. and {Ding}, X. and {Hilbert}, S. and {Marshall}, P.~J. and {Rumbaugh}, N. and {Sonnenfeld}, A. and {Tewes}, M. and {Tihhonova}, O. and {Agnello}, A. and {Blandford}, R.~D. and {Chen}, G.~C.-F. and {Collett}, T. and {Koopmans}, L.~V.~E. and {Liao}, K. and {Meylan}, G. and {Spiniello}, C.},
        title = "{H0LiCOW - I. H$_{0}$ Lenses in COSMOGRAIL's Wellspring: program overview}",
      journal = {\mnras},
         year = 2017,
        month = jul,
       volume = {468},
       number = {3},
        pages = {2590-2604},
          doi = {10.1093/mnras/stx483},
archivePrefix = {arXiv},
       eprint = {1607.00017},
 primaryClass = {astro-ph.CO},
       adsurl = {https://ui.adsabs.harvard.edu/abs/2017MNRAS.468.2590S}
}

@ARTICLE{Planck:2020,
       author = {{Planck Collaboration} and {Aghanim}, N. and {Akrami}, Y. and {Ashdown}, M. and {Aumont}, J. and {Baccigalupi}, C. and {Ballardini}, M. and {Banday}, A.~J. and {Barreiro}, R.~B. and {Bartolo}, N. and {Basak}, S. and {Battye}, R. and {Benabed}, K. and {Bernard}, J.-P. and {Bersanelli}, M. and {Bielewicz}, P. and {Bock}, J.~J. and {Bond}, J.~R. and {Borrill}, J. and {Bouchet}, F.~R. and {Boulanger}, F. and {Bucher}, M. and {Burigana}, C. and {Butler}, R.~C. and {Calabrese}, E. and {Cardoso}, J.-F. and {Carron}, J. and {Challinor}, A. and {Chiang}, H.~C. and {Chluba}, J. and {Colombo}, L.~P.~L. and {Combet}, C. and {Contreras}, D. and {Crill}, B.~P. and {Cuttaia}, F. and {de Bernardis}, P. and {de Zotti}, G. and {Delabrouille}, J. and {Delouis}, J.-M. and {Di Valentino}, E. and {Diego}, J.~M. and {Dor{\'e}}, O. and {Douspis}, M. and {Ducout}, A. and {Dupac}, X. and {Dusini}, S. and {Efstathiou}, G. and {Elsner}, F. and {En{\ss}lin}, T.~A. and {Eriksen}, H.~K. and {Fantaye}, Y. and {Farhang}, M. and {Fergusson}, J. and {Fernandez-Cobos}, R. and {Finelli}, F. and {Forastieri}, F. and {Frailis}, M. and {Fraisse}, A.~A. and {Franceschi}, E. and {Frolov}, A. and {Galeotta}, S. and {Galli}, S. and {Ganga}, K. and {G{\'e}nova-Santos}, R.~T. and {Gerbino}, M. and {Ghosh}, T. and {Gonz{\'a}lez-Nuevo}, J. and {G{\'o}rski}, K.~M. and {Gratton}, S. and {Gruppuso}, A. and {Gudmundsson}, J.~E. and {Hamann}, J. and {Handley}, W. and {Hansen}, F.~K. and {Herranz}, D. and {Hildebrandt}, S.~R. and {Hivon}, E. and {Huang}, Z. and {Jaffe}, A.~H. and {Jones}, W.~C. and {Karakci}, A. and {Keih{\"a}nen}, E. and {Keskitalo}, R. and {Kiiveri}, K. and {Kim}, J. and {Kisner}, T.~S. and {Knox}, L. and {Krachmalnicoff}, N. and {Kunz}, M. and {Kurki-Suonio}, H. and {Lagache}, G. and {Lamarre}, J.-M. and {Lasenby}, A. and {Lattanzi}, M. and {Lawrence}, C.~R. and {Le Jeune}, M. and {Lemos}, P. and {Lesgourgues}, J. and {Levrier}, F. and {Lewis}, A. and {Liguori}, M. and {Lilje}, P.~B. and {Lilley}, M. and {Lindholm}, V. and {L{\'o}pez-Caniego}, M. and {Lubin}, P.~M. and {Ma}, Y.-Z. and {Mac{\'\i}as-P{\'e}rez}, J.~F. and {Maggio}, G. and {Maino}, D. and {Mandolesi}, N. and {Mangilli}, A. and {Marcos-Caballero}, A. and {Maris}, M. and {Martin}, P.~G. and {Martinelli}, M. and {Mart{\'\i}nez-Gonz{\'a}lez}, E. and {Matarrese}, S. and {Mauri}, N. and {McEwen}, J.~D. and {Meinhold}, P.~R. and {Melchiorri}, A. and {Mennella}, A. and {Migliaccio}, M. and {Millea}, M. and {Mitra}, S. and {Miville-Desch{\^e}nes}, M.-A. and {Molinari}, D. and {Montier}, L. and {Morgante}, G. and {Moss}, A. and {Natoli}, P. and {N{\o}rgaard-Nielsen}, H.~U. and {Pagano}, L. and {Paoletti}, D. and {Partridge}, B. and {Patanchon}, G. and {Peiris}, H.~V. and {Perrotta}, F. and {Pettorino}, V. and {Piacentini}, F. and {Polastri}, L. and {Polenta}, G. and {Puget}, J.-L. and {Rachen}, J.~P. and {Reinecke}, M. and {Remazeilles}, M. and {Renzi}, A. and {Rocha}, G. and {Rosset}, C. and {Roudier}, G. and {Rubi{\~n}o-Mart{\'\i}n}, J.~A. and {Ruiz-Granados}, B. and {Salvati}, L. and {Sandri}, M. and {Savelainen}, M. and {Scott}, D. and {Shellard}, E.~P.~S. and {Sirignano}, C. and {Sirri}, G. and {Spencer}, L.~D. and {Sunyaev}, R. and {Suur-Uski}, A.-S. and {Tauber}, J.~A. and {Tavagnacco}, D. and {Tenti}, M. and {Toffolatti}, L. and {Tomasi}, M. and {Trombetti}, T. and {Valenziano}, L. and {Valiviita}, J. and {Van Tent}, B. and {Vibert}, L. and {Vielva}, P. and {Villa}, F. and {Vittorio}, N. and {Wandelt}, B.~D. and {Wehus}, I.~K. and {White}, M. and {White}, S.~D.~M. and {Zacchei}, A. and {Zonca}, A.},
        title = "{Planck 2018 results. VI. Cosmological parameters}",
      journal = {\aap},
         year = 2020,
        month = sep,
       volume = {641},
          eid = {A6},
        pages = {A6},
          doi = {10.1051/0004-6361/201833910},
archivePrefix = {arXiv},
       eprint = {1807.06209},
 primaryClass = {astro-ph.CO},
       adsurl = {https://ui.adsabs.harvard.edu/abs/2020A&A...641A...6P}
}

@ARTICLE{Riess:2022,
       author = {{Riess}, Adam G. and {Yuan}, Wenlong and {Macri}, Lucas M. and {Scolnic}, Dan and {Brout}, Dillon and {Casertano}, Stefano and {Jones}, David O. and {Murakami}, Yukei and {Anand}, Gagandeep S. and {Breuval}, Louise and {Brink}, Thomas G. and {Filippenko}, Alexei V. and {Hoffmann}, Samantha and {Jha}, Saurabh W. and {D'arcy Kenworthy}, W. and {Mackenty}, John and {Stahl}, Benjamin E. and {Zheng}, WeiKang},
        title = "{A Comprehensive Measurement of the Local Value of the Hubble Constant with 1 km s$^{-1}$ Mpc$^{-1}$ Uncertainty from the Hubble Space Telescope and the SH0ES Team}",
      journal = {\apjl},
         year = 2022,
        month = jul,
       volume = {934},
       number = {1},
          eid = {L7},
        pages = {L7},
          doi = {10.3847/2041-8213/ac5c5b},
archivePrefix = {arXiv},
       eprint = {2112.04510},
 primaryClass = {astro-ph.CO},
       adsurl = {https://ui.adsabs.harvard.edu/abs/2022ApJ...934L...7R}
}

@software{Oemler:2017,
       author = {{Oemler}, A. and {Clardy}, K. and {Kelson}, D. and {Walth}, G. and {Villanueva}, E.},
        title = "{COSMOS: Carnegie Observatories System for MultiObject Spectroscopy}",
 howpublished = {Astrophysics Source Code Library, record ascl:1705.001},
         year = 2017,
        month = may,
          eid = {ascl:1705.001},
archivePrefix = {ascl},
       eprint = {1705.001},
       adsurl = {https://ui.adsabs.harvard.edu/abs/2017ascl.soft05001O}
}

@ARTICLE{Werner:2004,
       author = {{Werner}, M.~W. and {Roellig}, T.~L. and {Low}, F.~J. and {Rieke}, G.~H. and {Rieke}, M. and {Hoffmann}, W.~F. and {Young}, E. and {Houck}, J.~R. and {Brandl}, B. and {Fazio}, G.~G. and {Hora}, J.~L. and {Gehrz}, R.~D. and {Helou}, G. and {Soifer}, B.~T. and {Stauffer}, J. and {Keene}, J. and {Eisenhardt}, P. and {Gallagher}, D. and {Gautier}, T.~N. and {Irace}, W. and {Lawrence}, C.~R. and {Simmons}, L. and {Van Cleve}, J.~E. and {Jura}, M. and {Wright}, E.~L. and {Cruikshank}, D.~P.},
        title = "{The Spitzer Space Telescope Mission}",
      journal = {\apjs},
         year = 2004,
        month = sep,
       volume = {154},
       number = {1},
        pages = {1-9},
          doi = {10.1086/422992},
archivePrefix = {arXiv},
       eprint = {astro-ph/0406223},
 primaryClass = {astro-ph},
       adsurl = {https://ui.adsabs.harvard.edu/abs/2004ApJS..154....1W}
}

@ARTICLE{Inada:2005,
       author = {{Inada}, Naohisa and {Oguri}, Masamune and {Keeton}, Charles R. and {Eisenstein}, Daniel J. and {Castander}, Francisco J. and {Chiu}, Kuenley and {Hall}, Patrick B. and {Hennawi}, Joseph F. and {Johnston}, David E. and {Pindor}, Bartosz and {Richards}, Gordon T. and {Rix}, Hans-Walter Rix and {Schneider}, Donald P. and {Zheng}, Wei},
        title = "{Discovery of a Fifth Image of the Large Separation Gravitationally Lensed Quasar SDSS J1004+4112}",
      journal = {\pasj},
         year = 2005,
        month = jun,
       volume = {57},
        pages = {L7-L10},
          doi = {10.1093/pasj/57.3.L7},
archivePrefix = {arXiv},
       eprint = {astro-ph/0503310},
 primaryClass = {astro-ph},
       adsurl = {https://ui.adsabs.harvard.edu/abs/2005PASJ...57L...7I}
}

@misc{Claude:2025,
  author = {Anthropic},
  title = {Claude 4.5 Haiku},
  year = {2025},
  publisher = {Anthropic},
  journal = {Anthropic Website},
  howpublished = {\url{https://www.anthropic.com/claude/haiku}},
  note = {Accessed: 2026}
}

@ARTICLE{Sweeney:2026,
       author = {{Sweeney}, David and {Krone-Martins}, Alberto and {Stern}, Daniel and {Tuthill}, Peter and {Scalzo}, Richard and {Djorgovski}, George and {Ducourant}, Christine and {Mahabal}, Ashish and {Teixeira}, Ramachrisna and {Graham}, Matthew},
        title = "{Semisupervised learning for lensed quasar detection}",
      journal = {\mnras},
         year = 2026,
        month = apr,
       volume = {547},
       number = {4},
          eid = {stag445},
        pages = {stag445},
          doi = {10.1093/mnras/stag445},
archivePrefix = {arXiv},
       eprint = {2504.00054},
 primaryClass = {astro-ph.IM},
       adsurl = {https://ui.adsabs.harvard.edu/abs/2026MNRAS.547ag445S}
}
\bibliographystyle{aasjournal}

\appendix

\section{ALMA Archival Imaging}
\label{app:alma}

\begin{figure}[ht!]
    \centering
    \includegraphics[width=\columnwidth]{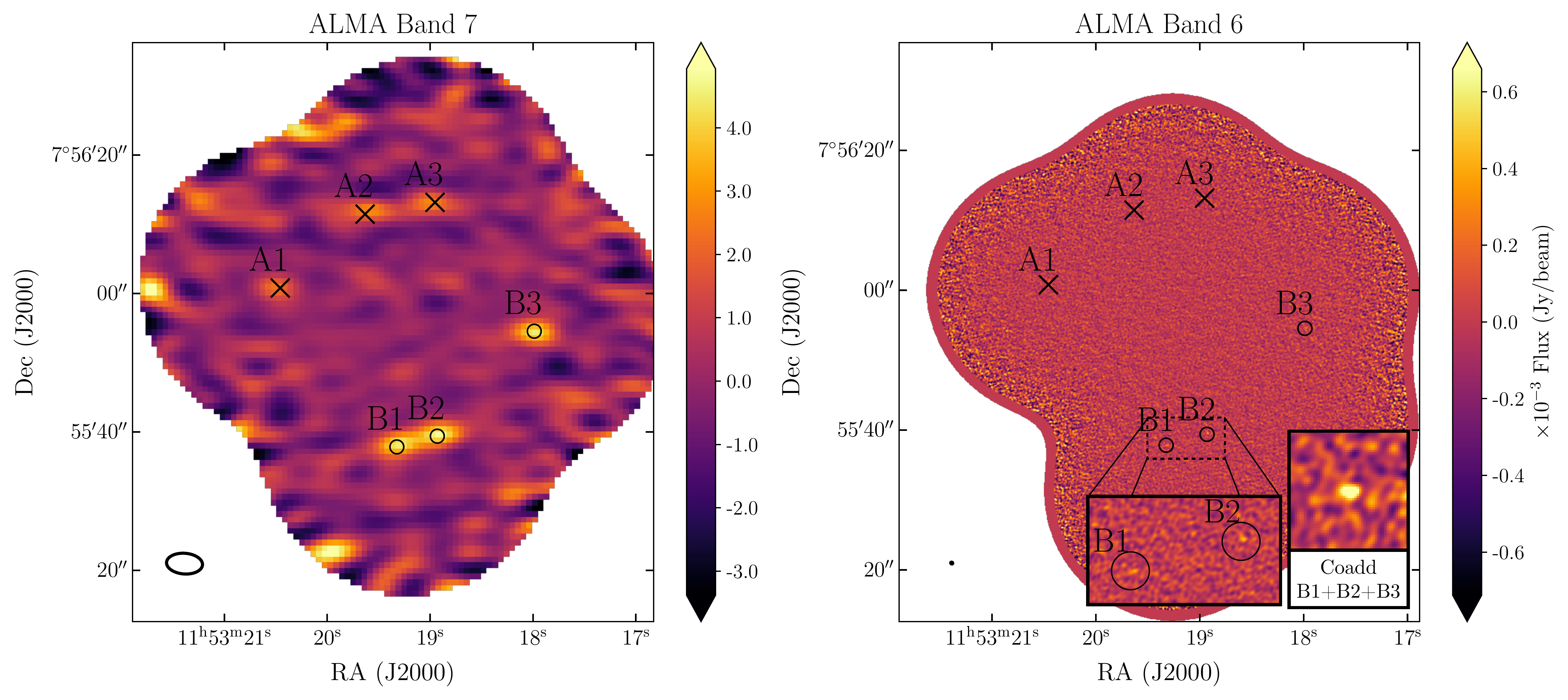}
    \caption{We report that \targ\ is detected in archival ALMA data. \textit{Left:} ALMA Band 7 shows $>4\sigma$ detections of \targA\ and \targB. \textit{Right:} ALMA Band 6 shows a detection of \targB\ and a non-detection of \targA. An enlarged inset panel is provided to show the Band 6 flux near the brightest two images, B1 and B2. We perform a spatial coaddition of images B1, B2 and B3 and show this in the lower right inset. In this coadded frame, we report a $>7\sigma$ Band 6 detection for \targB. We show the ALMA beam with a black ellipse in the bottom left corner of each panel.}
    \label{fig:alma}
\end{figure}

Archival ALMA data of \targ\ exists for Band 6 and 7. Band 6 is from 2016 with the 12-m array (Project Code: 2016.1.00372.S, PI: Egami, Eiichi) and Band 7 is from 2019 with the 7-m array (Project Code: 2019.2.00040.S, PI: Sun, Fengwu). \cite{Sun:2021} report a non-detection in Band 6 for this field and decided to abandon the field to avoid wrongly identifying sources with $\sim3\sigma$ spurious detections. The \targ\ sources are unresolved in the ALMA data in both bands. For Band 7, we choose to report the measured flux of the images A1 and B3 since these are most separated from the other images. Images A1 and B3 are detected at $4\sigma$ and $7\sigma$, respectively. \targA\ is not detected in Band 6 while all images of \targB\ are, with the strongest detection being image B2 ($2.7\sigma$). We measure the flux of image B2 and scale it by the ratio of the $K_s$ fluxes of B3 to B2. To measure fluxes in both ALMA bands, we use \texttt{CARTA} \citep[][]{Comrie:2021, Comrie:2021_zndo} to fit the signal with a two-dimensional Gaussian within an extraction region four times the size of the beam. We obtain the flux from the reported integrated flux in Jy/beam which is easily converted. We then estimate the uncertainty by calculating the RMS of large regions around the location of the signal. Photometry is reported in Table~\ref{tab:cigale_photometry}, and we show the continuum image of the ALMA data cubes in Fig.~\ref{fig:alma}.

\section{Model Fits of Spectra of \targA\ and \targB}
\label{app:fits}

\begin{figure*}[ht!]
    \centering
    \includegraphics[width=\linewidth]{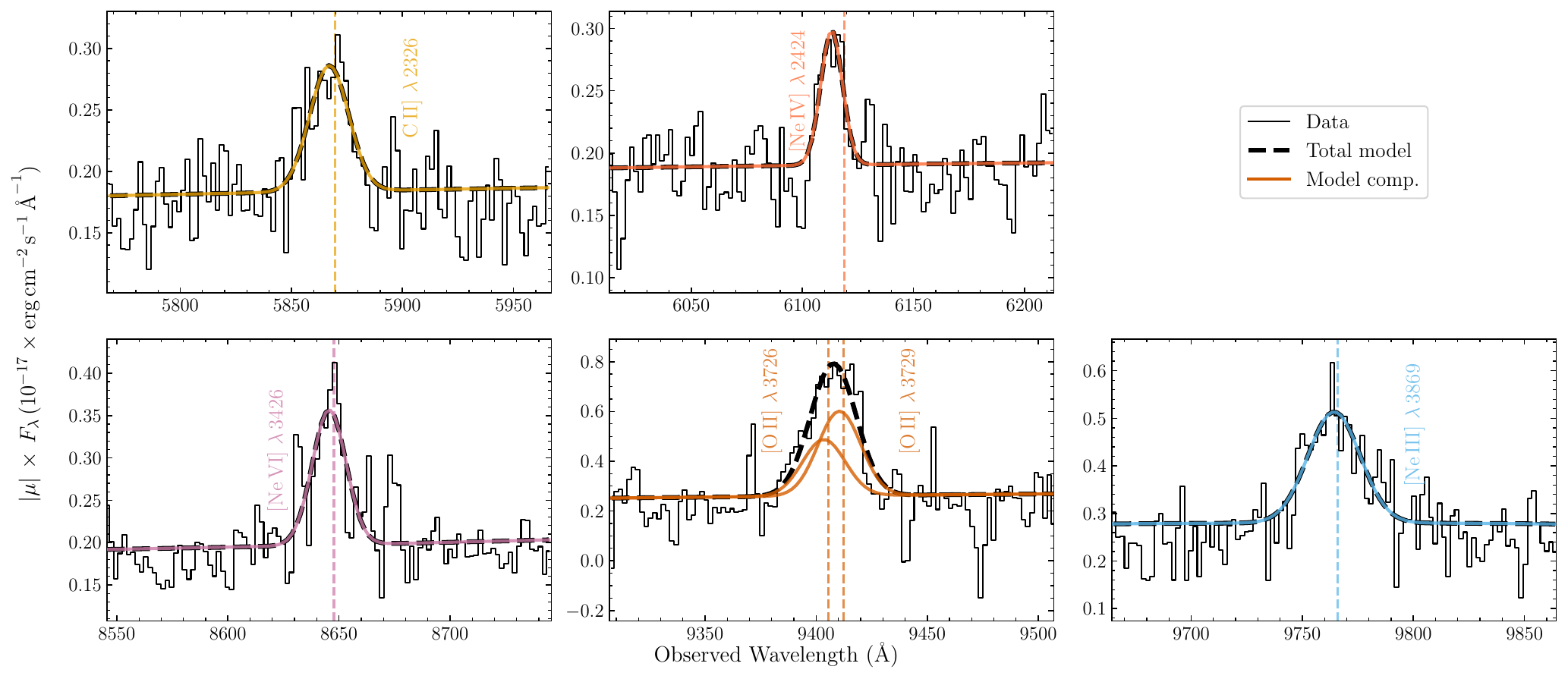}
    \caption{Enlarged panels of Fig.~\ref{fig:ldss3_spectra} showing the six fitted emission lines in \targA\ (here showing the fits for image A1), with individual lines displayed at the resolution of the full spectrum. Each panel displays the data (black step), total model (black dashed), and fitted Gaussian component (colored line) centered at the best-fit wavelength. The vertical dashed lines indicate the rest-frame wavelength positions of each transition, computed using the weighted mean of the redshift of quasar images A1-A3, $z_{\rm QSO} = 1.52352 \pm 0.00050$.}
    \label{fig:ldss3_COOLJ1153A1_panels}
\end{figure*}

\begin{figure*}[ht!]
    \centering
    \includegraphics[width=\linewidth]{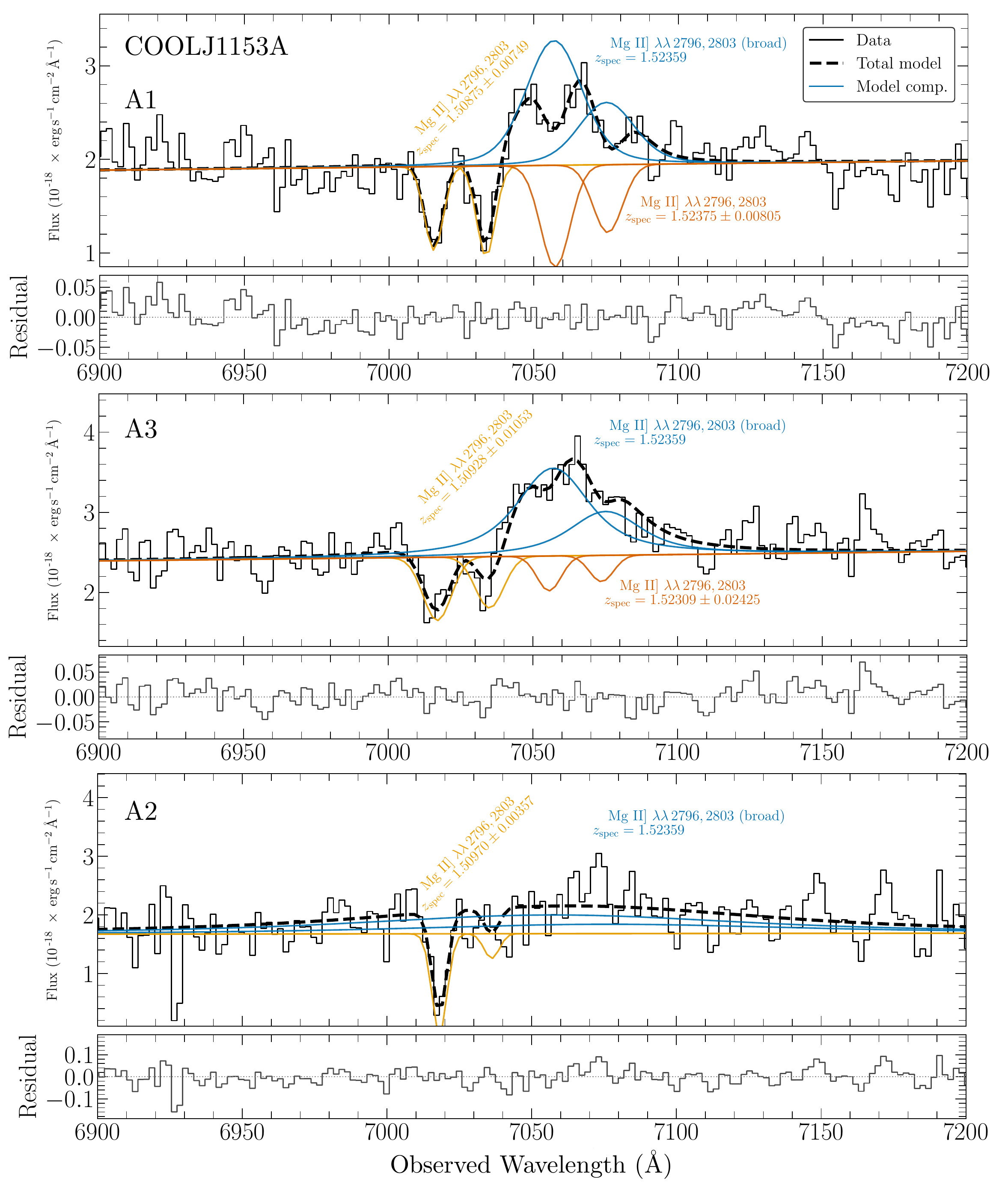}
    \caption{LDSS3 spectrum of \targA\ centered on the Mg \textsc{ii}] $\lambda\lambda\, 2796, \,2803$ doublet. Multiple absorbers are seen along the line of sight. We fit a broad doublet Voigt profile at the redshift of the quasar and the line-of-sight absorbers as Gaussian profiles. Each component of Mg \textsc{ii}] doublets is shown in colored lines, making up the total model (dashed black line). The second absorber system seen in A1 and A3 is not detected in A2, and in A2 the broad emission line is significantly suppressed.}
    \label{fig:ldss3_COOLJ1153A1_MgII}
\end{figure*}

\begin{figure*}[ht!]
    \centering
    \includegraphics[width=\linewidth]{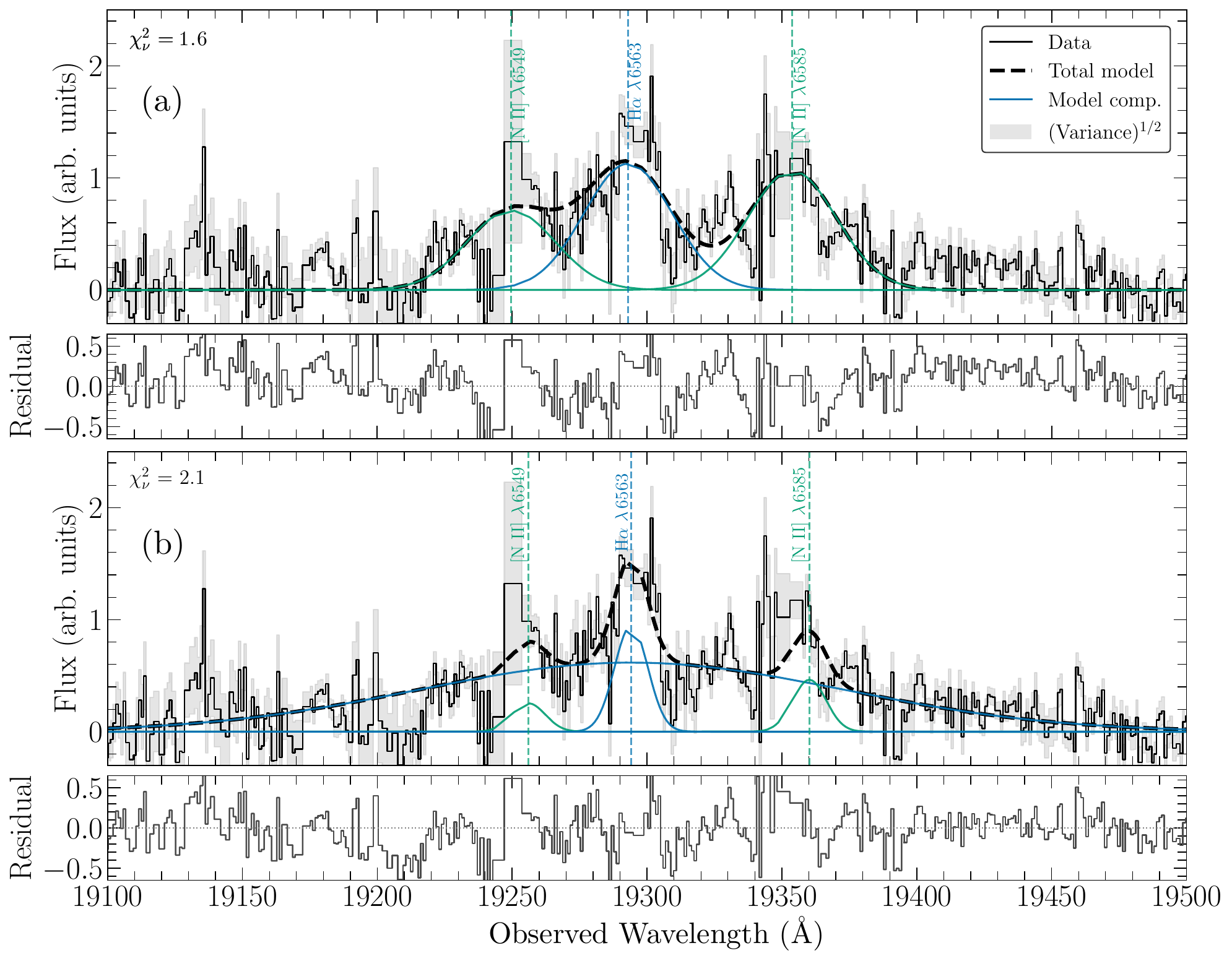}
    \caption{FIRE spectrum of \targB\ at $z=1.939$ (image B1) showing the H$\alpha$ and [N\,\textsc{ii}] emission lines. We simultaneously fit H$\alpha$, [N\,\textsc{ii}]~$\lambda\lambda$6548,6583, and [O\,\textsc{iii}]~$\lambda$5007 using two approaches: (a) single Gaussian profiles for each line with tied velocity widths, and (b) H$\alpha$ modeled with one narrow and one broad component while other lines' widths are tied to the narrow component. See Fig.~\ref{fig:fire_hbeta} for [O\,\textsc{iii}] and H$\beta$ fits.} \label{fig:fire_halpha}
\end{figure*}

\begin{figure*}[ht!]
    \centering
    \includegraphics[width=\linewidth]{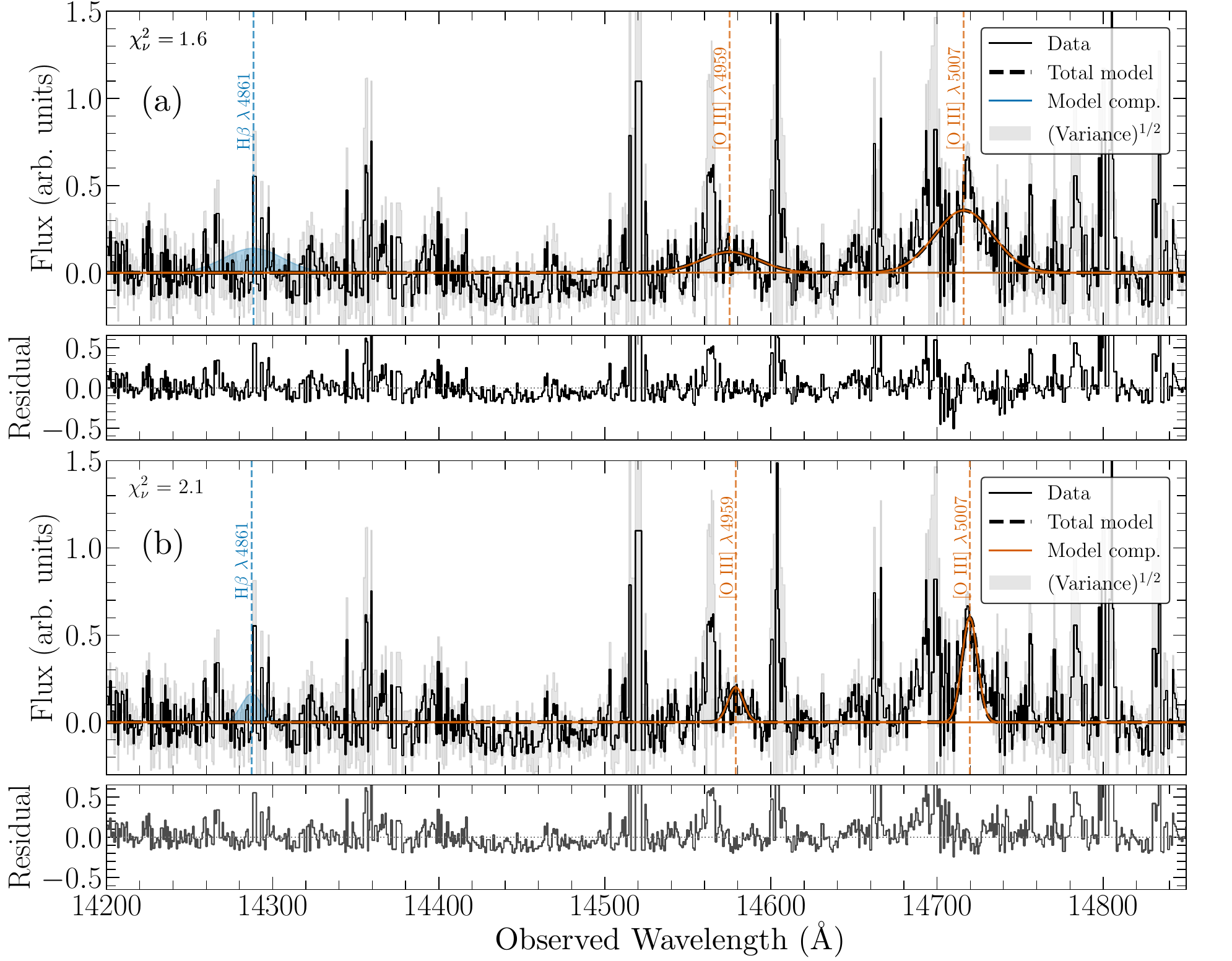}
    \caption{FIRE spectrum (same as in Fig.~\ref{fig:fire_halpha}) of the H$\beta$ and [O\,\textsc{iii}]~$\lambda$5007 region, analyzed using the (a) and (b) methods described in the caption of Fig.~\ref{fig:fire_halpha}. H$\beta$ is not detected; the displayed Gaussian represents a $3\sigma$ upper limit with amplitude set to the local noise level and velocity width matched to the [O\,\textsc{iii}] profile.}
    \label{fig:fire_hbeta}
\end{figure*}

We present the spectroscopic analyses and model fits for both lensed quasars. For \targA, we use LDSS3 spectroscopy to fit six emission lines in the UV rest frame. The Mg \textsc{ii}] doublet region is fitted separately due to multiple foreground absorbers along the line of sight, requiring simultaneous modeling of broad quasar emission and narrow absorption systems using Voigt and Gaussian profiles. For \targB, we model the H$\alpha$, [N \textsc{ii}], and [O \textsc{iii}] lines from the FIRE spectroscopy using two complementary fitting approaches: one with single Gaussian profiles for each line, referred to as method (a), and a second allowing H$\alpha$ to have both narrow and broad components, referred to as method (b). These fits are shown in Figs.~\ref{fig:ldss3_COOLJ1153A1_panels}--\ref{fig:fire_hbeta}.

\section{Additional LDSS3 Spectroscopic Redshifts}
\label{app:slit_redshifts}

Here, we tabulate the redshifts used for measuring the cluster redshift from LDSS3 multislit spectroscopy and other LDSS3 longslit redshifts acquired from the \targ\ field. These are tabulated in Tables~\ref{tab:multislit_redshifts} and \ref{tab:longslit_redshifts}.

\begin{table*}[ht!]
\centering
\renewcommand{\arraystretch}{0.9}
\caption{Spectroscopic Redshifts from LDSS3 Multislit Spectroscopy}
\label{tab:multislit_redshifts}
\begin{tabular}{ccc}
\hline
\hline
R.A. & Decl. & \textbf{$z_{\rm spec}$} \\
(J2000) & (J2000) & \\
\hline
\makebox[20mm][c]{178.3313667} & \makebox[20mm][c]{7.9351556} & \makebox[20mm][c]{0.4322} \\
\makebox[20mm][c]{178.3289333} & \makebox[20mm][c]{7.9273400} & \makebox[20mm][c]{0.4257} \\
\makebox[20mm][c]{178.3300917} & \makebox[20mm][c]{7.9312250} & \makebox[20mm][c]{0.4282} \\
\makebox[20mm][c]{178.3250708} & \makebox[20mm][c]{7.9285706} & \makebox[20mm][c]{0.4351} \\
\makebox[20mm][c]{178.3225958} & \makebox[20mm][c]{7.9280833} & \makebox[20mm][c]{0.4365} \\
\makebox[20mm][c]{178.3218542} & \makebox[20mm][c]{7.9186694} & \makebox[20mm][c]{0.4217} \\
\makebox[20mm][c]{178.3121833} & \makebox[20mm][c]{7.9211250} & \makebox[20mm][c]{0.4347} \\
\makebox[20mm][c]{178.3101667} & \makebox[20mm][c]{7.9197111} & \makebox[20mm][c]{0.5774} \\
\makebox[20mm][c]{178.2796542} & \makebox[20mm][c]{7.9109306} & \makebox[20mm][c]{0.5802} \\
\makebox[20mm][c]{178.2779500} & \makebox[20mm][c]{7.9084611} & \makebox[20mm][c]{0.5773} \\
\makebox[20mm][c]{178.2846042} & \makebox[20mm][c]{7.9058944} & \makebox[20mm][c]{0.5734} \\
\makebox[20mm][c]{178.2849167} & \makebox[20mm][c]{7.9123639} & \makebox[20mm][c]{0.1700} \\
\makebox[20mm][c]{178.2641167} & \makebox[20mm][c]{7.9062222} & \makebox[20mm][c]{0.5755} \\
\makebox[20mm][c]{178.2603708} & \makebox[20mm][c]{7.9093972} & \makebox[20mm][c]{0.4336} \\
\makebox[20mm][c]{178.2463792} & \makebox[20mm][c]{7.9077667} & \makebox[20mm][c]{0.1268} \\
\makebox[20mm][c]{178.2613708} & \makebox[20mm][c]{7.8867694} & \makebox[20mm][c]{0.4799} \\
\makebox[20mm][c]{178.2711708} & \makebox[20mm][c]{7.9110500} & \makebox[20mm][c]{0.5786} \\
\makebox[20mm][c]{178.2785208} & \makebox[20mm][c]{7.8987389} & \makebox[20mm][c]{1.2534} \\
\makebox[20mm][c]{178.2883917} & \makebox[20mm][c]{7.9218056} & \makebox[20mm][c]{0.4329} \\
\makebox[20mm][c]{178.3021125} & \makebox[20mm][c]{7.9252306} & \makebox[20mm][c]{0.4325} \\
\makebox[20mm][c]{178.3154292} & \makebox[20mm][c]{7.9194556} & \makebox[20mm][c]{0.2619} \\
\makebox[20mm][c]{178.3127833} & \makebox[20mm][c]{7.9034361} & \makebox[20mm][c]{0.7297} \\
\makebox[20mm][c]{178.3008250} & \makebox[20mm][c]{7.9060417} & \makebox[20mm][c]{0.4364} \\
\makebox[20mm][c]{178.2939583} & \makebox[20mm][c]{7.9323500} & \makebox[20mm][c]{0.3366} \\
\makebox[20mm][c]{178.3422250} & \makebox[20mm][c]{7.9405500} & \makebox[20mm][c]{0.4181} \\
\makebox[20mm][c]{178.3439542} & \makebox[20mm][c]{7.9285181} & \makebox[20mm][c]{0.4307} \\
\hline
\end{tabular}
\hspace{0.02\linewidth}
\begin{tabular}{ccc}
\hline
\hline
R.A. & Decl. & \textbf{$z_{\rm spec}$} \\
(J2000) & (J2000) & \\
\hline
\makebox[20mm][c]{178.3365708} & {7.9370028} & {1.5058} \\
\makebox[20mm][c]{178.3175167} & {7.9220472} & {0.4304} \\
\makebox[20mm][c]{178.2676083} & {7.9041028} & {0.4334} \\
\makebox[20mm][c]{178.3511083} & {7.9606750} & {0.4289} \\
\makebox[20mm][c]{178.3452000} & {7.9471889} & {0.4205} \\
\makebox[20mm][c]{178.3551833} & {7.9494556} & {0.4278} \\
\makebox[20mm][c]{178.3363292} & {7.9427611} & {0.4277} \\
\makebox[20mm][c]{178.3316375} & {7.9406917} & {0.4270} \\
\makebox[20mm][c]{178.3390917} & {7.9366958} & {0.4376} \\
\makebox[20mm][c]{178.3327375} & {7.9751472} & {0.5755} \\
\makebox[20mm][c]{178.3411375} & {7.9443153} & {0.6774} \\
\makebox[20mm][c]{178.3303625} & {7.9315297} & {0.4284, 0.4360} \\
\makebox[20mm][c]{178.3206208} & {7.9265528} & {0.4318} \\
\makebox[20mm][c]{178.3283167} & {7.9155806} & {0.4391} \\
\makebox[20mm][c]{178.3275750} & {7.9108583} & {0.4808} \\
\makebox[20mm][c]{178.3142833} & {7.9199028} & {0.4569} \\
\makebox[20mm][c]{178.3252292} & {7.9047333} & {0.4341} \\
\makebox[20mm][c]{178.3147583} & {7.9059444} & {0.4279} \\
\makebox[20mm][c]{178.2967333} & {7.9163528} & {0.4337, 0.5768} \\
\makebox[20mm][c]{178.2929000} & {7.9134056} & {0.5758, 0.5760} \\
\makebox[20mm][c]{178.3054583} & {7.9203556} & {0.7492} \\
\makebox[20mm][c]{178.2907125} & {7.9077889} & {0.2663} \\
\makebox[20mm][c]{178.2708958} & {7.9095056} & {0.1892} \\
\makebox[20mm][c]{178.2833292} & {7.9014797} & {0.4338} \\
\makebox[20mm][c]{178.2791667} & {7.8975444} & {0.4360} \\
\makebox[20mm][c]{178.3426667} & {7.9462833} & {0.2541} \\
\hline
\end{tabular}

\vspace{2mm}
\raggedright
\textbf{Notes.} The RA and DEC of the individual slits are shown, and for slits with two objects, the two corresponding redshifts are separated by commas.
\end{table*}

\begin{table*}[ht!]
\centering
\renewcommand{\arraystretch}{0.9}
\caption{Spectroscopic Redshifts from LDSS3 Longslit Spectroscopy}
\begin{tabular}{ccc}
\hline
\hline
R.A. & Decl. & \textbf{$z_{\rm spec}$} \\
(J2000) & (J2000) & \\
\hline
\makebox[20mm][c]{178.3451833} & {7.9281047} & {1.1663} \\
\makebox[20mm][c]{178.3424685} & {7.9294933} & {0.6286} \\
\makebox[20mm][c]{178.3304043} & {7.9366556} & {0.5725} \\
\makebox[20mm][c]{178.3272724} & {7.9377794} & {0.8009} \\
\makebox[20mm][c]{178.3298096} & {7.9326796} & {0.4309} \\
\makebox[20mm][c]{178.3285273} & {7.9292603} & {0.4254} \\
\hline
\end{tabular}
\label{tab:longslit_redshifts}
\end{table*}

\section{Photometric Measurements}
\label{app:photometry}

We report our measured photometric fluxes for all ground-based bands analyzed in this work in Table~\ref{tab:cigale_photometry}. Our photometry is extracted from measurements with \texttt{CARTA} (ALMA) and \texttt{GALFIT} (all other bands). A more detailed morphological analysis and photometry of the space-based imaging will be presented in a future paper.

\begin{table*}
\caption{Photometric Measurements}
\label{tab:cigale_photometry}
\renewcommand{\arraystretch}{1.2}
\centering
\begin{tabular}{lcc}
\hline
\hline
Bandpass & \targA1 & \targB3 \\[-1mm]
& ($\mu$Jy) & ($\mu$Jy) \\
\hline
$g$ & $4.16^{\pm 0.12}$ & $0.486^{\pm 0.134}$ \\
$r$ & $6.31^{\pm 0.36}$ & $0.668^{\pm 0.153}$ \\
$z$ & $11.4^{\pm 1.2}$ & $3.10^{\pm 0.59}$ \\
$J$ & $44.8^{\pm 3.7}$ & $10.7^{\pm 1.4}$ \\
$H$ & $49.71^{\pm 0.85}$ & $19.2^{\pm 2.9}$ \\
$K_s$ & $59.0^{\pm 1.7}$ & $31.99^{\pm 0.87}$ \\
${\rm I1}$ & $108.3^{\pm 2.3}$ & $90.9^{\pm 2.1}$ \\
\hline
\end{tabular}
\hspace{0.02\linewidth}
\begin{tabular}{lcc}
\hline
\hline
Bandpass & \targA1 & \targB3 \\[-1mm]
& ($\mu$Jy) & ($\mu$Jy) \\
\hline
${\rm I2}$ & $141.1^{\pm 2.6}$ & $145.5^{\pm 2.6}$ \\
${\rm PSW}$ & $5.1^{\pm 1.2} \times 10^{4}$ & $5.06^{\pm 0.72} \times 10^{4}$ \\
${\rm PMW}$ & $5.9^{\pm 1.2} \times 10^{4}$ & $5.86^{\pm 0.84} \times 10^{4}$ \\
${\rm PLW}$ & $2.99^{\pm 0.95} \times 10^{4}$ & $3.91^{\pm 0.56} \times 10^{4}$ \\
${\rm ALMA \, 7}$ & $2770^{\pm 620}$ & $4720^{\pm 660}$ \\
${\rm ALMA \, 6}$ & ... & $520^{\pm 190}$ \\
\hline
\end{tabular}

\vspace{2mm}
\raggedright
\textbf{Notes.} The measured flux values are reported in micro-Jansky ($\mu$Jy). Our SED fitting incorporates an additional 10\% uncertainty on top of the quoted flux uncertainties to account for systematics.
\end{table*}

\section{CIGALE SED Parameter Grids}
\label{app:cigale_grid}

We perform multi-wavelength SED fitting using the Code Investigating GALaxy Emission (CIGALE) \citep[v2025.0,][]{Boquien:2019}, employing the stellar population synthesis models of \citet{Charlot:2019}, dust attenuation prescriptions from \citet{Calzetti:2000}, and dust emission templates from \citet{Draine:2014}. AGN emission is modeled using the torus models of \citet{Stalevski:2012, Stalevski:2016}. Our parameter grids are detailed in Table~\ref{tab:cigale_params_A}. All fits include an additional $10\%$ systematic uncertainty on photometric measurements to account for calibration uncertainties.

\begin{table*}[ht!]
\renewcommand{\arraystretch}{1.2}
\caption{CIGALE SED Parameter Grids}
\label{tab:cigale_params_A}
\begin{center}
\centering
\begin{tabular}{llccc}
\hline
\hline
\makebox[40mm][l]{Module} & \makebox[41mm][l]{Parameter} & \makebox[19mm][l]{Symbol} & \makebox[34mm][l]{\targA} & \makebox[34mm][l]{\targB} \\
\hline
\makebox[40mm][l]{Star formation history;} & \makebox[41mm][l]{Stellar e-folding time} & \makebox[19mm][l]{$\tau_{\mathrm{star}}$} & \makebox[34mm][l]{500, 1000, 1500 Myr} & \makebox[34mm][l]{1000, 5000, 10000, 12500 Myr} \\
\makebox[40mm][l]{$\mathrm{SFR} \propto t \, \exp(-t/\tau)$;} & \makebox[41mm][l]{Stellar age} & \makebox[19mm][l]{$t_{\mathrm{star}}$} & \makebox[34mm][l]{200, 500, 1000 Myr} & \makebox[34mm][l]{2000, 2500, 3000, 5000 Myr} \\
\makebox[40mm][l]{\texttt{sfhdelayed}} & \makebox[41mm][l]{Mass fraction of late starburst} & \makebox[19mm][l]{$f_{\mathrm{burst}}$} & \makebox[34mm][l]{0.0} & \makebox[34mm][l]{0.0} \\
\hline
\makebox[40mm][l]{Stellar population; \texttt{cb19}} & \makebox[41mm][l]{Initial Mass Function} & \makebox[19mm][l]{...} & \makebox[34mm][l]{\citet{Chabrier:2003}} & \makebox[34mm][l]{\citet{Chabrier:2003}} \\
\makebox[40mm][l]{\citet{Charlot:2019}} & \makebox[41mm][l]{Stellar Metallicity} & \makebox[19mm][l]{$Z_{\rm star}$} & \makebox[34mm][l]{0.02} & \makebox[34mm][l]{0.02} \\
\hline
\makebox[40mm][l]{Nebular emission; \newline \texttt{nebular}} & \makebox[41mm][l]{Ionization parameter} & \makebox[19mm][l]{$\log U$} & \makebox[34mm][l]{$-2.0$} & \makebox[34mm][l]{$-2.0$} \\
& \makebox[41mm][l]{Gas metallicity} & \makebox[19mm][l]{$Z_{\mathrm{gas}}$} & \makebox[34mm][l]{0.02} & \makebox[34mm][l]{0.02} \\
& \makebox[41mm][l]{Electron density} & \makebox[19mm][l]{$n_e$} & \makebox[34mm][l]{100 cm$^{-3}$} & \makebox[34mm][l]{100 cm$^{-3}$} \\
\hline
\makebox[40mm][l]{Dust attenuation;} & \makebox[41mm][l]{Nebular line color excess} & \makebox[19mm][l]{$E(B-V)_{\mathrm{lines}}$} & \makebox[34mm][l]{0.5, 0.7, 0.9 mag} & \makebox[34mm][l]{0.0, 0.5, 0.9 mag} \\
\makebox[40mm][l]{\texttt{dustatt\_modified} \newline \texttt{\_starburst}} & \makebox[41mm][l]{Reduction factor} & \makebox[19mm][l]{$f$} & \makebox[34mm][l]{0.5, 0.7, 0.9 mag} & \makebox[34mm][l]{0.0, 0.5, 0.9 mag} \\
\makebox[40mm][l]{\citet{Calzetti:2000}} & \makebox[41mm][l]{Power-law modification} & \makebox[19mm][l]{$\delta$} & \makebox[34mm][l]{$-1.0, -0.5, 0.0$} & \makebox[34mm][l]{$-0.5, 0.0, 0.5$} \\
\hline
\makebox[40mm][l]{Dust emission; \texttt{dl2014}} & \makebox[41mm][l]{PAH mass fraction} & \makebox[19mm][l]{$q_{\mathrm{PAH}}$} & \makebox[34mm][l]{0.47, 1.0, 2.0} & \makebox[34mm][l]{0.47, 1.0, 2.0} \\
\makebox[40mm][l]{\citet{Draine:2014}} & \makebox[41mm][l]{Minimum radiation field} & \makebox[19mm][l]{$U_{\mathrm{min}}$} & \makebox[34mm][l]{10.0, 15.0, 20.0} & \makebox[34mm][l]{8.0, 10.0, 15.0} \\
& \makebox[41mm][l]{Slope in $dU/dM \propto U^\alpha$} & \makebox[19mm][l]{$\alpha$} & \makebox[34mm][l]{1.0, 2.0, 3.0} & \makebox[34mm][l]{2.8, 2.9, 3.0} \\
& \makebox[41mm][l]{Dust fraction (SF vs. diffuse)} & \makebox[19mm][l]{$\gamma$} & \makebox[34mm][l]{0.0, 0.1, 0.2} & \makebox[34mm][l]{0.9, 1.0} \\
\hline
\makebox[40mm][l]{AGN torus; \newline \texttt{SKIRTOR2016}} & \makebox[41mm][l]{Edge-on opt. depth at 9.7 $\mu$m} & \makebox[19mm][l]{$\tau_{9.7}$} & \makebox[34mm][l]{7, 9, 11} & \makebox[34mm][l]{3, 5, 7} \\
\makebox[40mm][l]{\citet{Stalevski:2012, Stalevski:2016}} & \makebox[41mm][l]{Radial density gradient} & \makebox[19mm][l]{$\mathrm{pl}$} & \makebox[34mm][l]{1.0} & \makebox[34mm][l]{1.0} \\
& \makebox[41mm][l]{Polar density gradient} & \makebox[19mm][l]{$q$} & \makebox[34mm][l]{1.0} & \makebox[34mm][l]{1.0} \\
& \makebox[41mm][l]{Half-opening angle} & \makebox[19mm][l]{$\mathrm{oa}$} & \makebox[34mm][l]{$40^\circ$} & \makebox[34mm][l]{$40^\circ$}\\
& \makebox[41mm][l]{Outer/inner radius ratio} & \makebox[19mm][l]{$R$} & \makebox[34mm][l]{20} & \makebox[34mm][l]{20} \\
& \makebox[41mm][l]{Clump mass fraction} & \makebox[19mm][l]{$M_{\mathrm{cl}}$} & \makebox[34mm][l]{0.97} & \makebox[34mm][l]{0.97}\\
& \makebox[41mm][l]{Inclination angle} & \makebox[19mm][l]{$i$} & \makebox[34mm][l]{$30^{\circ}$} & \makebox[34mm][l]{$60^\circ$} \\
& \makebox[41mm][l]{Disk spectrum type} & \makebox[19mm][l]{...} & \makebox[34mm][l]{\citet{Schartmann:2005}} & \makebox[34mm][l]{\citet{Schartmann:2005}} \\
& \makebox[41mm][l]{Disk deviation} & \makebox[19mm][l]{$\delta$} & \makebox[34mm][l]{$-1.5, -1.0, -0.5, 0.0, 0.5$} & \makebox[34mm][l]{$-1.5, -1.0, -0.5, 0.0, 0.5$} \\
& \makebox[41mm][l]{AGN fraction} & \makebox[19mm][l]{$\mathrm{frac}_{\mathrm{AGN}}$} & \makebox[34mm][l]{0.1, 0.15, 0.2, 0.25, 0.3 \tnote{a}$^{\rm a}$} & \makebox[34mm][l]{5e-4, 0.002, 0.005, 0.0125, 0.05} \\
& \makebox[41mm][l]{$\mathrm{frac}_{\mathrm{AGN}}$ wavelength range} & \makebox[19mm][l]{$\lambda_{\mathrm{min}}/\lambda_{\mathrm{max}}$} & \makebox[34mm][l]{$1.13$/$2.3~\mu{\rm m}$ \tnote{b}$^{\rm b}$}  & \makebox[34mm][l]{$0/0$\tnote{c}$^{\rm c}$}  \\
& \makebox[41mm][l]{Polar dust extinction law} & \makebox[19mm][l]{...} & \makebox[34mm][l]{SMC} & \makebox[34mm][l]{SMC} \\
& \makebox[41mm][l]{Polar dust $E(B-V)$} & \makebox[19mm][l]{$E(B-V)_{\mathrm{polar}}$} & \makebox[34mm][l]{0.03, 0.1, 0.3 mag} & \makebox[34mm][l]{0.03, 0.1, 0.3 mag} \\
& \makebox[41mm][l]{Polar dust temperature} & \makebox[19mm][l]{$T_{\mathrm{polar}}$} & \makebox[34mm][l]{10, 20, 100 K} & \makebox[34mm][l]{10, 20, 100 K} \\
\hline
\end{tabular}

\vspace{2mm}
\raggedright
\textbf{Notes.} Parameters represent the grid explored during model fitting. For parameters not listed, default CIGALE values are adopted.

$^{\rm a}$ $\mathrm{frac}_{\mathrm{AGN}}$ grid restricted for \targA\ based on measured photometric AGN fractions in the $J$, $H$, and $K_s$ bands for the modeled PSF and S\'ersic components.

$^{\rm b}$ Wavelength range covered by $J$-$H$-$K_s$ bands.

$^{\rm c}$ For \targB, the AGN fraction is integrated across the full SED.
\end{center}
\end{table*}

\section{Lens Model Parameters}
\label{app:cornerplot}

\begin{figure*}[ht!]
    \centering
    \includegraphics[width=\linewidth]{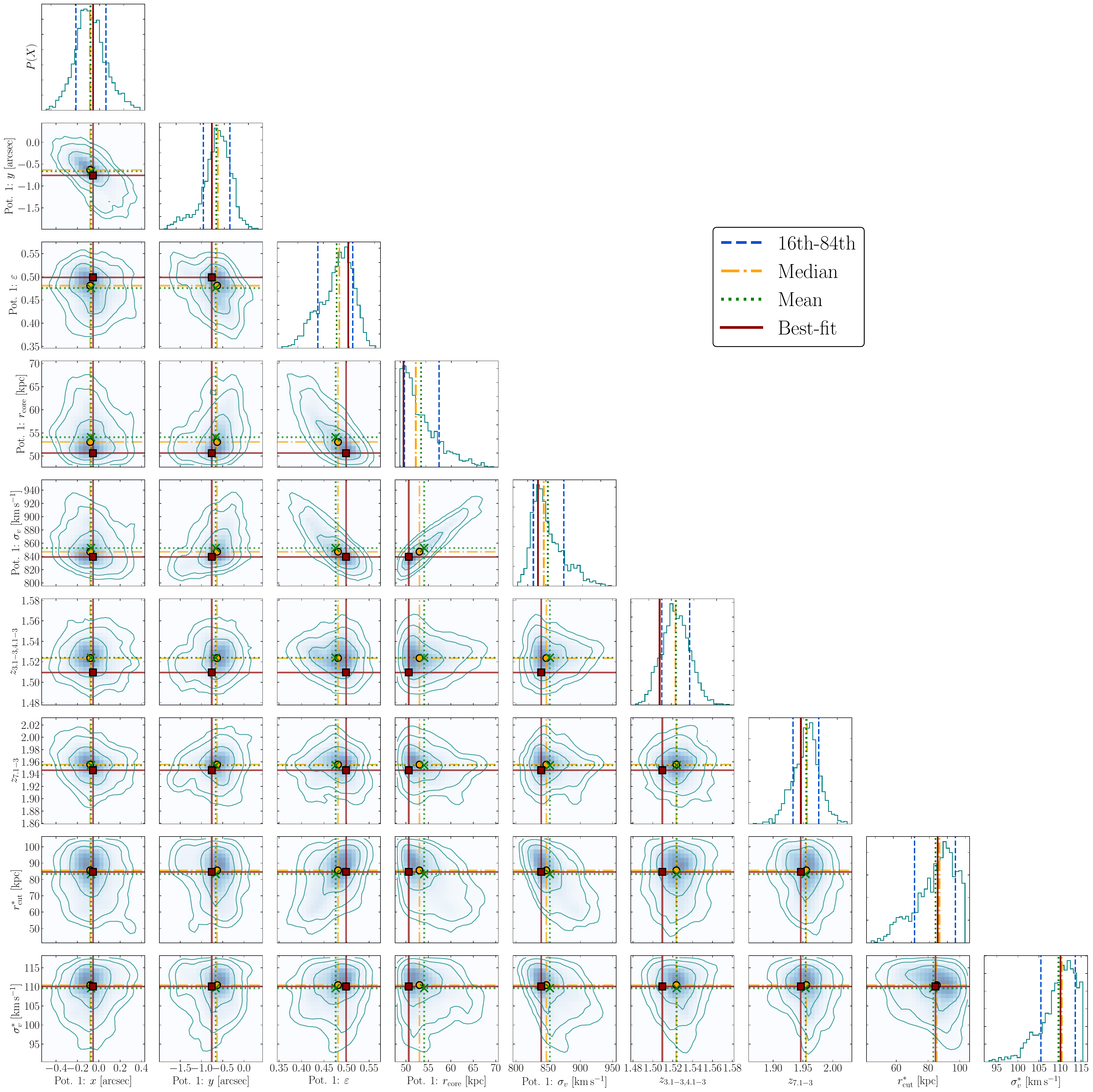}
    \caption{Corner plot of the ground-based lens model showing posterior distributions of nine of the parameters for the 2000 model samples derived from 10 Markov chains and 200 iterations.
    We indicate the best-fit model values (red) and median values (yellow) of the posterior distributions and the 1$\sigma$ confidence interval thereof. 
    The ground-based lens model includes 30 constraints and 29 free parameters and yields a reduced $\chi_{\nu}^2 \sim 0.95$.}
    \label{fig:cornerplot}
\end{figure*}

We present the lens model parameter constraints derived from our ground-based imaging data (``ground-based'' model). Our parametric lens model comprises six individually, but simultaneously, optimized potentials, of which the sixth (``Potential 6'') is centered on the secondary cluster. This model is optimized through \texttt{lenstool}'s Markov Chain Monte Carlo algorithm with 10 independent chains and 200 iterations each. Fig.~\ref{fig:cornerplot} displays the posterior distributions of nine key parameters. The best-fit parameter values and their confidence intervals (16th–84th percentiles) are listed in Table~\ref{tab:lensmodelparams}, providing an accounting of all 29 free parameters in the model.

\begin{table*}[ht!]
\caption{Ground-Based Lens Model Best-Fit and Posterior Parameter Values}
\label{tab:lensmodelparams}
\centering
\begin{tabular}{lcc}
\hline
\hline
\textbf{Parameter} & \textbf{Best} & \makebox[36mm][c]{\textbf{Median, 16th–84th \%}} \\
\hline
\multicolumn{3}{c}{\textbf{Potential O1}} \\
\hline
$x$ [arcsec] & -0.054 & \makebox[36mm][c]{-0.080$^{+0.13}_{-0.11}$} \\
$y$ [arcsec] & -0.76 & \makebox[36mm][c]{-0.63$^{+0.25}_{-0.30}$} \\
$\varepsilon$ & 0.50 & \makebox[36mm][c]{0.48$^{+0.03}_{-0.04}$} \\
$\theta$ & 117.7 & \makebox[36mm][c]{117.4$^{+1.1}_{-1.0}$} \\
$r_{\rm core}$ [kpc] & 51 & \makebox[36mm][c]{53$^{+4.6}_{-2.2}$} \\
$\sigma_v$ [$\rm{km} \, {\rm s}^{-1}$] & 839 & \makebox[36mm][c]{847$^{+27}_{-14}$} \\
\hline
\multicolumn{3}{c}{\textbf{Potential O2}} \\
\hline
$\varepsilon$ & 0.34 & \makebox[36mm][c]{0.40$^{+0.14}_{-0.24}$} \\
$\theta$ & 98.6 & \makebox[36mm][c]{79.6$^{+20.9}_{-23.0}$} \\
$r_{\rm core}$ [kpc] & 4.7 & \makebox[36mm][c]{3.5$^{+3.5}_{-2.1}$} \\
$r_{\rm cut}$ [kpc] & 5.4 & \makebox[36mm][c]{9.1$^{+8.5}_{-3.7}$} \\
$\sigma_v$ [$\rm{km} \, {\rm s}^{-1}$] & 181 & \makebox[36mm][c]{165$^{+24}_{-32}$} \\
\hline
\multicolumn{3}{c}{\textbf{Potential O3}} \\
\hline
$\varepsilon$ & 0.12 & \makebox[36mm][c]{0.34$^{+0.18}_{-0.22}$} \\
$r_{\rm core}$ [kpc] & 19 & \makebox[36mm][c]{33$^{+11}_{-17}$} \\
$r_{\rm cut}$ [kpc] & 76 & \makebox[36mm][c]{64$^{+20}_{-21}$} \\
$\sigma_v$ [$\rm{km} \, {\rm s}^{-1}$] & 84.0 & \makebox[36mm][c]{105$^{+52}_{-49}$} \\
\hline
\multicolumn{3}{c}{\textbf{Potential O4}} \\
\hline
$\varepsilon$ & 0.06 & \makebox[36mm][c]{0.29$^{+0.21}_{-0.19}$} \\
$r_{\rm core}$ [kpc] & 6.9 & \makebox[36mm][c]{23$^{+14}_{-16}$} \\
$r_{\rm cut}$ [kpc] & 51 & \makebox[36mm][c]{40$^{+34}_{-15}$} \\
$\sigma_v$ [$\rm{km} \, {\rm s}^{-1}$] & 56.4 & \makebox[36mm][c]{61.8$^{+23}_{-27}$} \\
\hline
\end{tabular}
\hspace{0.01\linewidth}
\begin{tabular}{lcc}
\hline
\hline
\textbf{Parameter} & \textbf{Best} & \textbf{Median, 16th–84th \%} \\
\hline
\multicolumn{3}{c}{\textbf{Potential O5}} \\
\hline
$\varepsilon$ & 0.52 & \makebox[36mm][c]{0.44$^{+0.12}_{-0.25}$} \\
$\theta$ & 173.4 & \makebox[36mm][c]{$147.4^{+24.8}_{-138.4}$} \\
$r_{\rm core}$ [kpc] & 7.5 & \makebox[36mm][c]{7.7$^{+4.3}_{-2.9}$} \\
$r_{\rm cut}$ [kpc] & 76 & \makebox[36mm][c]{80$^{+13}_{-12}$} \\
$\sigma_v$ [$\rm{km} \, {\rm s}^{-1}$] & 231 & \makebox[36mm][c]{221$^{+19}_{-20}$} \\
\hline
\multicolumn{3}{c}{\textbf{Potential O6}} \\
\hline
$\sigma_v$ [$\rm{km} \, {\rm s}^{-1}$] & 740 & \makebox[36mm][c]{757$^{+27}_{-30}$} \\
\hline
\multicolumn{3}{c}{\textbf{Redshifts}} \\
\hline
$z_{3.1-3, 4.1-3}$ & 1.51 & \makebox[36mm][c]{1.52$^{+0.01}_{-0.01}$} \\
$z_{7.1-3}$ & 1.95 & \makebox[36mm][c]{1.96$^{+0.02}_{-0.02}$} \\
\hline
\multicolumn{3}{c}{\textbf{Potfile0}} \\
\hline
$r_{\rm cut}^{*}$ [kpc] & 85 & \makebox[36mm][c]{86$^{+8.8}_{-14}$} \\
$\sigma_{v}^{*}$ [$\rm{km} \, {\rm s}^{-1}$] & 110 & \makebox[36mm][c]{110$^{+3}_{-5}$} \\
\hline
\end{tabular}

\vspace{2mm}
\raggedright
\textbf{Notes.} $x$ and $y$ denote offsets in arcseconds from the relative coordinate $({\rm RA}, {\rm DEC}) = (178.3302510, \, 7.9325182)$ degrees. \texttt{Potfile0} parameters define the scaling relations used to collectively constrain cluster halo potentials, which are not used for the individually optimized halo potentials (Potentials O1-6). Potential O6 corresponds to the secondary cluster ($z=0.5766$) which is about $\sim 3'$ away, but it is assigned the same redshift as the primary cluster ($z=0.4301$) despite not being physically associated with it.
\end{table*}



\end{document}